\documentclass[lettersize,journal]{IEEEtran}
\usepackage{float}
\usepackage{siunitx}
\usepackage{xcolor}
\usepackage[hyphens]{url}
\usepackage{algorithm}
\usepackage[noend]{algpseudocode}
\usepackage{dsfont}
\usepackage{bm}
\usepackage{amsmath,amssymb,amsfonts}
\usepackage{array}
\usepackage[caption=false,font=normalsize,labelfont=sf,textfont=sf]{subfig}
\usepackage{textcomp}
\usepackage{stfloats}
\usepackage{verbatim}
\usepackage{graphicx}
\usepackage{cite}
\usepackage{pgfplots}
\usepackage{hyperref}
\usepackage{refcount}

\pgfplotsset{compat=1.18}
\algrenewcommand\algorithmicrequire{\textbf{Inputs:}}
\algrenewcommand\algorithmicensure{\textbf{Outputs:}}
\algrenewcommand\algorithmiccomment[1]{\hfill\(\triangleright\)~#1}

\begin{document}

\title{RIFT: Entropy-Optimised Fractional Wavelet Constellations for Ideal Time–Frequency Estimation}

\author{James M. Cozens,~\IEEEmembership{Member,~IEEE,} Simon J. Godsill,~\IEEEmembership{Fellow,~IEEE}
\thanks{This work has been submitted to the IEEE for possible publication. Copyright may be transferred without notice, after which this version may no longer be accessible.}
\thanks{J. Cozens and S. Godsill are with the Engineering Department, University of Cambridge, CB2 1PZ Cambridge, U.K., E-mail: (jmc257@cam.ac.uk; sjg30@cam.ac.uk).
\textit{(Corresponding author: James M. Cozens)}}%
\thanks{The work of James M. Cozens was supported by EPSRC Doctoral Training Partnership (DTP) under Grant 2738835}}

%\markboth{ } {Shell \MakeLowercase{\textit{et al.}}: A Sample Article Using IEEEtran.cls for IEEE Journals}

\IEEEpubid{ }

\maketitle

\begin{abstract}
We introduce a new method for estimating the Ideal Time--Frequency Representation (ITFR) of complex nonstationary signals. The Reconstructive Ideal Fractional Transform (RIFT) computes a constellation of Continuous Fractional Wavelet Transforms (CFWTs) aligned to different local time–frequency curvatures. This constellation is combined into a single optimised time--frequency energy representation via a localised entropy-based sparsity measure, designed to resolve auto-terms and attenuate cross-terms. Finally, a positivity-constrained Lucy–Richardson deconvolution with total-variation regularisation is applied to estimate the ITFR, achieving auto-term resolution comparable to that of the Wigner–Ville Distribution (WVD), yielding the high-resolution RIFT representation. The required Cohen's class convolutional kernels are fully derived in the paper for the chosen CFWT constellations. Additionally, the optimisation yields an Instantaneous Phase Direction (IPD) field, which allows the localised curvature in speech or music extracts to be visualised and utilised within a Kalman tracking scheme, enabling the extraction of signal component trajectories and the construction of the Spline-RIFT variant. Evaluation on synthetic and real-world signals demonstrates the algorithm's ability to effectively suppress cross-terms and achieve superior time--frequency precision relative to competing methods. This advance holds significant potential for a wide range of applications requiring high-resolution cross-term-free time--frequency analysis.
\end{abstract}

\begin{IEEEkeywords}
Reconstructive Ideal Fractional Transform (RIFT), Ideal Time--Frequency Representation (ITFR), Fractional Wavelet Transform (FWT), Wigner--Ville Distribution (WVD), Entropic Filtering.
\end{IEEEkeywords}

\vspace{-3mm}\section{Introduction}

\vspace{-2mm}\subsection{Motivation}
\IEEEPARstart{T}{ime-frequency} (T--F) analysis is used extensively across many applications, owing to its ability to provide insights into nonstationary processes with time-varying frequency content \cite{Boashash2016, applications}. For example, in the biomedical industry, T--F methods utilised on physiological signals such as Electroencephalograms (EEGs) and Electrocardiograms (ECGs) aid in the diagnosis of neurological and cardiac conditions \cite{Zhang2023, Alazrai2018, Hussein2018, Escriva2018}. In the automotive and aerospace sectors, vibration and acoustic signal analysis facilitates fault detection and predictive maintenance \cite{Feng2013}. In the financial sector, T--F techniques applied to market data enable the inference of market volatility, trend identification, and anomaly detection, essential elements for risk management and algorithmic trading \cite{finance1, finance2}. Moreover, T--F analysis is a key component of the audio processing field for tasks such as speech and music classification, transcription, segmentation, visualisation, and restoration, in addition to facilitating both classical and deep-learning-based generative models for music and audio synthesis \cite{Cozens, classification, transcription, generation}.

\vspace{-2mm}\subsection{Overview of Related Work}
\label{overview}
\IEEEpubidadjcol
Numerous methods exist in the literature to provide application-appropriate T--F representations, such as spectrograms \cite{Cohen1989}, wavelet transforms \cite{wavelet}, Wigner--Ville Distribution functions \cite{wvd}, Cohen's class distributions \cite{Cohen1989}, and spectral reassignment algorithms \cite{reassignment}. In general, all methods aim to provide a compromise between (1) T--F resolution, (2) cross-term suppression, and (3) computational feasibility \cite{Boashash2016}.

The Wigner--Ville Distribution (WVD) \cite{wvd, ville} is a bilinear T--F representation known to provide high T--F resolution \cite{Flandrin1999, Cohen1989, Boashash1992} (see Section \ref{sec:wvd} for properties). However, by construction, the WVD is known to exhibit strong cross-term interference between components \cite{wvdinter}, a consequence of ensuring instantaneous power conservation. This interference renders the direct application of WVDs unpopular for many practical applications \cite{Boashash2016}. Cohen's class of bilinear T--F representations provides a unified framework of bilinear (quadratic) T--F methods \cite{Cohen1989}. By applying a smoothing kernel (Cohen's class kernel function) to the WVD, cross-term interference can be partially suppressed, such as within the Gabor Wigner Transform \cite{GaborWignerTransform} and the Choi-Williams Distribution \cite{ChoiWilliamsDistribution}, albeit at the expense of decreased T--F resolution \cite{Boashash2016}. Polynomial Wigner--Ville Distributions (PWVDs) provide optimal T--F resolution for high-order frequency modulated signals \cite{PWVD}. However, due to the direct convolutional relationship between bilinear representations (Cohen's class distributions), increased cross-term presence and computational complexity, PWVDs \cite{PWVD} are typically less favoured compared to WVDs in the context of high-resolution T--F analysis \cite{Boashash2016, Cohen1989}.

Optimisation of the WVD is typically performed by selecting an appropriate Cohen's class kernel \cite{Boashash2016}. Early methods focused on the development of globally applied kernels, such as the Gabor Wigner Transform \cite{GaborWignerTransform}, the Choi Williams Distribution \cite{ChoiWilliamsDistribution}, and the cone-shape distribution function \cite{cone}. While these methods considerably reduce cross-term interference, their global nature often results in suboptimal performance regardless of the kernel parameters selected \cite{jones2}, due to time and frequency-varying signal characteristics. However, certain classes of globally applied Cohen's class kernels have desirable properties; in particular, the spectrogram can be derived as a Cohen's class distribution, as shown in Section \ref{time-frequency relationships}. Due to the spectrogram's positivity, the corresponding Cohen's class kernel significantly suppresses cross terms resulting from oscillatory negative components \cite{Nuttall1988}, hence its popularity in T--F analysis \cite{Boashash2016, Nuttall1988}. 

Many methods have been proposed to account for time-varying signal characteristics by developing adaptive Cohen's class kernels, typically optimised through a T--F metric, such as energy concentration \cite{Jones, jones2, Awal2017, awal2, Mohammadi} (for instance, the Energy Concentration Measure (ECM) \cite{Awal2017}), Entropy (Shannon and Rényi Entropy) \cite{Saulig2016}, and Ratio of Norms \cite{Abed}. For instance, a method of inferring an optimised time-varying Gaussian Cohen's class kernel for an input signal has been proposed through maximisation of the energy concentration in the WVD plane with respect to the kernel variance along a generic axis \cite{Jones, jones2}. Similarly, adaptive kernel optimisation has been performed using Rényi Entropy \cite{Saulig2017, Saulig2016}. However, frequency-varying characteristics are typically not sufficiently accounted for, as optimisation is performed with respect to each time step \cite{Awal2017}. Conversely, the S-Transform provides a frequency-dependent (time-fixed) window function, offering a more suitable general T--F resolution compromise by adjusting the window width in relation to the signal's wavelengths \cite{Stockwell}.

To address time and frequency-varying signal curvature, methods have been proposed that maximise energy concentration through determining the optimal window function characteristics, such as Short-time Fractional Fourier Transform angles and window widths, for segmented sections of the T--F plane \cite{Awal2017, awal2}, additionally exploiting the positivity of the transform. The employment of Short-time Fractional Fourier (or similarly wavelet) Transforms is especially desirable in the context of time-varying signal curvature compared to their standard counterparts (for instance Short-Time Fourier Transforms), given their ability to align with signal trajectories and thus minimise cross term interference appropriately \cite{Awal2017}. Likewise, local and global optimisation has been performed for directional Gaussian filters in the T--F plane \cite{Mohammadi}. While entropic or energy-based optimisation schemes considerably reduce cross-term interference compared to the WVD, the discrete nature of segmentation schemes and optimised kernel assignments result in abrupt T--F transition characteristics and suboptimal spectrogram allocation when the chosen kernels, segments, or window widths do not align with the desired signal trajectories \cite{Awal2017, Mohammadi}. 

Similarly, as with the case for all Cohen's class methods, the resolution is fundamentally limited by the WVD of the window function being employed \cite{Cohen1989}, as illustrated in Section \ref{time-frequency relationships}. Several methods have been proposed to overcome these limitations; spectral reassignment is a popular algorithm that reallocates energy to the estimated instantaneous frequencies and times, enhancing both time and frequency localisation \cite{reassignment}. While it improves the representation, the algorithm is particularly susceptible to noise amplification and may struggle with overlapping spectral components (cross-terms) due to inaccuracies in instantaneous frequency estimates \cite{Flandrin2002}. Synchrosqueezing techniques \cite{Thakur2013} enhance frequency resolution by reassigning T--F coefficients based on instantaneous frequency estimates; however, the algorithm is similarly sensitive to noise and offers only minimal improvement in time resolution. Multi-resolution methods also exist that enhance time–frequency resolution by taking the geometric mean of representations obtained with different window functions \cite{ref:1}, or by employing PLCA across varying window lengths \cite{ref:2}. Whilst these approaches improve resolution and partially suppress cross-terms through the smoothing effect of Cohen's class kernels, they likewise remain susceptible to spectral leakage across overlapping components, and residual cross-terms persist \cite{ref:1,ref:2}.

Alternatively, decomposition techniques such as Empirical Mode Decomposition (EMD) \cite{Huang1998} and Variational Mode Decomposition (VMD) \cite{VMD} aim to handle non-linear and nonstationary signals by extracting intrinsic mode functions. However, EMD is susceptible to mode mixing and sensitivity to noise, whereas VMD, despite its more rigorous theoretical foundation \cite{VMD}, requires predefining the number of modes, can struggle with non-sinusoidal transients and silent intervals, and may still exhibit mode mixing, thus limiting its applicability for the types of generalised complex multi-component signals under consideration.

Likewise, matching pursuit algorithms that employ a dictionary of T--F atoms have been proposed, due to their ability to provide signal trajectories with optimised T--F resolution, if perfect T--F atom matches exist \cite{Mallat, Zhang}, eliminating the Cohen's class smoothing issue. However, matching pursuit requires a considerable discrete set of basis functions (like chirplets) as a dictionary, making optimisation through the Orthogonal Matching Pursuit (OMP) algorithm often computationally infeasible \cite{Mallat}. Additionally, prior knowledge of the number of components is required due to the residual fitting procedure \cite{Zhang}, making it increasingly impractical for highly complex multi-component signals. 

Some papers have alternatively proposed framing the problem as a direct deconvolution task, retrieving the desired high-resolution T--F representation from the Cohen's class distributions using the known Cohen's class kernels \cite{deconv_TF_multiple, lucy-richardson TF}. In particular, the iterative Lucy--Richardson Deconvolution Algorithm \cite{LR_TV} has been applied to spectrogram representations \cite{lucy-richardson TF}. Lucy--Richardson Deconvolution, and in particular the form with total variation regularisation \cite{LR_TV}, is especially promising given the incorporated statistical noise modelling, regularisation, and enforced positivity constraint.

\vspace{-2mm}\subsection{Overview of the Proposed Methodology}

\vspace{-0mm}\begin{figure*}
    \centering
    \subfloat{{\includegraphics[width=0.95\textwidth]{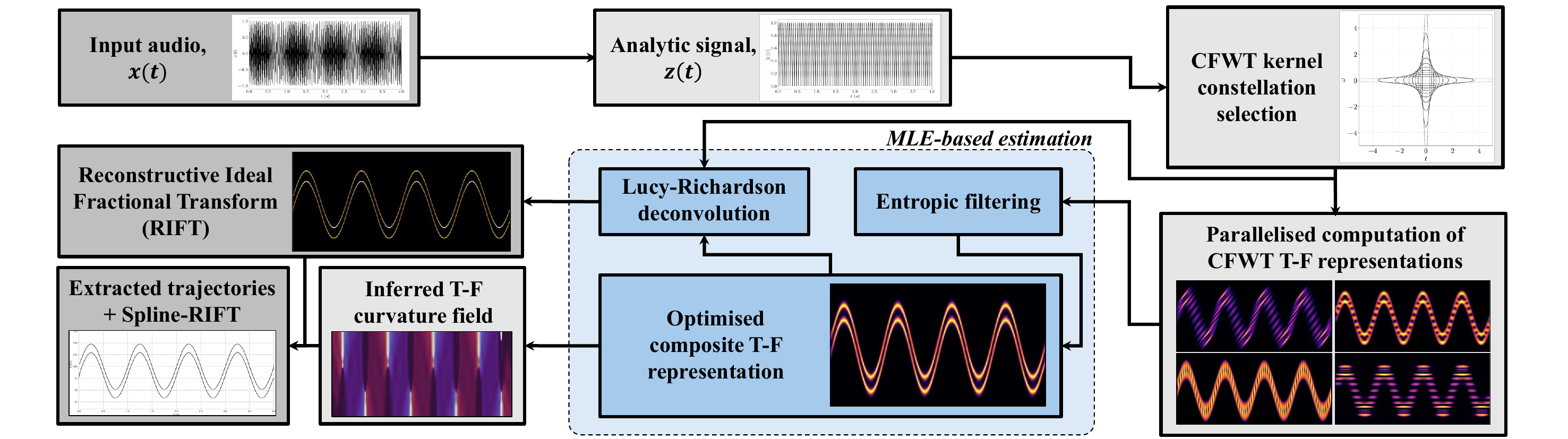}}}%=
    \vspace{-2mm}\caption{Overview of the proposed process, graphically illustrated for a parallel pair of sinusoidal frequency-modulated signals, $x_1(t)$ (Eq.~\eqref{x_1}). Each block corresponds to a stage in Algorithm~\ref{alg:rift}, where the workflow is formalised step-by-step.\vspace{-4mm}}%
    \label{fig:process}
\end{figure*}

Given the limitations discussed in the previous section, this paper proposes the Reconstructive Ideal Fractional Transform (RIFT), a transform that provides an estimated reconstruction of the cross-term-free Ideal Time--Frequency Representation (ITFR) with T--F resolution comparable to that of the WVD (see Section \ref{sec:itfr} for the mathematical form). First, the algorithm exploits the positivity and reduced oscillatory cross-term interference of wavelet-based methods, introducing a proposed parametric form for the Continuous Fractional Wavelet Transform (CFWT). A constellation of representative CFWTs is constructed, forming a set of Cohen's class kernels in the T--F plane. Unlike conventional Cohen's class methods, a probabilistic model is developed that estimates the signal ITFR from a weighted set of the CFWT representations, enabling signal components to be modelled as a combination of multiple CFWT representations. The model weights T--F representations by a proposed localised normalised entropy measure, which provides a continuous weighting distribution over the whole T--F plane to isolate auto-terms and attenuate cross-terms (interference), ensuring abruptly varying T--F characteristics are appropriately accounted for. Within the probabilistic model, we employ a spatially varying, positivity-constrained Lucy–Richardson deconvolution with total-variation regularisation to estimate the expected ITFR, promoting WVD-like resolution and providing robust noise mitigation, particularly for closely spaced or overlapping spectral components. The optimisation process additionally generates an Instantaneous Phase Direction (IPD) field, allowing for local curvature visualisation and use in Kalman tracking for the extraction of signal components. The overall workflow of the proposed method is summarised in Fig.~\ref{fig:process}, with detailed execution steps provided in Algorithm~\ref{alg:rift}.

\vspace{-2mm}\section{Preliminaries}

This section establishes the background used in Section \ref{sec:method}: we define the Ideal Time–Frequency Representation (ITFR), review the Wigner–Ville Distribution (WVD), and summarise Cohen’s class methods and their relation to the CFWT.

\vspace{-2mm}\subsection{Ideal Time--Frequency Representation (ITFR)}
\label{sec:itfr}

The concept of an Ideal Time--Frequency Representation (ITFR) for a signal is explored in several papers \cite{Mohammad, Boashash2016}. Let $z\left(t\right)$ be a multi-component complex signal \cite{Mohammad}:

\vspace{-1.85mm}  \small \begin{align}
\label{signal}
    z\left(t\right) &= \sum_{p=1}^P \mathds{1}_p\left(t\right)A_p(t)\exp{\left(j\phi_{p, 1} + j \int_{t_{p, 1}}^{t} \omega_p\left(\tau\right) \, d \tau\right)} \notag \\
    &= \sum_{p=1}^P \mathds{1}_p\left(t\right)A_p(t)\exp{\left(j \phi_{p}\left(t\right)\right)}
\end{align}\normalsize\vspace{-1.85mm}

\noindent where $A_p(t)$ and $\omega_p(t)$ are the amplitudes and instantaneous angular frequencies for component $p$ ($1 \le p \le P$) respectively, and $\phi_p\left(t\right)$ are the phases for each component. Here, $\phi_{p,1} \triangleq \phi_p(t_{p,1})$ denotes the initial phase at the onset $t_{p,1}$ of component $p$, so that $\phi_p(t) = \phi_{p,1} + \int_{t_{p,1}}^{t} \omega_p(\tau)\, d\tau$ and $\frac{d}{dt}\phi_p(t) = \omega_p(t)$. The term $\mathds{1}_p(t)$ is an indicator function that confines the component to its active interval between its onset $t_{p, 1}$ and offset $t_{p, 2}$:

\begingroup
\setlength{\abovedisplayskip}{-6pt}
\setlength{\belowdisplayskip}{3pt}
\vspace{-1.85mm}  \small  \begin{align}
\mathds{1}_{p}(t) = 
\begin{cases} 
1 & \text{if } t_{p, 1} \leq t < t_{p, 2} \\
0 & \text{otherwise}
\end{cases}.
\end{align}\normalsize\vspace{-0.85mm}
\endgroup

Conceptually, the ITFR consists of P trajectories with perfect (delta function) T--F resolution, with no cross terms \cite{Mohammad, Boashash2016}. As T--F analysis is typically concerned with energy distribution, the ITFR will refer to its squared modulus magnitude, which represents the ideal per-component energy distribution in the T--F domain. A suitable definition for the ITFR of signal $z\left(t\right)$ is thus:

\vspace{-3.85mm}  \small  \begin{align}
\label{ITFR_Eq}
    \text{ITFR}\left(\omega, \, t\right) = \sum_{p=1}^P \mathds{1}_p\left(t\right)A_p^2(t)\delta\left(\omega - \omega_p(t)\right).
\end{align}\normalsize\vspace{-2.85mm}

For example, let $x_1\left(t\right)$ be a parallel pair of sinusoidal frequency modulated signals:

\vspace{-2.85mm}  \small  \begin{align}
\label{x_1}
    x_1(t) &= \sin{\left(2 \pi \int_0^{t}100 + 50 \sin{\left(2 \pi \tau\right)} \, d \tau\right)} 
    \notag\\&\hspace{5mm}+ \sin{\left(2 \pi \int_0^{t}80 + 50 \sin{\left(2 \pi \tau\right)} \, d \tau\right)}.
\end{align} \normalsize\vspace{-2.85mm}

The ITFR can be subsequently evaluated employing Eq.~\eqref{ITFR_Eq}:

\vspace{-2.85mm}  \small  \begin{align}
    \text{ITFR}\left(\omega, \, t\right)
    &= \delta\left(\omega - \tfrac{\partial }{\partial t}\phi_1(t)\right) + \delta\left(\omega - \tfrac{\partial }{\partial t}\phi_2(t)\right) \notag\\
    &= \delta\left(\omega - 100 - 50 \sin{\left(2 \pi t\right)}\right) 
    \notag\\& \hspace{10mm} + \delta\left(\omega - 80 - 50 \sin{\left(2 \pi t\right)}\right).
\end{align}\normalsize\vspace{-4.85mm}

Note that whilst the exact form of the ITFR may be mathematically defined in the continuous T--F plane, discontinuities resulting from the use of point masses in a discretised version of the T--F plane may result in misrepresentative T--F representations \cite{Mohammad}. Thus, an extension to the ITFR could involve formulating an Ideal Spline Time--Frequency Representation \cite{Mohammad}. A suitable spline rasterisation scheme is the Xiaolin Wu's antialiased line algorithm \cite{wu1991aa}, which provides a continuous T--F trajectory in the discretised T--F plane; for the remainder of the paper, the target ITFR will be modelled in this form.

\vspace{-3mm}\subsection{Wigner--Ville Distribution properties}
\label{sec:wvd}

The Wigner--Ville Distribution (WVD) \cite{wvd, ville} is a bilinear T--F representation known to provide optimal T--F resolution for linear frequency-modulated signals due to its ability to achieve maximum energy concentration along the instantaneous frequency trajectory \cite{Flandrin1999, Cohen1989, Boashash1992}. It is defined as:

\vspace{-2.85mm}  \small  \begin{align}
    WVD_z\left(\omega, \, t\right) = \frac{1}{2\pi}\int_{\mathbb{R}} z\left(t + \frac{\tau}{2}\right) z^*\left(t - \frac{\tau}{2}\right) e^{-j \omega \tau} \, d\tau,
\end{align} \normalsize \vspace{-3.85mm} 

\noindent where $z(t)$ is the analytic signal evaluated for the input signal, $x(t)\in \mathbb{R}$:

 \vspace{-3.85mm}\small  \begin{align}
    z(t) &= x(t) + j\mathcal{H}\{x(t)\}.
    \intertext{\normalsize $\mathcal{H}\{x(t)\}$ is the Hilbert transform \cite{ref:hilbert} of the signal $x(t)$:\small}
    \mathcal{H}\{x(t)\}&= x(t) * \frac{1}{\pi t}= \frac{1}{\pi} \int_{\mathbb{R}}\frac{x(t-\tau)}{\tau} \,d \tau.
\end{align} \normalsize \vspace{-3.85mm} 

Note that, whilst Polynomial Wigner--Ville Distributions (PWVDs) \cite{PWVD} provide optimal T--F resolution for high-order frequency modulated signals, given the employment of linear frequency-modulated wavelets, the useful convolutional relationship between the WVD and the wavelet window functions (derived in Section \ref{time-frequency relationships}), and the increased cross-term presence and computational complexity in PWVDs \cite{PWVD}, the WVD is particularly suitable in this context. Despite the optimality of WVDs for linear frequency-modulated signals, the WVD is known to exhibit strong cross-term interference between components \cite{wvdinter}, a consequence of ensuring instantaneous power conservation. The negative cross-term components in the WVD compensate for interference-induced beating frequencies, ensuring that integrating over frequency at any time yields the correct instantaneous power. 

\vspace{-3mm}\subsection{Relationship between time--frequency representations}

\label{time-frequency relationships}

Cohen's class \cite{Cohen1989} of bilinear T--F representations can be generalised in the T--F domain as:

\vspace{-2.85mm} \small \begin{align}
\label{cohen_eq}
    C_x\left(\omega, t\right) &=  2 \pi\int_{\mathbb{R}^2} WVD_x(u, v) \Pi \left(\omega - u, t - v\right) \, d u \, d v
    \notag \\&=  2 \pi \left[WVD_x * \Pi \right] \left(\omega, t\right),
\end{align} \normalsize \vspace{-3.85mm} 

\noindent where:

\vspace{-3.85mm}\small \begin{align*}
    \Pi\left(\omega, t\right)\! &= \!\mathcal{F}_t \mathcal{F}_{\omega}^{-1}\!\left\{\Gamma\left(\eta, \tau\right)\right\}\!\left(\omega, t\right)\!
= \!\frac{1}{2 \pi}\!\!\int_{\mathbb{R}^2} \!\Gamma\left(\eta, \tau\right) e^{j \left(t\eta - \omega\tau\right)}  d\eta \, d\tau,
\end{align*} \normalsize \vspace{-2.85mm} 

\noindent and $\Gamma\left(\eta, \, \tau\right)$ is the conventionally defined Cohen's kernel function \cite{Cohen1989}. Note that this paper will refer to the T--F definition of Cohen's class of distributions, as per Eq.~\eqref{cohen_eq}, with $\Pi\left(\omega, t\right)$ referred to as the (T--F) Cohen's class kernel. Cohen's class of T--F distributions provides a unified representation of bilinear (quadratic) T--F distributions, with Cohen's kernel function often being employed as a smoothing kernel to partially suppress cross-term interference, such as within the Gabor Wigner Transform \cite{GaborWignerTransform} and the Choi Williams Distribution \cite{ChoiWilliamsDistribution}, however at the expense of decreased T--F resolution. Note that for $\Pi\left(\omega, t\right)=\frac{1}{2\pi}\delta\left(\omega, \, t\right)$, $C_x\left(\omega, t\right) = WVD_x\left(\omega, t\right)$. Note also that the following convention is employed here for Fourier transforms:

\vspace{-2.85mm} \small \begin{align}
    \bar{Z}\left(\omega\right) &= \mathcal{F}\left\{z(t)\right\} = \frac{1}{\sqrt{2 \pi}} \int_{\mathbb{R}} z(t) e^{j\omega t} \, dt, \\
    z\left(t\right) &= \mathcal{F}^{-1}\left\{\bar{Z}\left(\omega\right)\right\} = \frac{1}{\sqrt{2 \pi}} \int_{\mathbb{R}} \bar{Z}\left(\omega\right) e^{-j\omega t} \, d \omega.
\end{align} \normalsize\vspace{-2.85mm}

\vspace{-3mm}\subsection{Wavelet Transform T--F properties}

Let $\Phi{\left(\omega, t\right)}$ be a Continuous Wavelet Transform (CWT) \cite{wavelet} with a \textit{complex-valued} window function, such that:

\vspace{-2.85mm} \small \begin{align}
    \Phi{\left(\omega, t\right)} &= \left |\left[z * W_{\omega}\right]{\left(t\right)} \right |^2
    = \left |\left[z\left(\tau\right) * \Omega^*{\left(\tau\right)} e^{j \omega \tau}\right](t) \right |^2,
\end{align} \normalsize \vspace{-2.85mm} 

\noindent where $W_{\omega}{\left(t\right)}$ and $\Omega{\left(t\right)}$ are the CWT wavelet and window functions respectively for angular frequency $\omega$. From this formulation, the CWT can be expressed as: 

\vspace{-1.85mm}\small  \begin{align}
\label{WVD_CWT_REL}
    \Phi{\left(\omega, t\right)}
    &=  2 \pi \int_{\mathbb{R}^2} WVD_z\left(\omega_1, T\right) WVD_{\Omega}\left(\omega - \omega_1, t - T\right) \, d T \, d \omega_1 \notag \\
    &= 2 \pi \left[WVD_z * WVD_{\Omega} \right] \left(\omega, t\right),
\end{align} \normalsize \vspace{-3.85mm} 

\noindent where the Cohen's class kernel may be identified as:

\vspace{-1.85mm} \small  \begin{align}
\label{kernel eq}
    \Pi\left(\omega, t\right) &= WVD_{\Omega}\left(\omega, t\right).
\end{align} \normalsize\vspace{-3.85mm}

A derivation of this result is provided in the Supplementary Material (SM), Section \ref{sm:A}. Thus, significantly, the CWT representation, $\Phi{\left(\omega, t\right)}$, is equivalent to a 2D convolution of the WVD of the signal $z(t)$ with a Cohen's class kernel $\Pi = WVD_{\Omega}$ in the T--F domain. We will shortly utilise this relationship in algorithms for reconstruction of the ITFR. The kernel is thus equal to the WVD of the complex conjugate of the complex-valued window function employed in the CWT wavelet. A notable consequence of this result is the positivity, resulting in the elimination of negative components in the WVD, significantly reducing oscillatory WVD cross-terms \cite{Nuttall1988}. Specific forms of $\Omega(t)$ designed for our work are given in the following section. 

\vspace{-1.5mm}\section{Proposed Methodology}
\label{sec:method}

\subsection{Proposed Wavelet Transform}
\label{section:transform_derivation}

In order to generate a suitable T--F representation that jointly exploits the reduced computational complexity and increased oscillatory cross-term suppression properties \cite{Nuttall1988} of CWTs, consider first the following generalised Continuous Fractional Wavelet Transform (CFWT):

\vspace{-2.85mm} \small   \begin{align}
\label{ex:Phi}
   \Phi_{\sigma,\theta}{\left(\omega, t\right)} &= \left | \left[z * W_{\sigma, \, \theta, \, \omega}\right]{\left(t\right)} \right |^2 \\
   &= \left |\int_{\mathbb{R}} z\left(\tau\right) W_{\sigma, \, \theta, \, \omega}{\left(t-\tau\right)} \, d \tau \right |^2,
\end{align}\normalsize \vspace{-2.85mm} 

\noindent where $\Phi_{\sigma, \theta}{\left(\omega, t\right)}$ is the magnitude squared of the T--F representation evaluated at angular frequency $\omega$, and time $t$. The corresponding wavelet function, $W_{\sigma, \, \theta, \, \omega}{\left(t\right)}$, is given by:

\vspace{-2.85mm} \small
\begin{align}
    W_{\sigma, \, \theta, \, \omega}{\left(t\right)} &= \Omega^*_{\sigma, \theta}{\left(t\right)} e^{j \omega t},
\end{align}
\normalsize \vspace{-2.85mm}

\noindent where $\Omega_{\sigma, \theta}(t)$ is the wavelet window function. Firstly, let:

\vspace{-2.85mm} \small
\begin{align}
\label{omega_simple}
\Omega_{\sigma, \theta}(t) &=
\frac{1}{\sqrt[4]{\pi\sigma^2}}e^{-\tfrac{1}{2\sigma^2}t^2}e^{j\tfrac{\tan{\theta}}{2}t^2}.
\end{align}
\normalsize \vspace{-2.85mm}

$\theta$ is the \textit{intended} angle of the instantaneous angular frequency relative to the horizontal axis (time axis) in the T--F plane, and $\sigma$ is the \textit{intended} standard deviation of the major axis of the corresponding Cohen's class kernel. As explored previously in the overview (Section~\ref{overview}), the employment of Short-time Fractional Fourier (or wavelet) Transforms is especially desirable in the context of time-varying signal curvature compared to their standard counterparts (for instance Short-Time Fourier Transforms or Wavelet Transforms), given their ability to align with signal trajectories and thus minimise cross term interference appropriately \cite{Awal2017}. Employing the result of Eq.~\eqref{kernel eq}, the exact Cohen's class smoothing kernel, $\bar{\Pi}_{\sigma, \theta}\left(\omega, \, t\right)$, resulting from the wavelet convolutional process can be obtained as: 

\vspace{-2.85mm} \small
\begin{align}
    \bar{\Pi}_{\sigma, \theta}\left(\omega, \, t\right) &= WVD{\left\{\Omega_{\sigma, \theta}{\left(t\right)}\right\}}\left(\omega, \, t\right)
    \notag \\&=\frac{1}{\pi}e^{-\left[\frac{t^2}{\sigma^2} +\sigma^2\left(\omega -\tan(\theta)\, t\right)^2\right]} \notag \\
    &=\frac{1}{\pi}e^{-\frac{1}{2}\begin{bmatrix}
        t & \omega 
    \end{bmatrix}\begin{bmatrix}
        2a & b \\
        b & 2c
    \end{bmatrix}\begin{bmatrix}
        t \\
        \omega
    \end{bmatrix}}\label{pi_form}
\end{align}
\normalsize \vspace{-2.85mm}

\noindent where:

\vspace{-2.85mm} \small
\begin{align}
\label{a, b, c}
    a = \frac{1}{\sigma^2} + \sigma^2\tan^2{\theta}, \quad b =  -2\sigma^2\tan{\theta}, \quad c = \sigma^2.
\end{align} \normalsize \vspace{-2.85mm}

For the derivation, see the SM, Section \ref{sm:kernel_derivation}. Note that due to the skewed form of the kernel in Eq.~\eqref{pi_form}, the intended angle and standard deviation of the principal axis of the kernels may not necessarily coincide with the actual values. The required mapping between the intended pair $(\sigma,\theta)$ (principal-axis angle and standard deviation) to the implemented parameters $(\sigma_0,\kappa)$ is given by:

\vspace{-2.85mm} \small
\begin{align}
        \sigma_{0}(\sigma,\kappa)
  \!&=\! \left[\!
      \frac{ (\sigma^{2}\!+\!\sigma^{-2})
        +\operatorname{sgn}(\sigma\!-\!1) \sqrt{(\sigma^{2}\!+\!\sigma^{-2})^{2}\! -\! 4\,\sec^{2}\kappa}
      }{ 2\,\sec^{2}\kappa }
     \!\right]^{\!\tfrac{1}{2}} \label{sigma_{kappa}}
\end{align}
\normalsize \vspace{-2.85mm}

\noindent valid when:

\vspace{-1.85mm} \small
\begin{align}
\label{theta_limit}
    \left(\frac{1}{\sigma^{2}}+\sigma^{2}\right)^{2}-4\sec^{2}\kappa \ge 0\,,
    \, \text{i.e.} \,
    \left| \kappa\right| \leq \arccos\left(\frac{2\sigma^{2}}{\sigma^{4}+1}\right).
\end{align}
\normalsize \vspace{-1.85mm}

\noindent Note that $\operatorname{sgn}(x)$ denotes the sign function, with $\operatorname{sgn}(x) = -1\ (x<0),\ 0\ (x=0),\ 1\ (x>0)$. Likewise, the principal-axis angle satisfies:

\vspace{-1.25mm} \small
\begin{align}
\label{phi_core}
    \varphi\left(\sigma, \kappa\right)
    &= -\frac{1}{2}\arctan{\!\left(\frac{2\tan{\kappa}}{\sigma_{0}{\left(\sigma, \kappa\right)}^{-4}\! +\! \tan^2{\kappa}\! - \!1}\right)}\! + \!K_\sigma\left(\kappa\right),
\end{align}
\normalsize \vspace{-1.85mm}

\noindent where $K_{\sigma}\!\left(\kappa\right)$ is the axis switching function provided in Eq.~\eqref{axis-switch} in the SM. Thus, $\kappa(\sigma,\theta)=\varphi^{-1}(\sigma,\theta)$, the inverse function \emph{with respect to $\theta$}. Note that in the limiting case $\sigma \to \infty$, we obtain $\varphi(\sigma,\theta)\to \theta$ as expected. Equivalently, the mapping $(\sigma, \theta) \mapsto (\sigma_0, \kappa)$ can be obtained by performing covariance matching with the intended Cohen's class kernel (introduced in Eq.~\eqref{final_cohen_kernel}); further details and derivations are given in the SM, Sections \ref{sm:kernel_parameters_offset_sd} and \ref{sm:kernel_parameters_offset_angle}.

Now, accounting for the derived mappings, the proposed CFWT is reformulated with the window function:

\vspace{-1.25mm} \small  \begin{align}
\label{window_function_CFWT}
\Omega_{\sigma, \theta}(t) &=
\begin{cases}
\frac{1}{\sqrt[4]{\pi\sigma_0^2}}
e^{-\tfrac{t^2}{2}\left[\tfrac{1}{\sigma_0^2} - j\tan\kappa\right]}
&\hspace{-2mm}
\text{if }\begin{array}[t]{l}
|\theta| \le \theta_L(\sigma),\\
\sigma > 0,\ \sigma \neq 1
\end{array}
\\
\frac{1}{\sqrt[4]{\pi\sigma_0^2}}
e^{-\tfrac{t^2}{2}\left[\frac{1}{\sigma_0'^2} - j\tan\kappa' \right]}
&\hspace{-2mm}
\text{if }\begin{array}[t]{l}
\theta_L(\sigma) <|\theta| \le \tfrac{\pi}{2},\\
\sigma > 0,\ \sigma \neq 1
\end{array}
\\
\frac{1}{\sqrt[4]{\pi\sigma^2}}
e^{-\tfrac{t^2}{2\sigma^2}}
&\hspace{-2mm}
\text{if }
\sigma = 1
\end{cases}
\end{align} \normalsize \vspace{-2.85mm}

\noindent where $\sigma_0 \equiv \sigma_0(\sigma,\kappa(\sigma,\theta))$, $\kappa \equiv \kappa(\sigma,\theta)$, $\sigma_0' \equiv \sigma_0(1/\sigma,\kappa(1/\sigma,\theta'))$, $\kappa' \equiv \kappa(1/\sigma,\theta')$, and $\theta' \triangleq \theta - \operatorname{sgn}(\theta)\,\frac{\pi}{2}$. Note that when $|\theta|$ reaches $\theta_L(\sigma)$, the implementation switches to an equivalent parametrisation in which the wavelet’s instantaneous-frequency direction follows the axis orthogonal to $\theta$, allowing the Cohen’s class kernel to cover the full range of $\theta$. Note also that, due to the rotational symmetry of the wavelet, $\Omega_{\sigma,\theta}(t)=\Omega_{\sigma,\theta \pm \pi}(t)$, for $|\theta| > \tfrac{\pi}{2}$.

Thus, the corresponding Cohen's class kernel is now:

\vspace{-2.85mm} \small
\begin{align}
    \Pi_{\sigma, \theta}\left(\omega, \, t\right) &= WVD{\left\{\Omega_{\sigma, \theta}{\left(t\right)}\right\}}\left(\omega, \, t\right)
    =\frac{1}{\pi}e^{-\frac{1}{2}\mathbf{x}^T\mathbf{\Sigma}^{-1}\mathbf{x}}, \label{final_cohen_kernel}
\end{align}
\normalsize \vspace{-2.85mm}

\noindent where:

\vspace{-2.85mm} \small
\begin{align}
    \mathbf{x}=\begin{bmatrix}
        t \\ \omega 
    \end{bmatrix}, \, \mathbf{\Sigma}=\mathbf{R_{\theta}}\begin{bmatrix}
        \frac{\sigma^2}{2} & 0 \\
        0 & \frac{1}{2\sigma^2}
    \end{bmatrix}\mathbf{R_{\theta}}^T, \,  \mathbf{R_{\theta}} = \begin{bmatrix}
        \cos{\theta} & -\sin{\theta} \\
        \sin{\theta} & \cos{\theta}
    \end{bmatrix},
\end{align}
\normalsize \vspace{-2.85mm}

\noindent and, in relation to the skewed form in Eq.~\eqref{pi_form}:

\vspace{-1.85mm} \small  \begin{align}
\label{pi_relationship}
\Pi_{\sigma, \theta}\left(\omega, \, t\right) &=
\begin{cases}
\bar{\Pi}_{\sigma_0, \kappa}\left(\omega, \, t\right)
&
\text{if }
|\theta| \le \theta_L(\sigma),\, \sigma > 0,\ \sigma \neq 1
\\
\bar{\Pi}_{\sigma_0', \kappa'}\left(\omega, \, t\right)
&
\text{if }
\theta_L(\sigma) <|\theta| \le \tfrac{\pi}{2},\, \sigma > 0,\ \sigma \neq 1
\\
\bar{\Pi}_{\sigma_, 0}\left(\omega, \, t\right)
&
\text{if }
\sigma = 1
\end{cases}
\end{align} \normalsize \vspace{-2.85mm}

To derive the upper limit  $\theta_L\left(\sigma\right)$ on $|\theta|$, the exact range can be determined from Eq.~\eqref{phi_core} and the limits imposed in Eq.~\eqref{theta_limit}:

\vspace{-1.85mm} \small  \begin{align}
\label{varphi_limit}
    \theta_L(\sigma) = \left | \varphi\!\left(\sigma,\; \arccos\left(\frac{2\sigma^{2}}{\sigma^{4}+1}\right)\right)\right |.
\end{align}\normalsize\vspace{-1.85mm}

Figure~\ref{fig:max_phi} in the SM plots Eq.~\eqref{varphi_limit} for a range of $\sigma$. Note that in the case where $\theta = 0$, the wavelet function collapses to a generalised Morlet wavelet, defined as a complex exponential with a Gaussian envelope \cite{morlet}:

\vspace{-1.85mm} \small  \begin{align}
\label{ex:1a}
    W_{\sigma, \omega}\left(t\right) &= \frac{1}{\sqrt[4]{\pi \sigma^2}}e^{-\tfrac{1}{2 \sigma^2} t^2} e^{j \omega t}.
\end{align} \normalsize\vspace{-1.85mm}

Also, $W_{\sigma, \, \theta, \, \omega}{\left(t\right)}$ is constructed such that the Energy of the wavelet, $E_W$, is normalised to 1:

\vspace{-1.85mm} \small  \begin{align}
\label{eq:wavelet}
    E_W &= \int_{\mathbb{R}}W_{\sigma, \, \theta, \, \omega}{\left(t\right)}W_{\sigma, \, \theta, \, \omega}{\left(t\right)}^*\, dt  = 1.
\end{align} \normalsize\vspace{-1.85mm}

Note that the DC offset correction required by the admissibility criterion \cite{wavelet} is omitted from Eq.~\eqref{window_function_CFWT}, as it is negligible for audio signals where typically $\sigma \gg \omega$. Note also, from Eq.~\eqref{final_cohen_kernel}:

\vspace{-1.85mm} \small \begin{align}
    &\int_{\mathbb{R}^2}\Pi_{\sigma, \theta}\left(\omega, \, t\right) \, d\omega \, dt 
    =\int_{\mathbb{R}^2}\frac{1}{\pi}e^{-\frac{1}{2}\mathbf{x}^T\mathbf{\Sigma}^{-1}\mathbf{x}}\, d\omega \, dt = 1.
\end{align} \normalsize\vspace{-1.85mm} 

Therefore, $\Pi\left(\omega, t\right)$ is normalised with respect to the T--F plane as expected given the energy property of WVDs and  (\ref{eq:wavelet}):

\vspace{-2.85mm} \small  \begin{align}
\label{energy}
    E_W = \int_{\mathbb{R}}\left | z(t)\right |^2 \, dt = \int_{\mathbb{R}}\left | \bar{Z}(\omega)\right |^2 \, d\omega = \int_{\mathbb{R}^2} WVD\left(\omega, t\right) \, d\omega \, dt.
\end{align} \normalsize\vspace{-2.85mm} 

\vspace{-2mm}\subsection{Reconstructing the ITFR}
\label{section:probability}

 Eq.~\eqref{WVD_CWT_REL} demonstrates that the proposed CFWT can be expressed as a convolution of the WVD of the signal with the proposed Cohen's class kernel, according to:
 
\vspace{-2.85mm}\small  \begin{align}
    \label{conv_rel}
    \Phi_{\sigma, \theta}{\left(\omega, t\right)}&= 2\pi\left[WVD_z * \Pi_{\sigma, \theta} \right] \left(\omega, t\right).
\end{align}\normalsize\vspace{-2.85mm}

Thus, if suitable Cohen's class kernels are selected that predominantly pass auto-terms, then the ITFR can be estimated by deconvolving the CFWT with respect to these Cohen's class kernels (Eq.~\eqref{conv_rel}). To promote a cross-term free estimate, a suitable T--F concentration measure must thus be designed to align signal components locally with the appropriate kernel (in terms of $\sigma$ and $\theta$); such a sparsity measure is derived and detailed in Section \ref{section:entropy}. As discussed in the overview, Section~\ref{overview}, although current concentration-based optimisation methods reduce cross-term interference compared to the WVD, the discrete nature of segmentation schemes and optimised kernel assignments can still result in abrupt T--F transition characteristics and suboptimal spectrogram allocation when the chosen kernels, segments, or window widths do not align with the desired signal trajectories \cite{Awal2017, Mohammadi}. In contrast, we propose a probabilistic model that alternatively estimates the signal ITFR (as defined in Section~\ref{sec:itfr}) through modelling signal components in the T--F plane as a linear combination of CFWTs, weighted by the proposed sparsity measure, thereby ensuring that abruptly varying T--F characteristics are more effectively accounted for. Furthermore, the probabilistic model frames the optimisation process as a deconvolution task, enabling the original WVD resolution to be estimated from the optimised aligned representation. 

The model is defined on a regular T--F grid with steps $\Delta \omega$ and $\Delta t$, with indices denoted by $i, j$ for frequency and time pixel coordinates, respectively. We re-use $*$ to denote the discrete time--frequency convolution in this context. The discretisation of Eq.~\eqref{conv_rel} is implemented via a Riemann-sum quadrature, leading to the discrete forward model in \eqref{eq:discrete_forward_model}; the factor $\Delta \omega \Delta t$ absorbed into the discrete kernel so that the same convolutional form holds up to this constant and edge effects (handled by zero-padding). Likewise, the orientations and standard deviations of the kernels are evaluated on a discrete grid $(\theta(m),\,\sigma(n)),\,m\in \{1,\,...\,, M\},\, n\in \{1,\,...\,, N\}$, where proposed forms for $\theta(m)$ and $\sigma(n)$ are presented in Section~\ref{kernel selection}. The resulting discretised variables are denoted as follows:

\vspace{-2.85mm} 
\small  \begin{align}
  \text{ITFR}_{i, j} &= \text{ITFR}\left(i\Delta \omega, j\Delta t\right),  \notag\\
    \Pi_{i, j}^{(n, m)} & = \Delta \omega \Delta t \, \Pi_{\sigma(n), \theta(m)}\left(i\Delta \omega, j\Delta t\right), \notag \\
    \Phi_{i, j}^{(n, m)} &= \Phi_{\sigma(n), \theta(m)}{\left(i\Delta \omega, j\Delta t\right) }, \notag\\
    &= \left |\left[z * W_{\sigma(n), \,
    \theta(m), \, i\Delta \omega}\right]{\left(j\Delta t\right)} \right |^2, \notag 
    \\
    R_{i, j}^{(n, m)} &= \left[\text{ITFR} * \Pi^{(n, m)}\right]_{i, j}, \notag\\
    \bar{P}_{i, j}^{(n, m)} &= \bar{P}_{\sigma(n), \theta(m)}\left(i\Delta \omega, j\Delta t\right),\label{eq:discrete_forward_model} 
\end{align}\normalsize  \vspace{-2.85mm}

\noindent where $R_{i, j}^{(n, m)}$ is the forward-modelled  CFWT corresponding to the RIFT estimate of the $\text{ITFR}$, defined in Section~\ref{sec:itfr}, $z(j \Delta t)$ is the discretised input analytic signal, and $\bar{P}_{i, j}^{(n, m)}$ is the entropic weighting function (defined later in Section \ref{section:entropy}). Note that $z(t)$ is sampled such that Nyquist frequency is above the supported frequency range, and arrays are appropriately zero-padded to realise linear (non-circular) convolution.

Motivated by the idea that $\Phi_{i, j}^{(n, m)}$ will not perfectly fit the forward-modelled CFWT $R_{i, j}^{(n, m)}$ in \eqref{eq:discrete_forward_model}, we adopt a likelihood function based on an independent Gaussian noise assumption for the model mismatch, with precision parameters obtained from the entropic weighting terms $\bar{P}_{i, j}^{(n, m)}$: 

\vspace{-2.85mm} \small  \begin{align}
    \label{eq:prob_model}
    p\left(\mathbf{\Phi} \mid \text{\bf{ITFR}}\right) &\propto \prod_{n=1}^{N} \prod_{m=1}^{M} \prod_{i=1}^{I} \prod_{j=1}^{J} e^{-\bar{P}_{i, j}^{(n, m)} \left(R_{i, j}^{(n, m)} - \Phi_{i, j}^{(n, m)}\right)^2} \notag\\
    &=\prod_{n, m, i, j} e^{-\bar{P}_{i, j}^{(n, m)} \left(\left[\text{ITFR} * \Pi^{(n, m)}\right]_{i, j} - \Phi_{i, j}^{(n, m)}\right)^2},
\end{align}\normalsize\vspace{-2.85mm}

\noindent where we denote $\mathbf{\Phi}$ and $\text{\bf{ITFR}}$ as the collection of elements $\Phi_{i, j}^{(n, m)}$ and $\text{ITFR}_{i, j}$ respectively, for $i\in\{1,...,I\},\,j\in\{1,...,J\},\,n\in\{1,...,N\},\,m\in\{1,...,M\}$. Taking the negative log of the likelihood in \eqref{eq:prob_model} and differentiating with respect to $\text{ITFR}_{u,v}$ yields the following normal equations for the unconstrained maximum likelihood solution for this model:

\vspace{-2.85mm} \small  \begin{align}
    \left[\hat{\text{ITFR}} *_{SV} \Psi\right]_{u, v} &= \Phi_{T, u, v}, \label{spatial_varying_sol}
\end{align} \normalsize \vspace{-2.85mm} 

\noindent where $*_{SV}$ denotes a spatially varying convolution, with:

\vspace{-2.85mm} \small  \begin{align}
\label{kernels}
\left[\hat{\text{ITFR}} *_{SV} \Psi\right]_{u, v} &= \sum_{i,j}\hat{\text{ITFR}}_{u - i, v - j} \, \Psi_{i, j; u, v} \\
    \Psi_{i, j; u, v} &= \sum_{n, m}\bar{P}_{u, v}^{(n, m)}\left[\Pi^{(n, m)} * \Pi^{(n, m)}\right]_{i, j} \\ 
    \Phi_{T, u, v} &= \sum_{n, m}\bar{P}_{u, v}^{(n, m)}\left[\Phi^{(n, m)} * \Pi^{(n, m)}\right]_{u, v}\,.
\end{align}\normalsize\vspace{-2.85mm}

See SM, Section \ref{sm:deriving_RIFT}, for the full derivation. However, it is known that the solution for  $\hat{\text{ITFR}}$ needs to be strictly non-negative from Eq.~\eqref{conv_rel}. Thus, we propose a nonnegativity-constrained solution to these normal equations (Eq.~\eqref{spatial_varying_sol}) via Lucy--Richardson Deconvolution with total variation regularisation (LR-TV) \cite{LR_TV}. TV regularisation, implemented via the multiplicative divergence term in Eq.~\eqref{rift_solution} below, penalises the sum over T–F pixels of the Euclidean ($\ell_2$) norm of the local T–F gradient (isotropic TV). TV suppresses isolated pixel-scale speckle and deconvolution ringing while preserving edge-like ridges \cite{LR_TV}. This standard LR-TV form \cite{LR_TV} yields the following iterative update for the RIFT estimate of the ITFR, with respect to a spatially varying PSF, $\mathbf{\Psi}$:

\vspace{-2.85mm} \small  \begin{align}
\label{rift_solution}\hat{\mathbf{ITFR}}_{k+1}\! &= \!\left(\frac{\mathbf{\Phi_T}}{\hat{\mathbf{ITFR}}_{k}\! *_{SV}\! \mathbf{\Psi}}\! *_{SV}\! \tilde{\mathbf{\Psi}}\right)\! \odot\! \frac{\hat{\mathbf{ITFR}}_{k}}{1\! -\! \lambda \operatorname{div}\left(\frac{\nabla \hat{\mathbf{ITFR}}_{k}}{\|\nabla \hat{\mathbf{ITFR}}_{k}\|_2}\right)}.
\end{align}\normalsize\vspace{-2.85mm}

We denote $\boldsymbol{\Phi_T}$, $\hat{\mathbf{ITFR}}$, and $\mathbf{\Psi}$ as the matrices
$\boldsymbol{\Phi_T} := [\Phi_{T,u,v}]_{1\le u\le U,\,1\le v\le V}$,
$\hat{\mathbf{ITFR}} := [{\hat{\text{ITFR}}}_{u,v}]_{1\le u\le U,\,1\le v\le V}$, and $\boldsymbol{\Psi} := [\Psi_{i, j; u,v}]_{1\le i\le I_{\Psi}, 1\le j\le J_{\Psi}; 1\le u\le U,\,1\le v\le V}$, where $(I_{\Psi}, J_{\Psi})$ is the support of the kernel. Likewise, $\tilde{\mathbf{\Psi}}$ is space-variant adjoint flip of $\mathbf{\Psi}$ defined by $\tilde{\Psi}_{i, j; u, v} := \Psi_{-i, -j; u-i, v-j}$, $\odot$ is the Hadamard (element-wise) product, and $k$ represents the $k$th iteration. $\lambda=0.002$ is employed here, as recommended in \cite{LR_TV}. Here, we implement isotropic TV by taking finite-difference gradients, $\partial$, along rows and columns, normalising by $\|\nabla \hat{\mathbf{ITFR}}_{k}\|_2=((\partial_u \hat{\mathbf{ITFR}}_{k})^2+(\partial_v \hat{\mathbf{ITFR}}_{k})^2)^{1/2}$. The discrete divergence is computed as $\operatorname{div}(\mathbf{p})=\partial_u \mathbf{p}_u + \partial_v \mathbf{p}_v$. Figure~\ref{fig:LR_figure} in the SM demonstrates the iterative behaviour of this Lucy--Richardson deconvolution algorithm for a simple Gaussian PSF example. 

To efficiently compute the spatially varying Lucy--Richardson (LR) process, the overlap-save method can be employed \cite{Trussell1978, Rabiner1975}. This approach involves performing parallel LR deconvolutions on $L \times H$ partitioned blocks of the 2D array. The results are then ``stitched'' together (with appropriate padding) to approximate the spatially varying process while maintaining computational efficiency. Full implementation details are provided in Section~\ref{block-wise implementation} in the SM.
Likewise, a breakdown of the global RIFT process is provided in Algorithm \ref{alg:rift}.

\vspace{-2mm}\subsection{Selection of the Kernel Constellation}
\label{kernel selection}

\begin{figure}
    \centering
    \vspace{-2mm}\subfloat[]{{\includegraphics[width=0.45\columnwidth, trim={0cm 3mm 0cm 2cm}, clip]{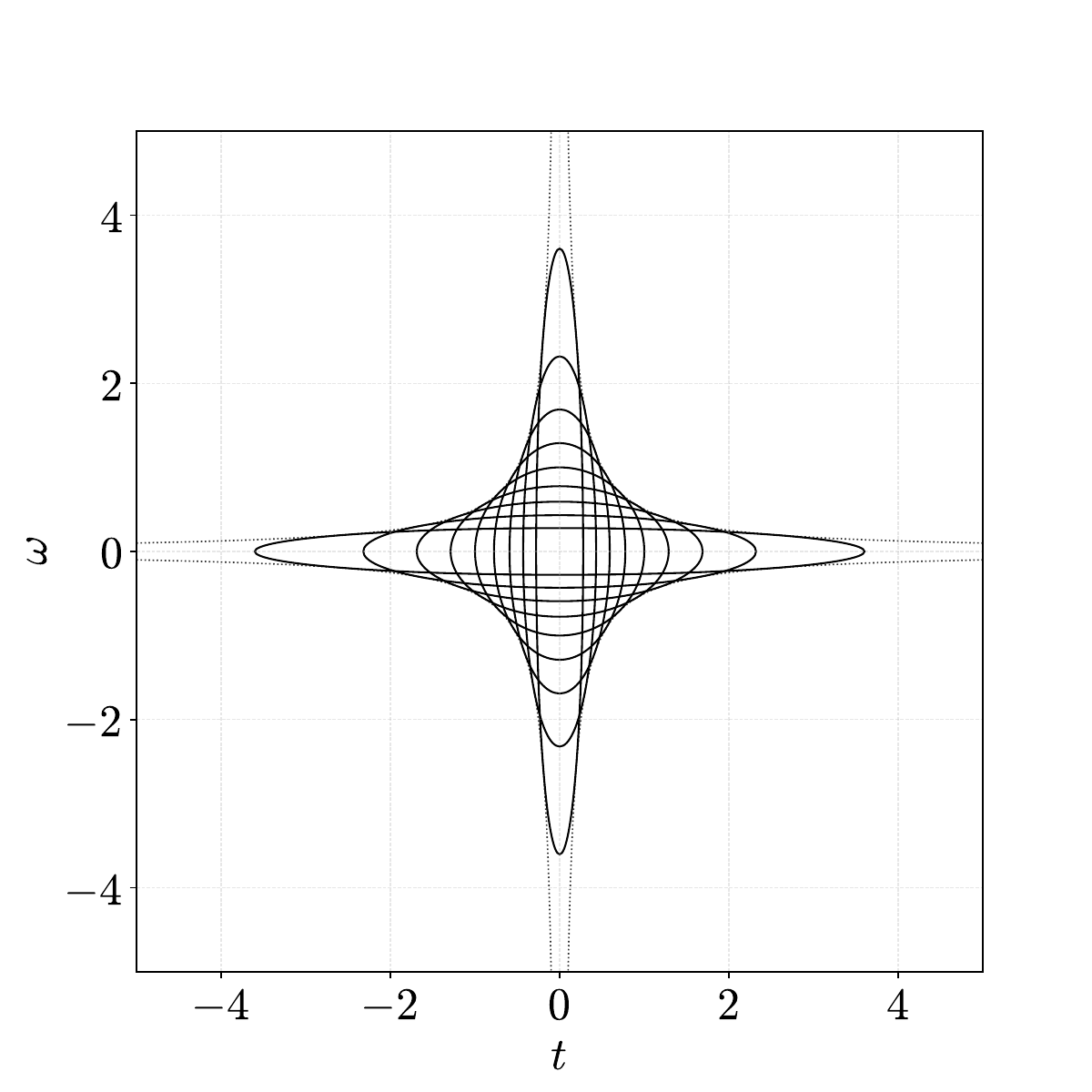}}}%=
    \subfloat[]{{\hspace{3mm}\includegraphics[width=0.45\columnwidth, trim={0cm 3mm 0cm 2cm}, clip]{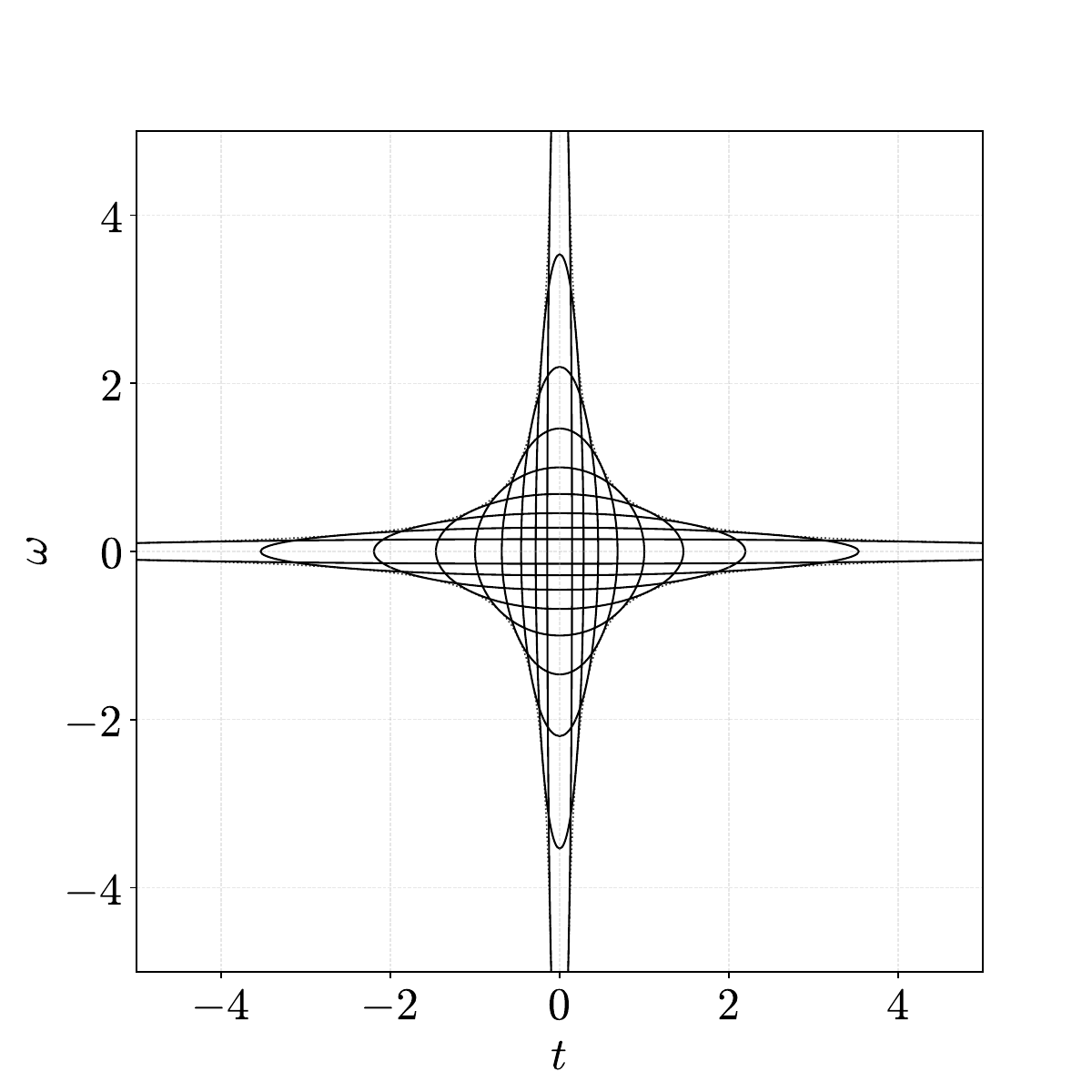}}}%=
    \vspace{-1mm}\caption{Visualisation of the log-normal constellation kernel selection for $\theta=0$ and $N=7$ (and $\sigma_{\text{iso}}=1$), with (a) $\sigma_l=1$, (b) $\sigma_l=1.5$. The ellipses correspond to the isodensity contours for each T--F kernel, given by $\frac{1}{2}\mathbf{x}^T\mathbf{\Sigma}^{-1}\mathbf{x}=1$, for $\sigma\left(1 \le n \le N\right)$.\vspace{-4mm}}%
    \label{fig:sigma}
\end{figure}

The selection of the kernel parameter grid, ${\sigma(n),\theta(m)}$ (the `constellation'), is crucial for ensuring that the resulting family of T--F kernels spans the full range of time--frequency uncertainty trade-offs (via $\sigma$) and angular orientations (via $\theta$), thereby providing sufficient flexibility both for isolating curved T--F components and for optimally capturing auto-terms across a wide variety of signal structures. Thus, in order to provide suitable coverage of the positive $\sigma > 0$ range, the proposed constellation consists of a grid of $\sigma(n)$ values that are distributed deterministically according to a log-normal distribution with variance $\sigma_l^2$ in the log-domain:

\vspace{-2.85mm} \small  \begin{align}
    f\left(\sigma\right) &= \frac{1}{\sigma \sqrt{2 \pi \sigma_l^2}}\exp{\left(-\frac{\left(\ln{\sigma}\right)^2}{2 \sigma_l^2}\right)}, \label{lognormal}
\end{align} \normalsize

\noindent which can be generated in a standard fashion using the inverse CDF method \cite{Devroye} via the formula:

\vspace{-2.85mm} \small  \begin{align}
\label{sigma(n)}
    \sigma\left(n\right) &= \exp{\left(\sigma_l \text{F}^{-1}\left(\frac{n}{N + 1}\right)\right)},\,n\in\{1,...,N\}
\end{align} \normalsize \vspace{-2.85mm} 

\noindent where $\text{F}^{-1}\left(z\right)$ is the inverse of the standard Normal Cumulative Distribution Function $F$. Note that $N$ must be odd to ensure the isotropic kernel case ($\sigma=1$) is included. Figure~\ref{fig:sigma} presents a visualisation of two different selections of T--F kernels for $\theta=0$. Although the proposed wavelet (Eq.~\eqref{window_function_CFWT}) is only defined for $\sigma \ge 1$, the rotational symmetry of the corresponding Cohen's class kernel (Eq.~\eqref{final_cohen_kernel}), $\Pi_{\sigma,\theta} = \Pi_{1/\sigma,\theta\pm\pi/2}$, implies that the kernels with $0<\sigma<1$ are simply $\pm\pi/2$-rotated versions of those with $\sigma>1$. Because the log--normal grid satisfies $\sigma(n)\,\sigma(N-n+1)=1$, it is therefore sufficient to evaluate only the kernels with $\lceil N/2\rceil \le n\le N$; the remaining kernels are obtained by a rotation of $\theta$ by $\pm\pi/2$.

Thus to ensure both adequate coverage of $\theta$ and the remaining $1 \le n\le \lfloor N/2\rfloor$, a uniform grid of $\theta$ values is used:

\vspace{-2.85mm} \small  \begin{align}
\label{actual theta(m)}
    \theta\left(m\right) &= \pi \left(\frac{m - 1}{2M(\sigma(n))} - \frac{1}{2}\right),\,m\in\{1,...,2M(\sigma(n))\},
\end{align} \normalsize\vspace{-2.85mm}
 
\noindent where $M(\sigma(n))$ is the number of $\theta$ steps dependent on $\sigma(n)$; note that as $\sigma$ increases, the radial ``width'' of the kernels decreases (see Fig.~\ref{fig:sigma}), thus to ensure suitable coverage of $\theta$, a denser grid of kernels in large $\sigma$ regions is recommended. 

Note also that the provided grid is with respect to the $(\omega, t)$ axis. To provide an appropriate scaling with respect to the discretised T--F grid, the properties of $\Pi_{i, j}$ in the discretised T--F plane must be considered; for a specific value  $\sigma=\sigma_{\text{iso}}$, the kernel $\Pi_{i, j}$ collapses to an isotropic 2D Gaussian relative to the discretised T--F grid. Centering the set of kernels around this $\sigma_{\text{iso}}$ value provides an isotropic representative constellation of kernels (see Fig.~\ref{fig:sigma}). To determine this value, consider the discretised kernel for $\theta = 0$: 

\vspace{-2.85mm} \small  \begin{align}
\label{discrete}
    \Pi_{i, j} &=\frac{\Delta \omega \Delta t}{\pi}e^{-\left[\frac{\left(j\Delta t\right)^2}{\sigma^2} +\sigma^2\left(i\Delta \omega\right)^2\right]}.
\end{align} \normalsize\vspace{-2.85mm}

\noindent Thus, for the kernel to be isotropic: 

\vspace{-2.85mm} \small  \begin{align}
    \frac{\left(\Delta t\right)^2}{\sigma_{\text{iso}}^2} &= \sigma_{\text{iso}}^2\left(\Delta \omega\right)^2 \quad
    \Rightarrow \sigma_{\text{iso}} = \sqrt{\frac{\Delta t}{\Delta \omega}}. \label{log_mean}
\end{align} \normalsize\vspace{-2.85mm}

Therefore, $\theta_{\text{iso}}\left(m\right)$, the angle relative to the discretised grid, can be determined simply by scaling with respect to $\sigma_{\text{iso}}$:

\vspace{-2.85mm} \small  \begin{align}
\label{theta(m)}
    \theta_{\text{iso}}\left(m\right) &= \arctan\left(\tfrac{\Delta \omega}{\Delta t}\tan{\left(\theta\left(m\right)\right)}\right) = \arctan\left(\tfrac{\tan{\left(\theta\left(m\right)\right)}}{\sigma_{\text{iso}}^2}\right).
\end{align} \normalsize\vspace{-2.85mm}

Likewise, to centre the $\sigma(n)$ values about $\sigma_{\text{iso}}$, $\sigma_0 \equiv \sigma_{\text{iso}} \, \sigma_0(\sigma(n),\kappa_{\text{iso}}(\sigma(n),\theta\left(m\right)))$ can be substituted in the wavelet function (Eq.~\eqref{window_function_CFWT}), in addition to $\kappa \equiv \kappa_{\text{iso}}(\sigma(n),\theta\left(m\right))$, where $\kappa_{\text{iso}} \equiv \arctan\left(\tan{\left(\kappa\right)}/\sigma_{\text{iso}}^2\right)$. Similarly , $\sigma_0' \equiv \sigma_{\text{iso}} \,\sigma_0(1/\sigma,\kappa_{\text{iso}}(1/\sigma,\theta'))$, $\kappa' \equiv \kappa_{\text{iso}}(1/\sigma,\theta')$, and for the isometric case, $\sigma \equiv \sigma_{\text{iso}}$. Likewise, the corresponding Cohen's class kernels can be determined from Eq.~\eqref{pi_relationship} using the same substitutions.

\vspace{-0mm}

\vspace{-3mm}\subsection{Entropic Filtering for Cross-term Suppression}
\label{section:entropy}
\begin{figure*}
    \centering
    \subfloat[]{{\includegraphics[width=0.25\textwidth, trim={0cm 3mm 2cm 0cm}, clip]{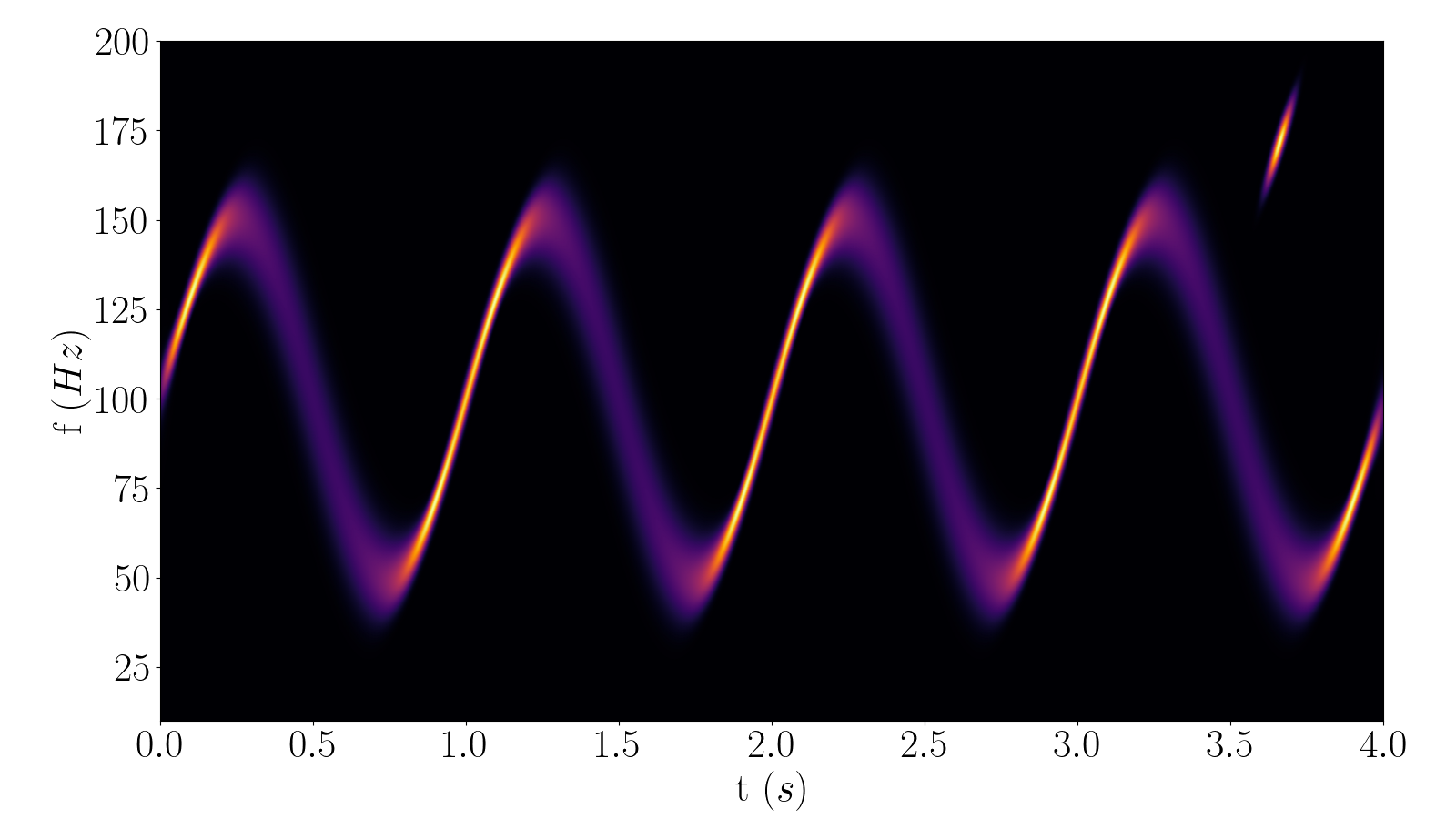}}}%=
    \subfloat[]{{\includegraphics[width=0.25\textwidth, trim={0cm 3mm 2cm 0cm}, clip]{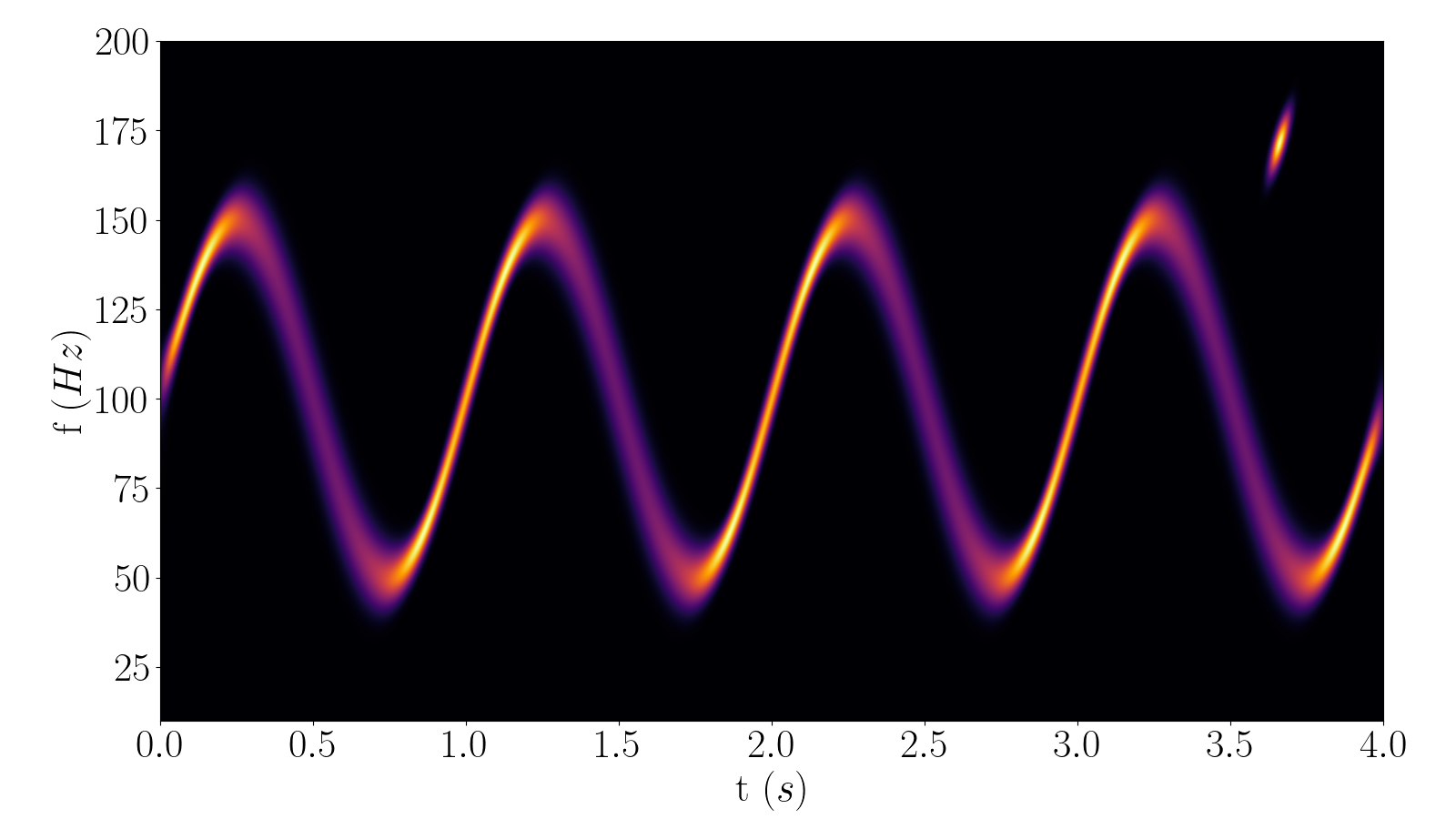}}}%=
    \subfloat[]{{\includegraphics[width=0.25\textwidth, trim={0cm 3mm 2cm 0cm}, clip]{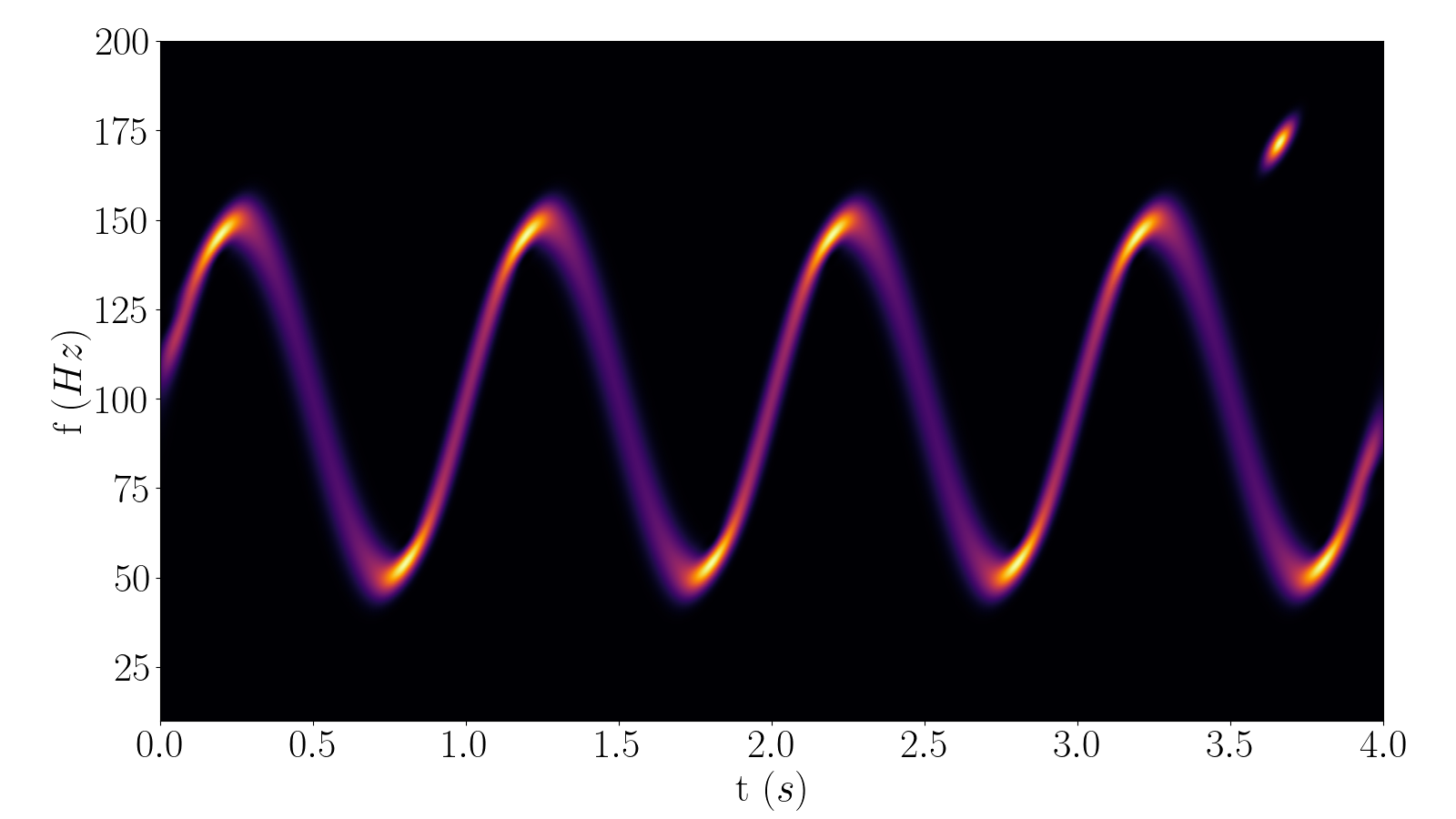}}}%=
    \subfloat[]{{\includegraphics[width=0.25\textwidth, trim={0cm 3mm 2cm 0cm}, clip]{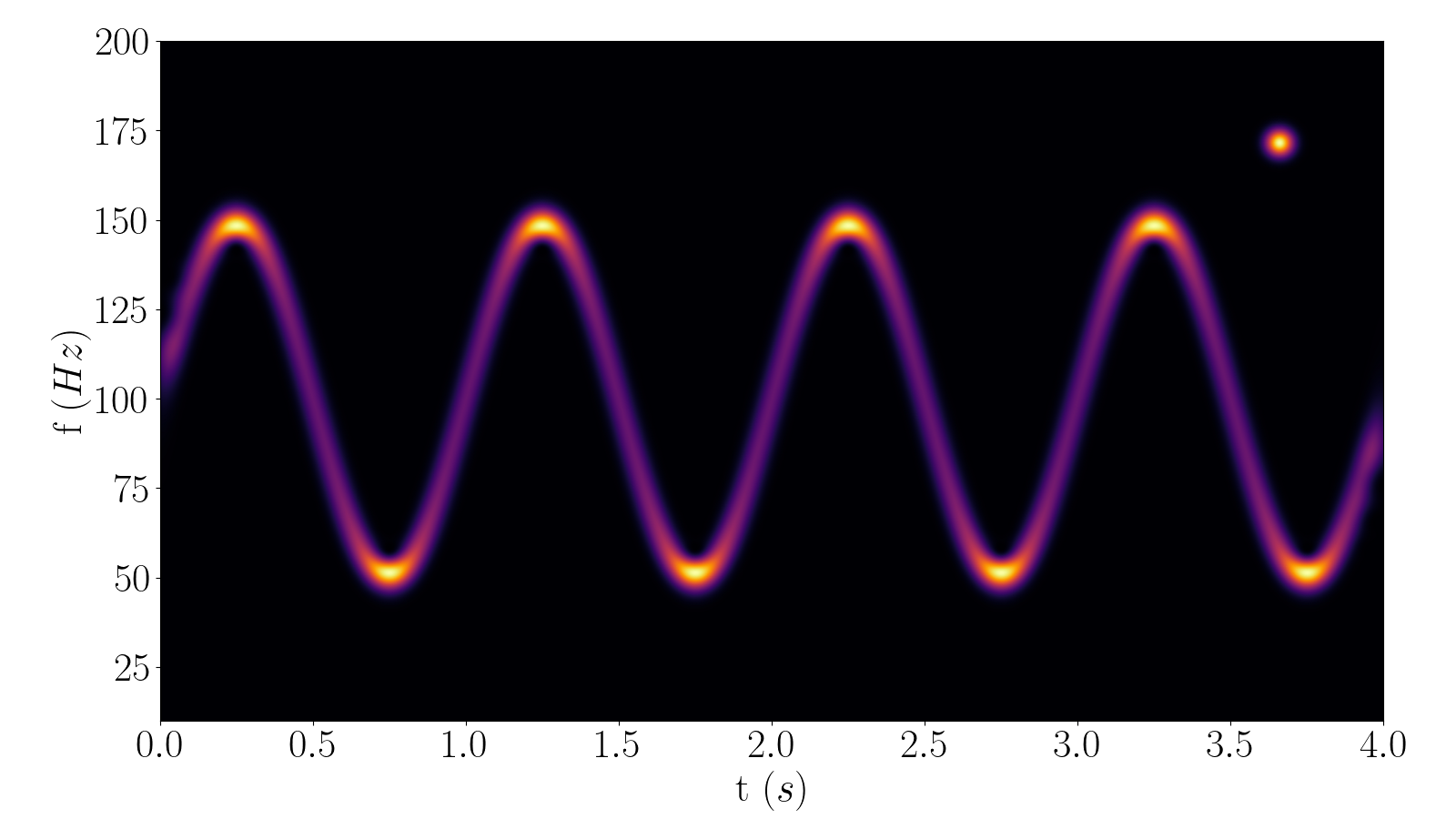}}}%=
    \\
    \vspace{-3mm}\subfloat[]{{\includegraphics[width=0.25\textwidth, trim={0cm 3mm 2cm 0cm}, clip]{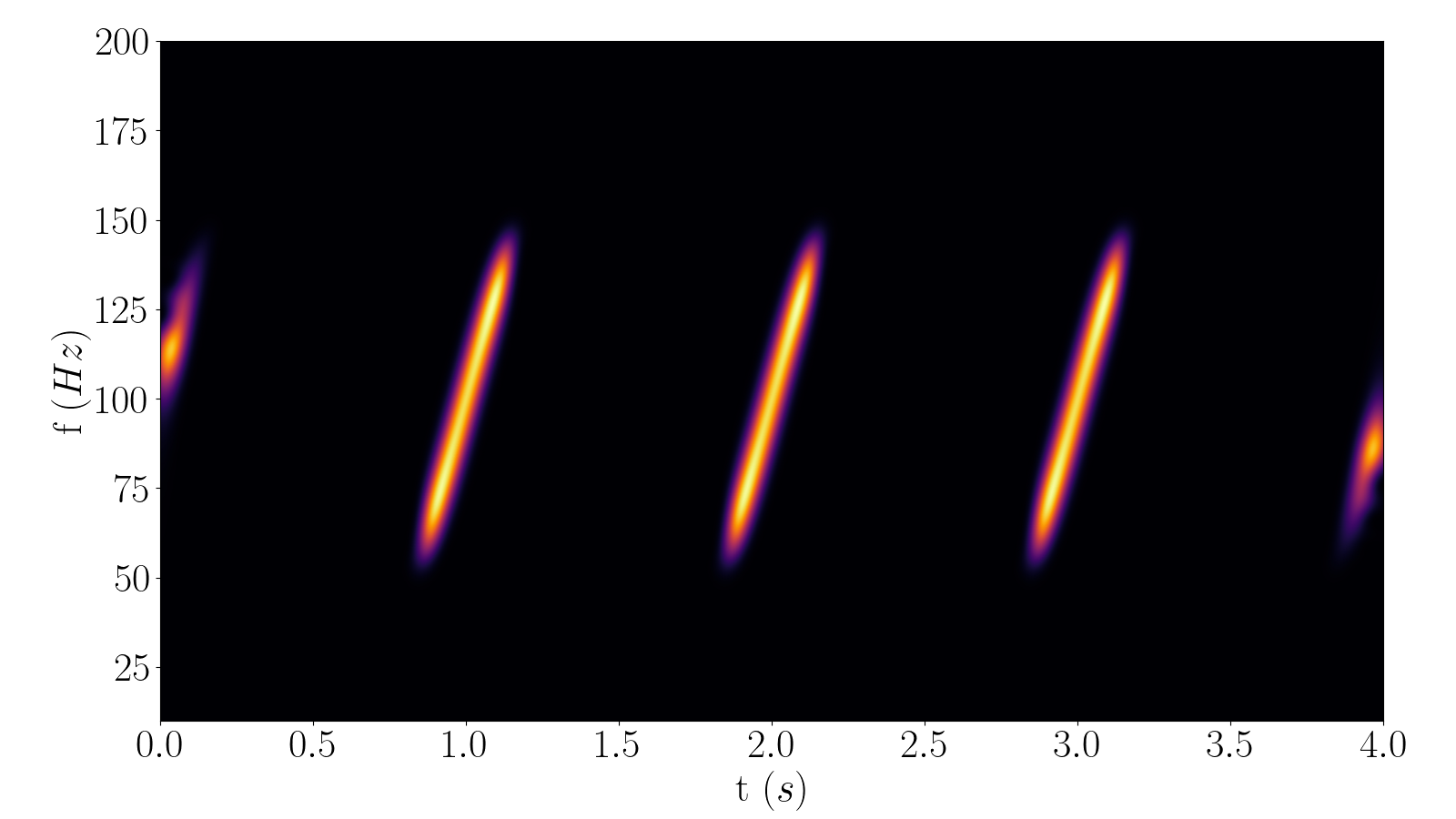}}}%=
    \subfloat[]{{\includegraphics[width=0.25\textwidth, trim={0cm 3mm 2cm 0cm}, clip]{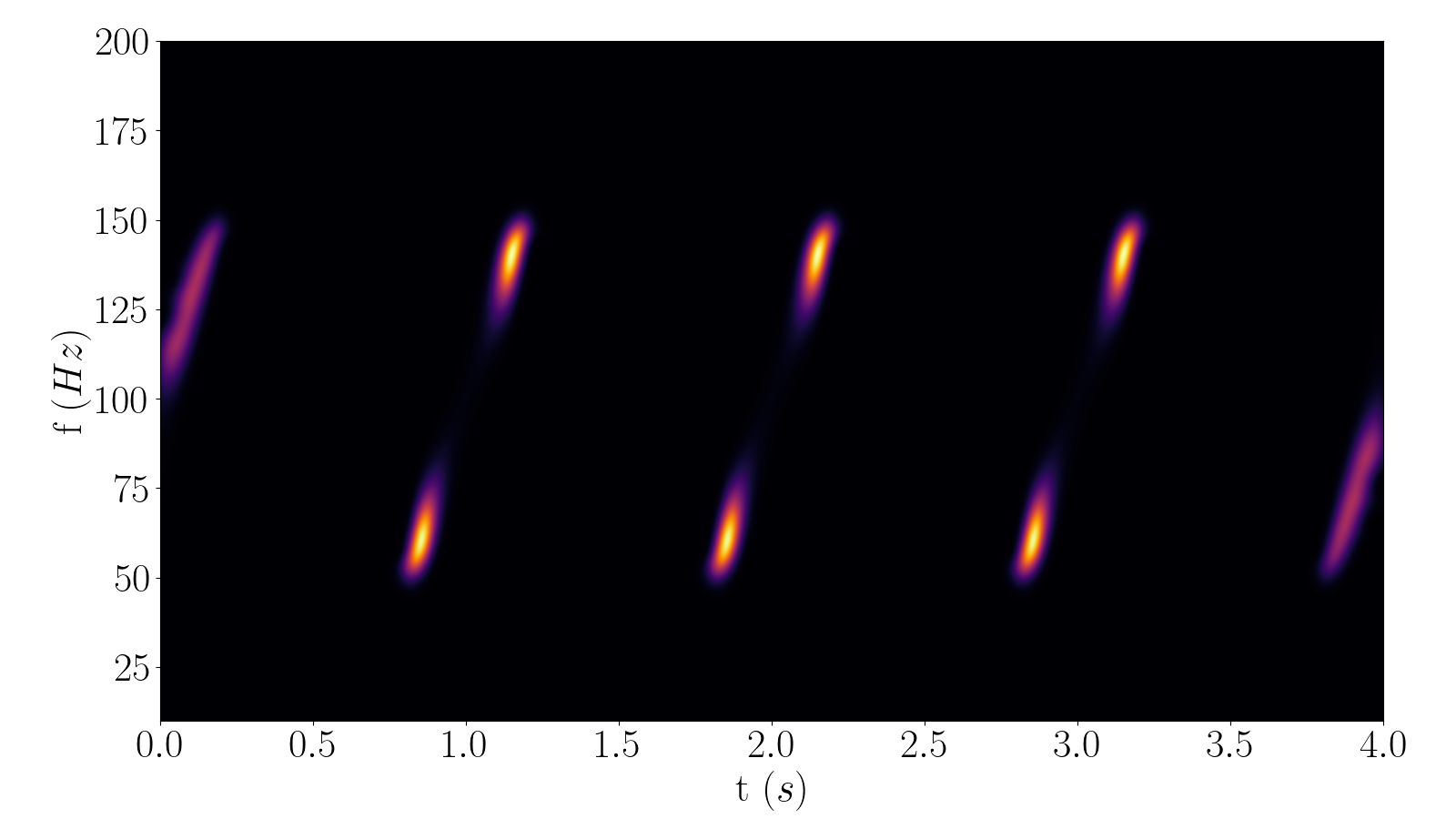}}}%=
    \subfloat[]{{\includegraphics[width=0.25\textwidth, trim={0cm 3mm 2cm 0cm}, clip]{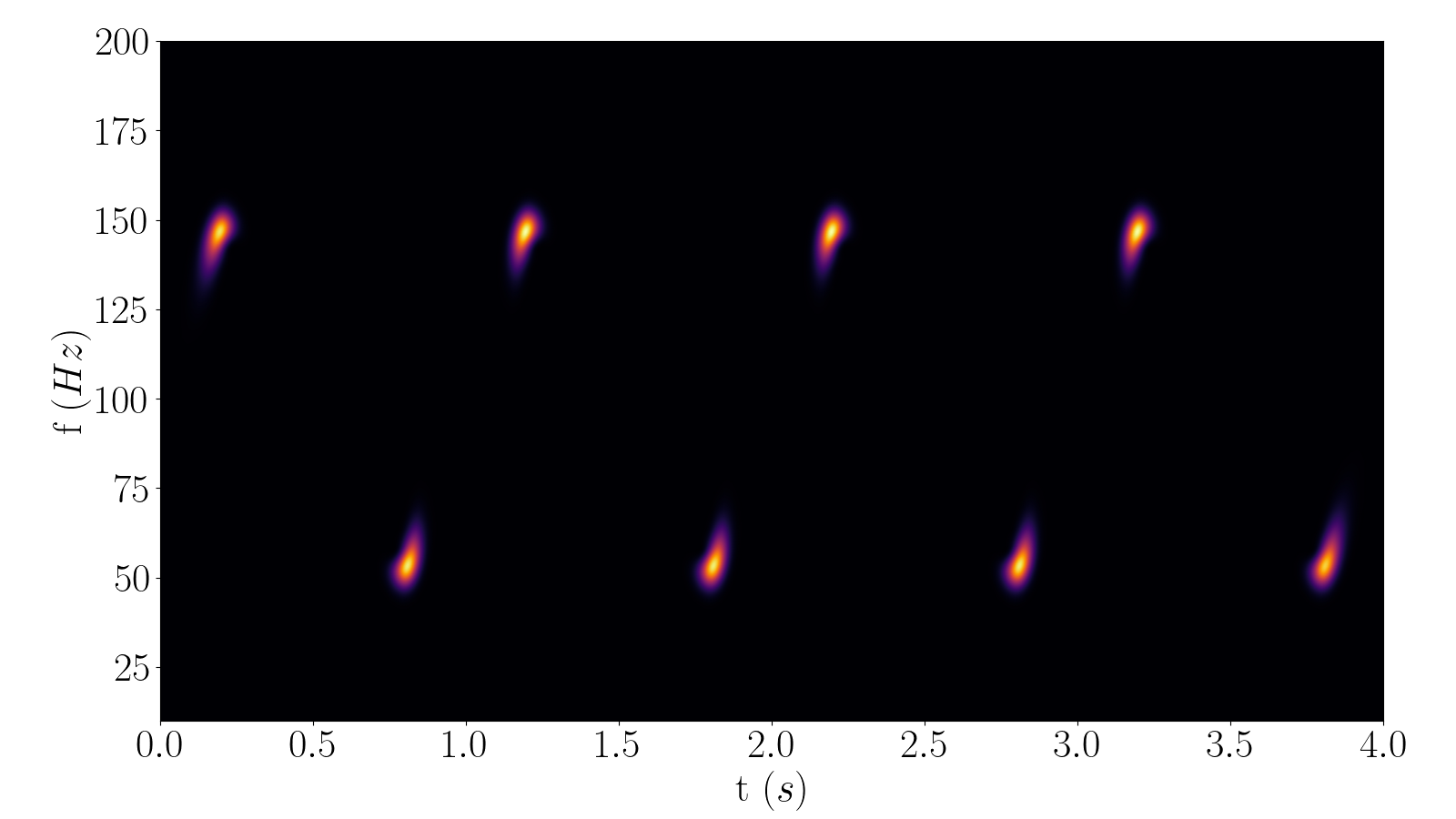}}}%=
    \subfloat[]{{\includegraphics[width=0.25\textwidth, trim={0cm 3mm 2cm 0cm}, clip]{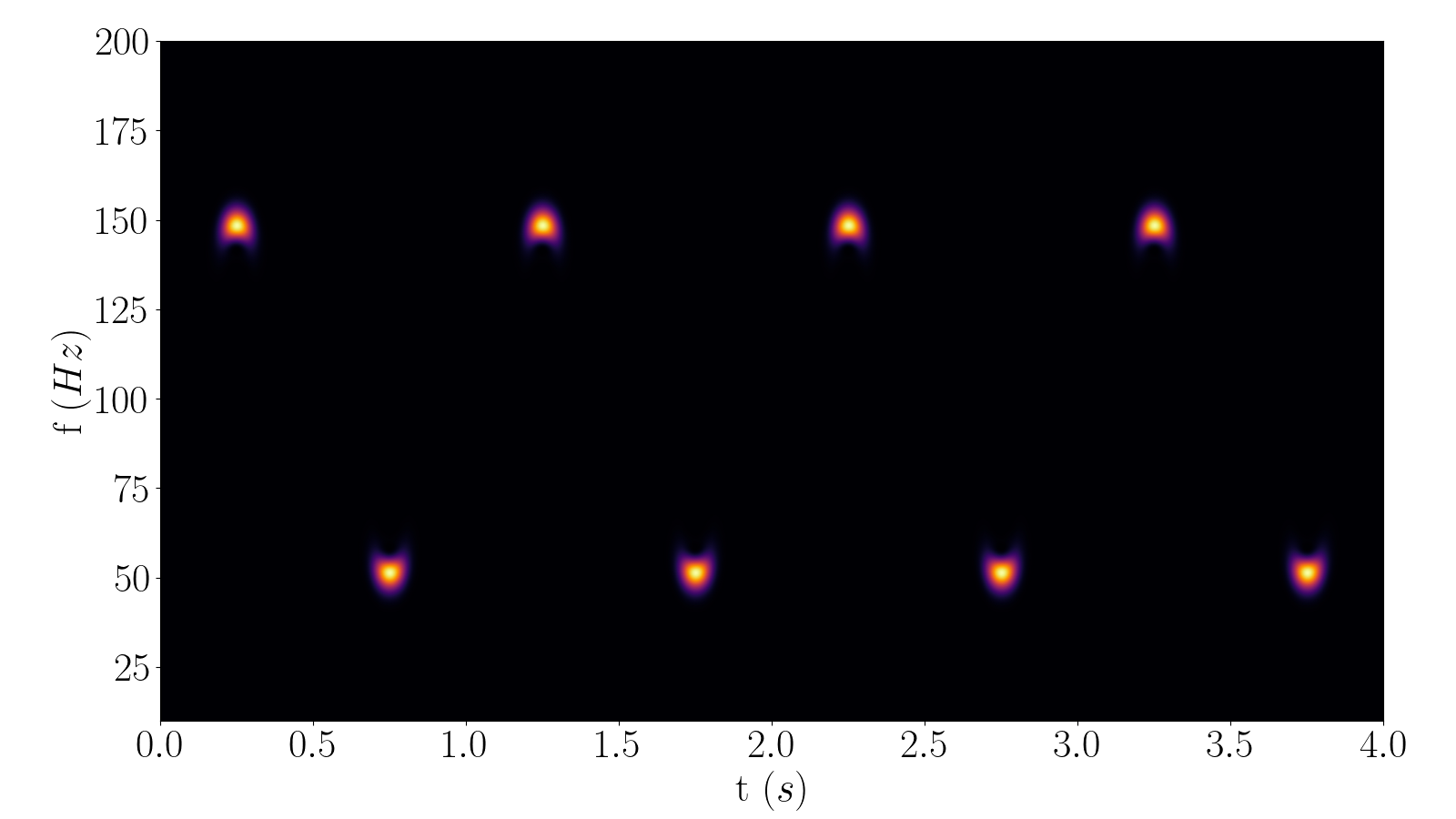}}}%=
    \\
    \vspace{-3mm}\subfloat[]{{\includegraphics[width=0.25\textwidth, trim={0cm 3mm 2cm 0cm}, clip]{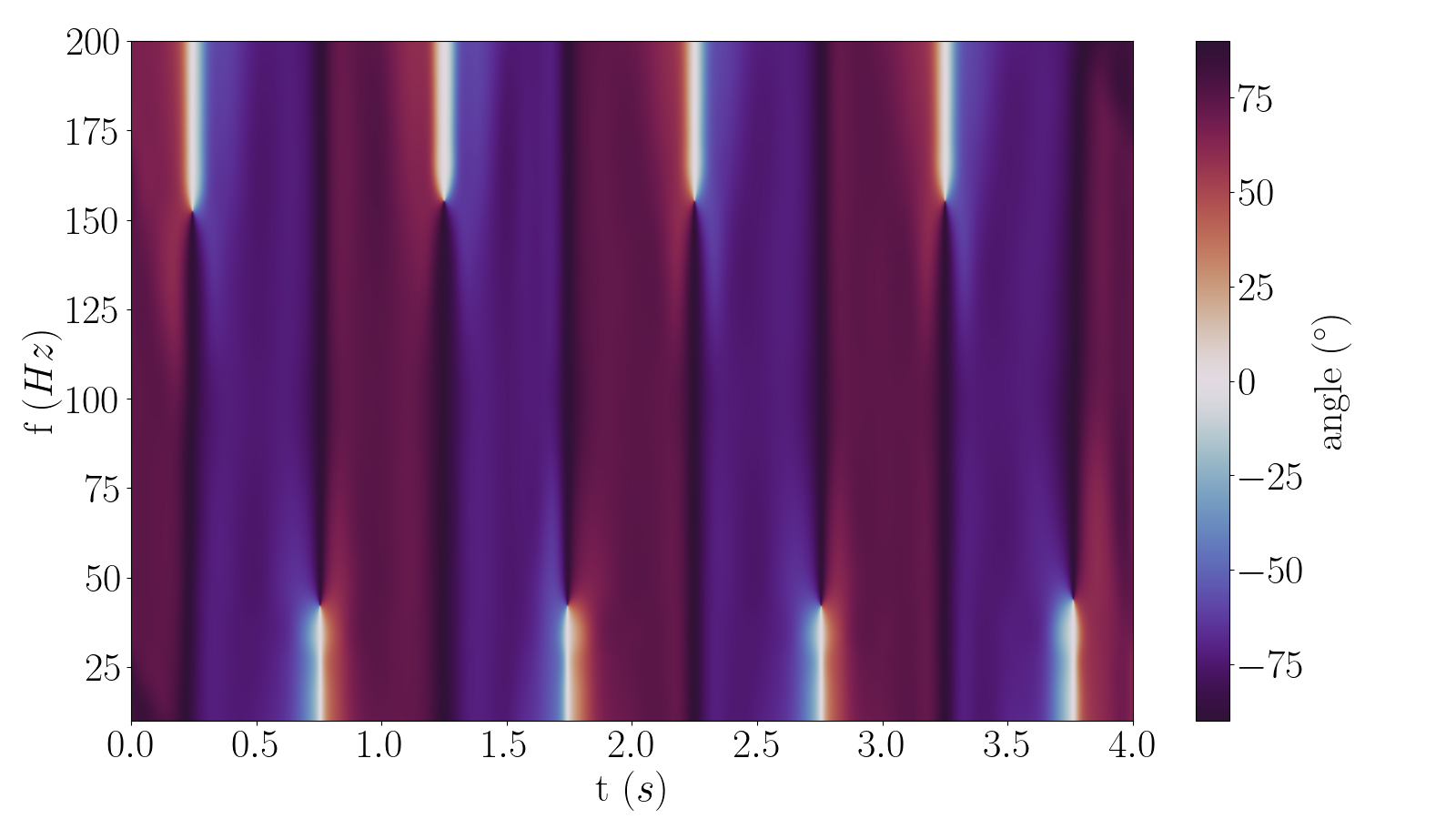}}}%=
    \subfloat[]{{\includegraphics[width=0.25\textwidth, trim={0cm 3mm 2cm 0cm}, clip]{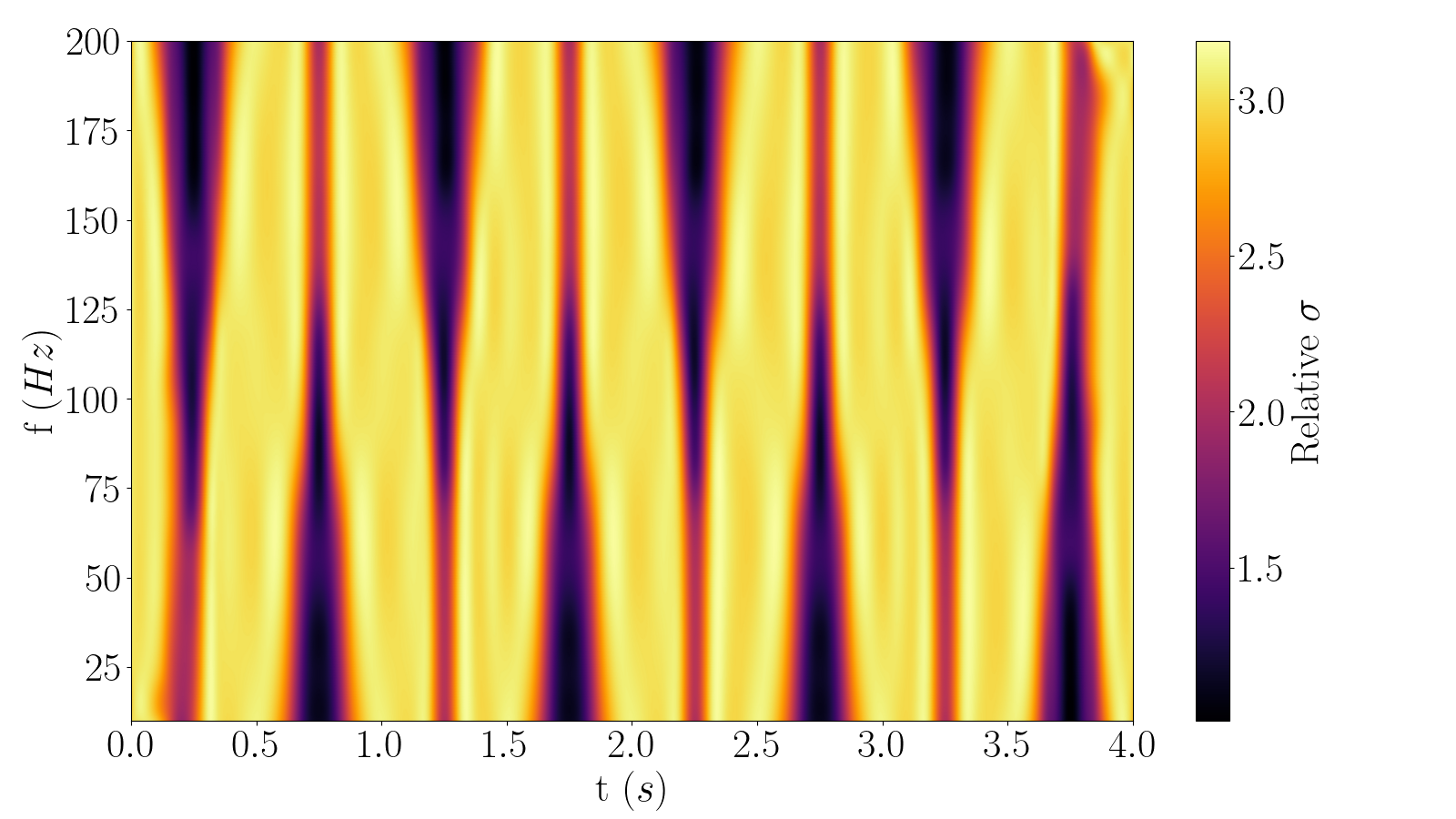}}}%=
    \subfloat[]{{\includegraphics[width=0.25\textwidth, trim={0cm 3mm 2cm 0cm}, clip]{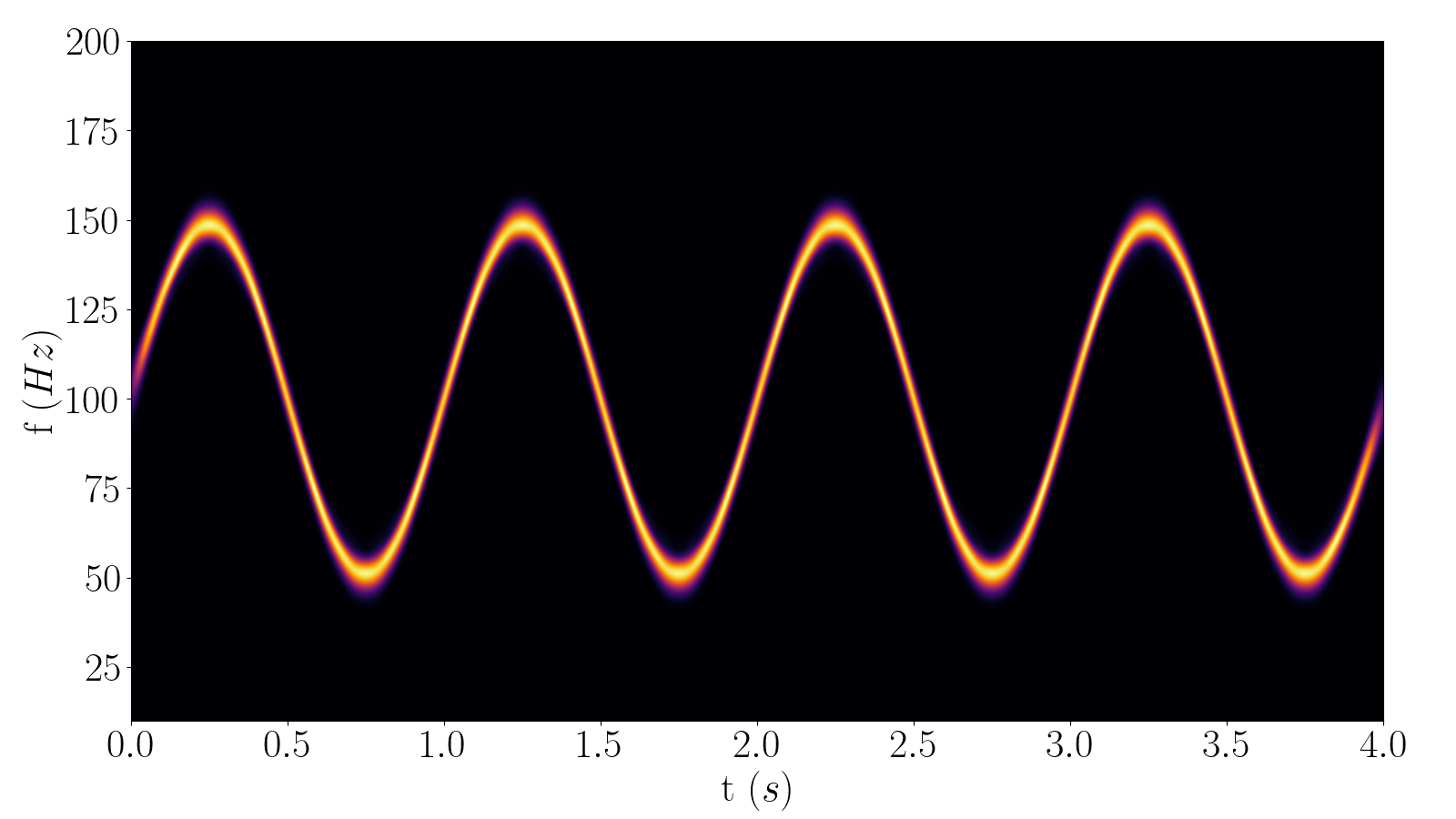}}}%=
    \subfloat[]{{\includegraphics[width=0.25\textwidth, trim={0cm 3mm 2cm 0cm}, clip]{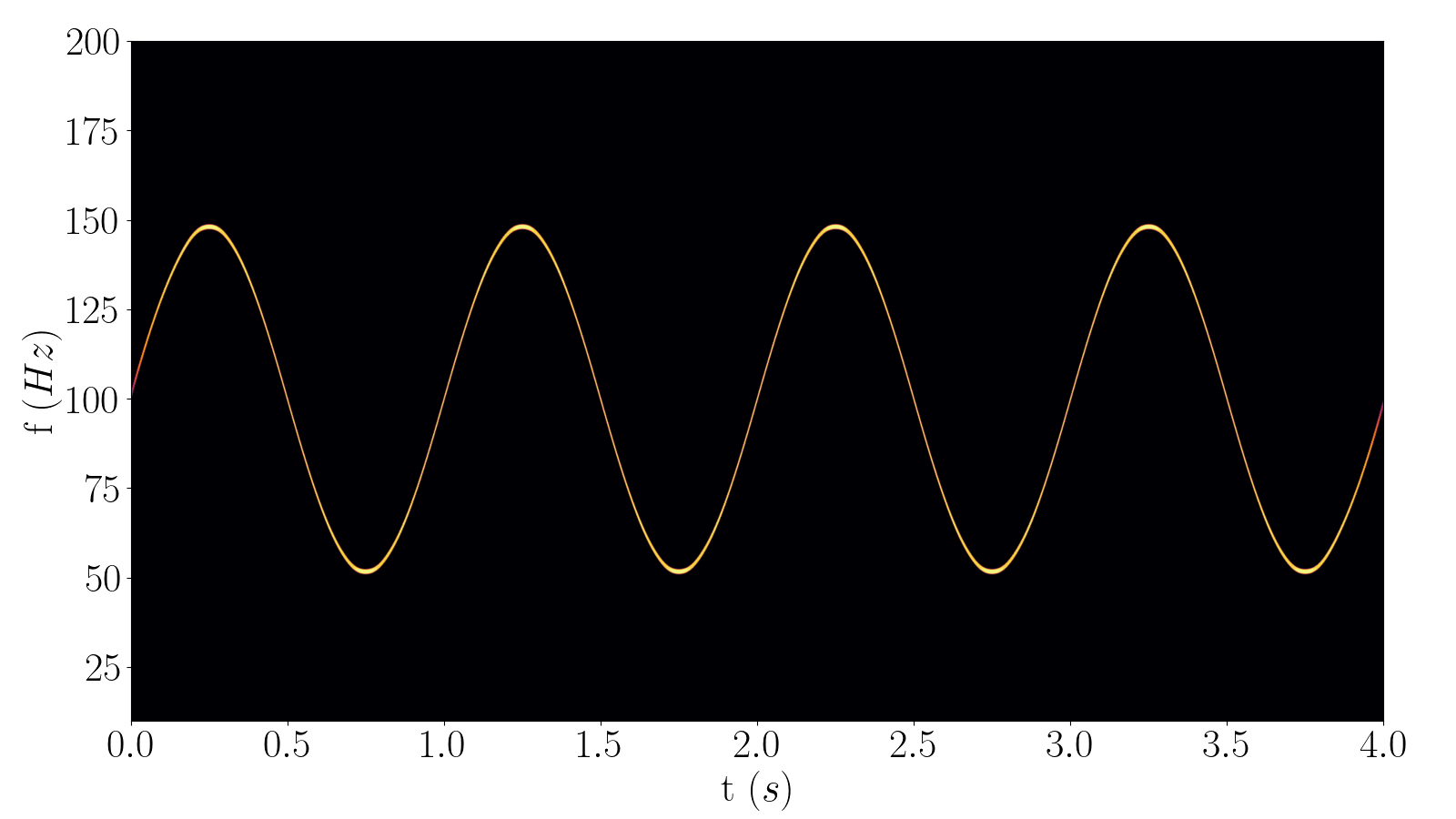}}}%=
    \vspace{-2mm}\caption{Visualisation of the Entropic filtering process for a simple sinusoidally varying sine wave, $x_5(t) = \sin{\left(2 \pi \int_0^{t}100 + 50 \sin{\left(2 \pi \tau\right)} \, d \tau\right)}$. Sub-figures (a) - (d) present the CFWT for various angles and standard deviations corresponding to the optimal fit for different sections of the sinusoid, with the corresponding Cohen's class T--F kernel for each provided in the right-hand top corner. The corresponding localised inverse perplexity (evaluated along the signal path) for each CFWT is presented in (e)-(h). The spline smoothed optimal angle (Instantaneous Phase Direction field) and relative standard deviation (with respect to $\sigma_{\text{iso}}$) are presented in (i) and (j) respectively. The corresponding optimally weighted distribution is presented in (k), and the inferred ITFR is shown in (l).\vspace{5mm}}%
    \label{fig:PROCESS}
\end{figure*}

As discussed in Section \ref{section:transform_derivation}, if suitable parameters for the CFWT defined in Eq.~\eqref{window_function_CFWT} are selected ($\theta$ and $\sigma$), cross-terms are substantially suppressed and auto-terms accentuated in the T--F representation. As examined in the overview (Section~\ref{overview}), numerous potential T--F measures exist in the literature designed to estimate the optimally aligned kernels in the time and T--F domains. For instance, T--F measures such as energy concentration \cite{Jones, jones2, Awal2017, awal2, Mohammadi} (in particular, the Energy Concentration Measure (ECM) \cite{Awal2017}), Entropy (Shannon and Rényi Entropy) \cite{Saulig2016}, and Ratio of Norms \cite{Abed} are investigated. However, as discussed, Entropic measures have only been applied in the context of time domain optimisation. Thus, as implied from the probabilistic model derived in Section \ref{section:probability}, this paper alternatively proposes a filtering scheme that provides a continuous T--F domain localised entropic weighting function, weighting individual T--F representations according to their scaled \textit{inverse perplexity}. Crucially, this enables signal components to be represented as a linear combination of several representations such that optimal kernel assignments are smooth and continuous along T--F trajectories. Entropy is particularly suitable because it provides a measure for data \textit{sparsity}. In the context of T--F analysis, an appropriately aligned T--F kernel results in an increased energy concentration along the instantaneous frequency trajectory with cross-terms suppressed, thus resulting in a local entropy minimum, and consequently, local inverse perplexity maximum. Figure~\ref{fig:PROCESS} in the SM illustrates the entropic filtering process and subsequent ITFR reconstruction to demonstrate this phenomenon. 

$\bar{P}_{u, v}^{(n,m)}$, the entropic weighting coefficient employed in Eq.~\eqref{kernels}, is defined as:

\vspace{-2.85mm} \small 
\begin{align}
    \bar{P}_{u, v}^{(n,m)} &= \frac{1}{\sum_{n'=1}^N\sum_{m'=1}^M P_{u, v}^{(n',m')}}P_{u, v}^{(n,m)}
\end{align}\normalsize \vspace{-2.85mm} 

\noindent where:

\vspace{-3.85mm} \small 
\begin{align}
    P_{u, v}^{(n,m)} &= 2^{- \alpha H_{u, v}^{(n,m)}} \\
    H_{u, v}^{(n,m)} &= -\sum_i \sum_j p_{u, v, i, j}^{(n,m)}\log_2{\left(p_{u, v, i, j}^{(n,m)}\right)}\,.\label{eq:H}
\end{align}\normalsize \vspace{-2.85mm} 

$\alpha$ is a constant that determines the ``sharpness'' of the filtering scheme, and: 

\vspace{-3.85mm} \small 
\begin{align}
    p_{u, v, i, j}^{(n,m)} &= \frac{w_{i, j}\Phi_{u - i, v - j}^{(n,m)}}{Z_{u, v}^{(n,m)}},
\end{align}\normalsize \vspace{-2.85mm} 

\noindent where $w_{i,j}^{(n,m)}$ is a weighting function, and $Z_{u, v}^{(n,m)}$ is the normalising factor:

\vspace{-2.85mm} \small 
\begin{align}
Z_{u, v}^{(n,m)} &= \sum_i \sum_j w_{i, j}\Phi_{u - i, v - j}^{(n,m)} = \left[\Phi^{(n,m)} * w\right]_{u, v}
    \end{align}
\normalsize \vspace{-3.85mm}

Note that for $\alpha \to \infty$, the resulting T--F representation approaches a hard-assigned optimal kernel fitting scheme. As discussed previously, a hard assignment is not recommended given potential discrete transitions between kernel assignments in the T--F plane. A parameter sweep of $\alpha$, provided in Fig. \ref{fig:ALPHA_SWEEP} in the SM, reveals the expected trade-off between cross-term suppression and overly sharp transitions, resulting in a peak in the combined score at $\alpha = 15$, the recommended value. Note also that $\alpha=0$ corresponds to a reconstruction of a \textit{positively-constrained} WVD, and $\alpha < 0$ provides cross-term isolation (auto-term reduction). Figure~\ref{fig:ALPHA_SWEEP} in the SM also presents a visualisation of the effect of $\alpha$ on the RIFT representations. 

In order to compute $H_{u, v}^{(n,m)}$ efficiently, the expression can be written in terms of convolution operations:

\vspace{-2.85mm} \small \begin{align}
    H_{u, v}^{(n,m)} &=\log_2{\left(\left[\Phi^{(n,m)} * w\right]_{u,v}\right)} \notag\\
    &\hspace{5mm}-\frac{1}{\left[\Phi^{(n,m)} * w\right]_{u,v}}\left(\left[\Phi^{(n,m)} * w \log_2{\left(w\right)}\right]_{u,v}\right. \notag\\
    &\hspace{10mm}
    \left.+ \left[\Phi^{(n,m)} \log_2{\left(\Phi^{(n,m)}\right)} * w\right]_{u,v} \right)
\end{align}
\normalsize\vspace{-3.85mm}

It is recommended that $w_{u,v}$ is a bivariate Gaussian, with kernel variances selected to optimise performance for the desired application. For instance, for an instrumental or speech extract with known linearly spaced harmonics, the $\omega$ axis variance can be set high to enforce harmonic trajectory alignment in the presence of noise, otherwise, the kernel can be set to an isotropic Gaussian of variance greater than that of the derived variance of the isotropic Gaussian kernel in Eq.~\eqref{log_mean} to ensure valid entropic weightings.

\vspace{-2mm} \subsection{Fundamental Properties of the RIFT}
Note that, by construction, the RIFT output, $\widehat{I}(\omega,t)$, is positive ($\widehat{I}(\omega,t)\!\ge\!0$), time–shift and frequency–shift equivariant up to discretisation ($\widehat{I}_{x_\tau}(\omega,t)=\widehat{I}_{x}(\omega,t-\tau)$ for $x_\tau(t)=x(t-\tau)$ and $\widehat{I}_{x_\nu}(\omega,t)=\widehat{I}_{x}(\omega-\nu,t)$ for $x_\nu(t)=x(t)\,e^{j2\pi\nu t}$). See Section~\ref{sm:time/freq_shift} in the SM for more details. It is also non-linear, specifically quadratic, due to the magnitude–squared of the CFWT coefficients (corresponding to energy). Likewise, by construction, the RIFT estimates the squared modulus of the ideal time–frequency representation (no phase information) and is thus non-invertible. Also, note that whilst the RIFT substantially suppresses cross-terms, it does not guarantee their elimination in all cases. Likewise, the target is a cross-term-free, WVD-resolution energy estimate closely related to, but not identical with, the conceptual ITFR (note that exact WVD “optimality” holds for linear FM components as discussed in Section \ref{sec:wvd}). Regardless, these residual WVD ``artificial'' auto terms are substantially mitigated in the RIFT through the employment of the entropic weighting scheme, CFWT wavelet basis, and the positivity-constrained LR--TV process, which suppress sign-changing (positive and negative) oscillatory artefacts (as observed in the synthetic signal results).

In terms of computational complexity, let $V$ be the number of time samples, $U$ the number of frequency bins in the T--F grid, and
$N_{k}=\sum_{n=1}^{N} M(\sigma_n)$, the total number of CFWT kernels. Given also that the LR deconvolution process runs for $K_{\mathrm{LR}}$ iterations, the RIFT complexity can be derived to be $\approx \mathcal{O}\!\left((N_{k} + K_{\mathrm{LR}})\,U\,V\log(UV)\right)$. Given that the WVD costs $\mathcal{O}\!\big(V^2 + V\,U\log U\big)$, with Cohen's class methods requiring an additional $\mathcal{O}(U\,V\log(UV))$, for $U\!\approx\!V$, the RIFT, WVD, and Cohen's class methods are all $\mathcal{O}(UV\log U)$ up to constants. The RIFT trades factors $N_k$ and $K_{\mathrm{LR}}$ against WVD’s $V^2$ and Cohen’s extra 2-D smoothing. For long signals ($V\!\gg\!U$), the $V^2$ term dominates WVD-based methods, so for small-$N_k$, the RIFT can be cheaper; for frequency-heavy grids ($U\!\gg\!V$), costs are comparable up to logarithmic factors.

\vspace{-4mm}\subsection{Kalman Filtering for Signal Component Extraction}
\label{tracking}

Whilst the RIFT provides an estimate of the ITFR, to extract individual signal trajectories from the RIFT output, this paper proposes a multi-track Kalman tracking scheme for the extraction of signal components and the subsequent construction of the Spline-RIFT variant. The following second-order model in terms of the angular velocity and acceleration (phase curvature) is proposed:

\subsubsection{Prediction Step (for trajectory \(i\), time step \(k\))}

\small  \begin{align}
    \hat{\bm{Y}}_k^{(i)} &= \bm{F} \bm{Y}_{k-1}^{(i)}, \, \text{\normalsize where: \small} \, \bm{Y}_k^{(i)} = 
\begin{bmatrix}
y_k^{(i)} \\ \dot{y}_k^{(i)}
\end{bmatrix}, \,
\mathbf{F} = \begin{bmatrix}
1 & \Delta t \\
0 & 1
\end{bmatrix},
\end{align} \normalsize \vspace{-2.85mm} 

\noindent and $y_k^{(i)}$ is the angular frequency for time step $k$ and component $i$. The predicted covariance matrix:

\vspace{-2.85mm} \small  \begin{align}
\hat{\bm{P}}_k^{(i)} = \bm{F} \bm{P}_{k-1}^{(i)} \bm{F}^T + \bm{Q}, \, \text{\normalsize where: \small} \, \bm{Q} = 
\begin{bmatrix}
\frac{\Delta t^4}{4} & \frac{\Delta t^3}{2} \\
\frac{\Delta t^3}{2} & \Delta t^2
\end{bmatrix} \varepsilon^2.
\end{align} \normalsize\vspace{-2.85mm}

\subsubsection{Data Association}

Global Nearest Neighbour (GNN) \cite{GNN} assignment can be achieved through the construction of a probability association matrix, $\mathbf{L}$, such that each $(j, i)$th entry corresponds to the probability of the \(i\)-th trajectory assigment to the \(j\)-th detection. Assuming a uniform prior over associations, $a_k^{(j \to i)}$:

\vspace{-2.85mm} \small  \begin{align}
l_{j,i} = p(a_k^{(j \to i)} \mid \bm{z}_k^{(j)}, \bm{z}_{1:n-1}^{(i)}) \propto \mathcal{N}\left(\bm{z}_k^{(j)} - \hat{\bm{Y}}_k^{(i)}, \bm{S}_k^{(i)}\right)
\end{align} \normalsize \vspace{-2.85mm} 

\noindent where:

\vspace{-2.85mm} \small  \begin{align}
\bm{S}_k^{(i)} =   \hat{\bm{P}}_k^{(i)} + \bm{R}, \, \text{\normalsize with: \small} \, \bm{R} = 
\begin{bmatrix}
\sigma_y^2 & 0 \\
0 & \sigma_{\dot{y}}^2
\end{bmatrix},
\end{align} \normalsize \vspace{-2.85mm} 

\noindent and with $\bm{z}_k^{(j)}$, the observation, obtained by Eq.~\eqref{observations}. Assignment is subsequently performed through maximisation with respect to the associations, determined by the Hungarian Sorting Algorithm \cite{kuhn1955hungarian}, with padded rows or columns of zeros employed if required to ensure the matrix is square. If there are fewer detections than current tracks, resulting tracks with no assignments are assigned to their predicted values. Likewise, tracks are initiated if multiple adjacent detections are made with no assignments for a threshold number of time steps. Likewise, if the assignment probability for a specific track falls below a certain threshold for a user-specified number of frames, the track terminates. 

\subsubsection{Update Step}

The Kalman gain is:

\vspace{-2.85mm} \small  \begin{align}
\bm{K}_k^{(i)} = \hat{\bm{P}}_k^{(i)} (\hat{\bm{P}}_k^{(i)} + \bm{R})^{-1}.
\end{align} \normalsize\vspace{-2.85mm}

The state estimate update equation:

\vspace{-2.85mm} \small  \begin{align}
\bm{Y}_k^{(i)} = \hat{\bm{Y}}_k^{(i)} + \bm{K}_k^{(i)} (\bm{z}_k^{(i)} - \hat{\bm{Y}}_k^{(i)}).
\end{align} \normalsize\vspace{-2.85mm}

The updated covariance matrix:

\vspace{-2.85mm} \small  \begin{align}
\bm{P}_k^{(i)} = (\bm{I} - \bm{K}_k^{(i)}) \hat{\bm{P}}_k^{(i)} (\bm{I} - \bm{K}_k^{(i)})^T + \bm{K}_k^{(i)} \bm{R} \bm{K}_k^{(i) \,T}.
\end{align} \normalsize\vspace{-2.85mm}

The observations are determined through:

\vspace{-2.85mm} \small  \begin{align}
\label{observations}
\bm{z}_k^{(i)} = 
\begin{bmatrix}
y_{peak}^{(i)} \\ \frac{1}{\Delta t} \tan{\left(\tilde{\theta}_{k, \,\hat{y}_k^{(i)}}\right)}
\end{bmatrix}.
\end{align} \normalsize\vspace{-2.85mm}

$y_{peak}^{(i)}$ values are computed by extracting peaks from each column of the derived ITFR array, corresponding to each time step. Likewise, $\tilde{\theta}_{u, v}$ is determined through extracting the $\theta$ value for each pixel that maximises the entropic weighting coefficient, $\bar{P}_{u, v}^{(n,m)}$:

\vspace{-2.85mm} \small  \begin{align}
    \left(\tilde{m}_{u, v}, \, \tilde{n}_{u, v}\right) &= \arg\max_{\left(m, \, n\right)}\left\{\bar{P}_{u, v}^{(n,m)}\right\} \\
    \Rightarrow \left(\tilde{\theta}_{u, v}, \, \tilde{\sigma}_{u, v}\right) &= \left(\theta\left(\tilde{m}_{u, v}\right), \, \sigma\left(\tilde{n}_{u, v}\right)\right)
\end{align} \normalsize \vspace{-2.85mm} 

\noindent where $\tilde{\sigma}_{u, v}$ is the $\sigma$ value corresponding to the obtained maximum and $\theta\left(m\right)$ and $\sigma\left(n\right)$ are given by Eq.~\eqref{actual theta(m)} and Eq.~\eqref{sigma(n)} respectively. $\tilde{\theta}_{u, v}$ can be considered the \textit{Instantaneous Phase Direction} (IPD) field. The IPD field provides insights into the localised curvature of the signal at any region relative to the discretised T--F plane, as shown in the results Section \ref{results}. Likewise, an \textit{Instantaneous Phase Curvature} (IPC) field, $\frac{\partial^2 \phi_{u, v}}{\partial t^2}$, can also be obtained:

\vspace{-2.85mm} \small  \begin{align}
\label{curvature}
\frac{\partial^2 \phi_{u, v}}{\partial t^2} &= \tan{\left(\theta_{\text{iso}}\left(\tilde{m}_{u,v}\right)\right)}.
\end{align} \normalsize \vspace{-2.85mm} 

\noindent where $\theta_{\text{kernel}}\left(m\right)$ is given by Eq.~\eqref{theta(m)}. Note that $\theta\left(m\right)$, Eq.~\eqref{actual theta(m)}, can alternatively be used in Eq.~\eqref{curvature} to provide an IPC field defined relative to the discretised grid, as employed in Eq.~\eqref{observations} as the observed velocity. 

Given that the optimised extracted value for $\tilde{\theta}_{u, v}$ will be discretised with respect to the kernel $\theta\left(m\right)$ values, 2D spline interpolation can be employed to provide a smoothed function. Given that the IPD field is ``wrapped'' about $(-\pi/2, \pi/2)$, the following formulation for the smoothed IPD field, $\text{IPD}_{u,v}$ can be applied:

\vspace{-2.85mm} \small 
\begin{align}
    \text{IPD}_{u,v} &= \frac{1}{2} \arctan{\left(\frac{S\left[\sin{\left(2\tilde{\theta}_{u, v}\right)}\right]}{S\left[\cos{\left(2\tilde{\theta}_{u, v}\right)}\right]}\right)},
\end{align}
\normalsize \vspace{-2.85mm} 

\noindent where $S\left[X_{u,v}\right]$ is the rectangle bivariate spline \cite{deBoor1978} of $X_{u,v}$.

To provide the Spline-RIFT array, the extracted trajectories can then be rasterised on the discretised grid using the Xiaolin Wu's antialiased line algorithm \cite{wu1991aa}, as described in Section~\ref{sec:itfr}. 

\begin{algorithm}[t]
\begingroup
\footnotesize
\caption{Pseudocode for the RIFT / Spline-RIFT}
\label{alg:rift}
\begin{algorithmic}[1]
\Require audio $x[p]$; grid $(\omega_i, t_j)$ with steps $(\Delta \omega,\Delta t)$; CFWT constellation size $N \times M$; log-normal width $\sigma_{\ell}$; entropic filtering sharpness $\alpha$; LR iterations $K_{LR}$; number LR block partitions $L,H$; TV weight $\lambda$; entropic window $w(\omega, t)$
\Ensure RIFT $\widehat{I}(\omega_i, t_j)$; IPD field $\Theta(\omega_i, t_j)$; IPC field $\Gamma(\omega_i, t_j)$; trajectories $\mathcal{T}$; Spline-RIFT $\tilde{I}(\omega_i, t_j)$
\State $z[p]\gets x[p]+j\,\mathcal{H}\{x[p]\}$ \Comment{analytic signal}
\State $\sigma_{\mathrm{iso}}\gets\sqrt{\Delta t/\Delta\omega}$; select deterministic range of $\{\sigma(n)\}_{n=1}^{N}$
\For{$n=1$ \textbf{to} $N$}
  \State choose $M(\sigma(n))$; select uniform range of $\{\theta(m)\}_{m=1}^{M(\sigma(n))}$
  \State build normalised wavelets $W_{\sigma(n),\theta(m),\omega}$ and corresponding Cohen's class kernels $\Pi^{(n,m)}$
\EndFor
\ForAll{$(n,m)$ \textbf{in parallel}}
  \State $\Phi^{(n,m)}(\omega_i, t_j)\gets\big|\,z \ast W_{\sigma(n),\theta(m),\omega}\,\big|^2$
  \State $H^{(n,m)}\gets \mathrm{localEntropy}\!\left(\Phi^{(n,m)},w\right)$;\quad $P^{(n,m)}\gets 2^{-\alpha H^{(n,m)}}$
\EndFor
\State $\widehat{P}^{(n,m)}\gets P^{(n,m)}\Big/\sum_{n',m'} P^{(n',m')}$
\State $\Phi \gets \sum_{n,m}\widehat{P}^{(n,m)}\left(\Phi^{(n,m)} \ast \Pi^{(n,m)}\right)$
\State $\Psi \gets \sum_{n,m}\widehat{P}^{(n,m)}\left(\Pi^{(n,m)} \ast \Pi^{(n,m)}\right)$
\State Initialise $\widehat{I}_{0}\ge 0$
\For{$k=1$ \textbf{to} $K_{LR}$} \Comment{Lucy--Richardson with TV}
  \ForAll{$(l,h)$ \textbf{in parallel}} \Comment{blockwise update}
    \State update block $\widehat{I}_{k}^{(l,h)}$ using the blockwise LR--TV
  \EndFor
\EndFor
\State stitch $\widehat{I}^{(l,h)}$ blocks into $\widehat{I}$
\State $(\tilde m,\tilde n)(\omega_i, t_j)\gets \arg\max_{m,n}\widehat{P}^{(n,m)}(\omega_i, t_j)$
\State $\Theta(\omega_i, t_j)\gets \tfrac{1}{2}\,\mathrm{arctan}\!\big(\mathcal{S}[\sin(2\theta(\tilde m)]/\,\mathcal{S}[\cos(2\theta(\tilde m)]\big)$
\State $\Gamma(\omega_i, t_j)\gets \tan\!\big(\theta^{\mathrm{ker}}_{\tilde m}\big)$
\For{each $t_j$} \Comment{trajectory extraction}
  \State detect peaks $\{y_r^{(j)}\}$ in $\widehat{I}(:,t_j)$
  \State track with second order Kalman filter; associate by global nearest neighbour (GNN)
\EndFor
\State $\mathcal{T}\gets\{\tau_i\}$ \Comment{set of tracks}
\State \textbf{Spline-RIFT:} rasterise each $\tau_i$ on the T--F grid via the anti-aliased Wu line algorithm, using $\widehat{I}$ along the curve
\State \Return $\widehat{I},\,\Theta,\,\kappa,\,\mathcal{T}, \, \tilde{I}$
\end{algorithmic}
\endgroup
\end{algorithm}\vspace{-4mm}

\normalsize

\vspace{-2.5mm}\section{Results}
\label{results}

\begin{figure*}[!t]
  \centering
  \subfloat[]{
    \begin{minipage}[t]{0.245\textwidth}\centering
      \includegraphics[width=\linewidth,trim={0cm 4mm 0cm 0cm},clip]{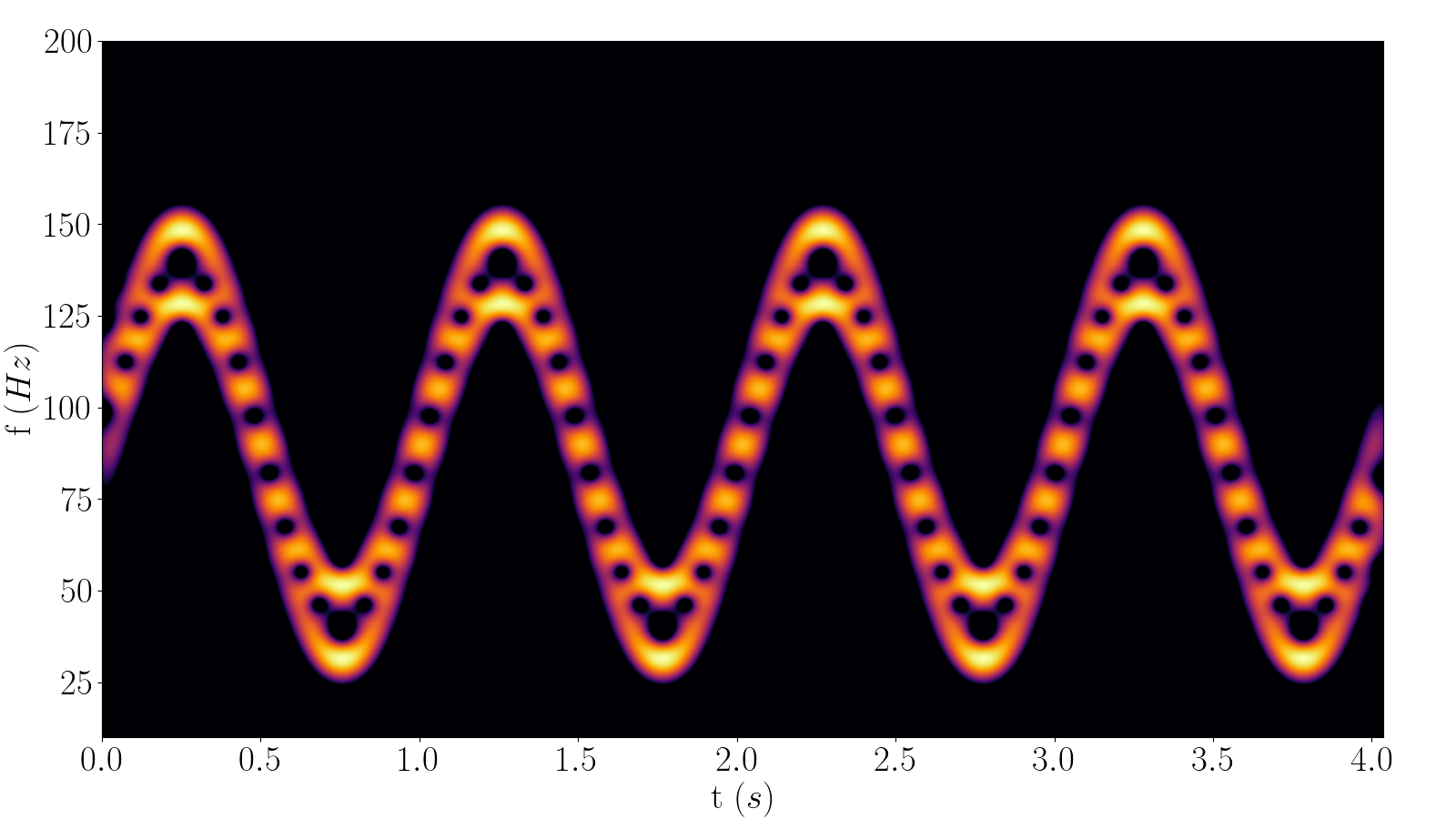}\\[-2mm]
      \includegraphics[width=\linewidth,trim={0cm 4mm 0cm 0cm},clip]{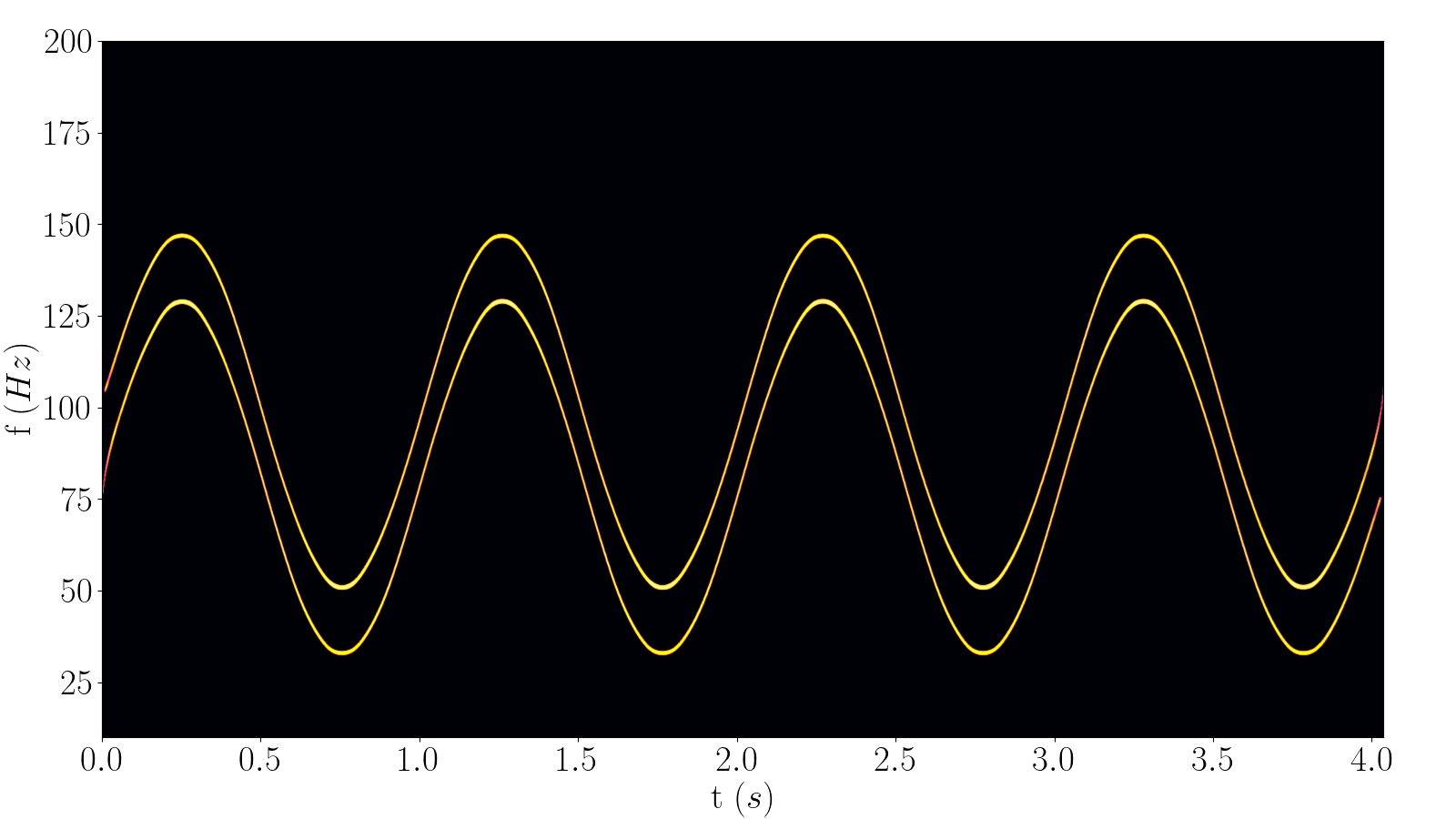}\\[-2mm]
      \includegraphics[width=\linewidth,trim={0cm 4mm 0cm 0cm},clip]{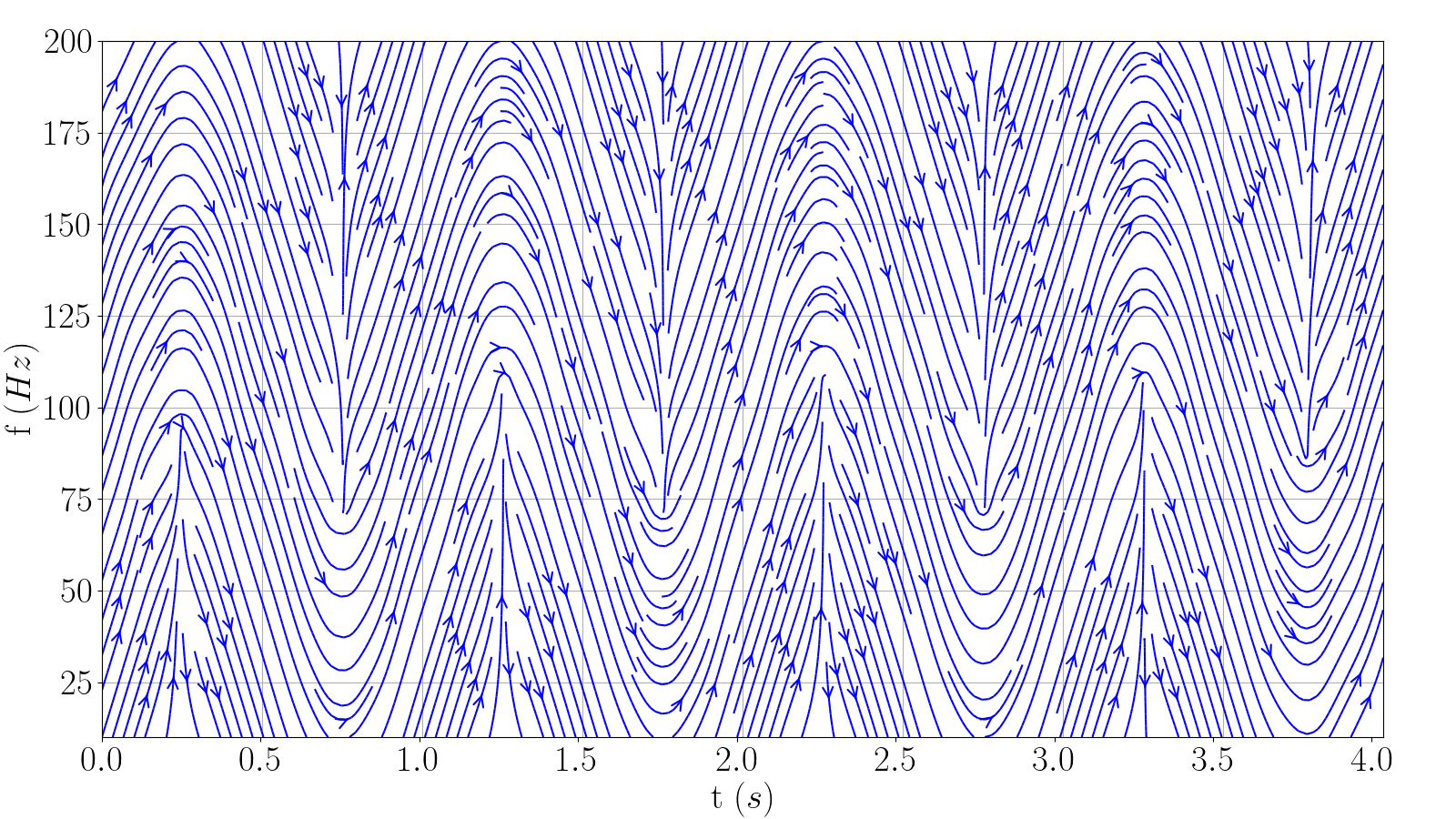}\\[-2mm]
      \includegraphics[width=\linewidth,trim={0cm 0mm 0cm 0cm},clip]{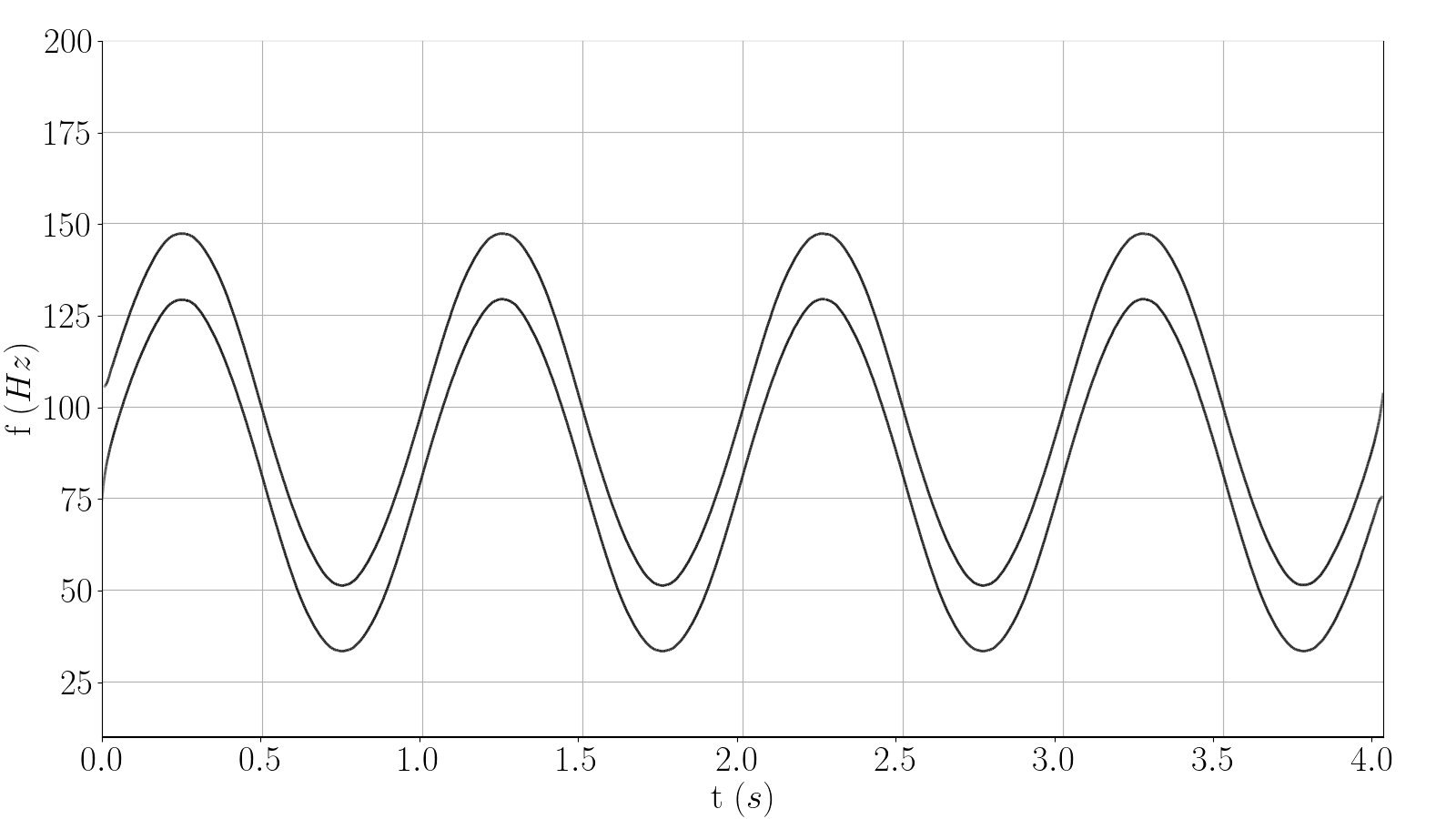}\\[-0mm]
      \includegraphics[width=\linewidth,trim={0cm 2mm 0cm 0cm},clip]{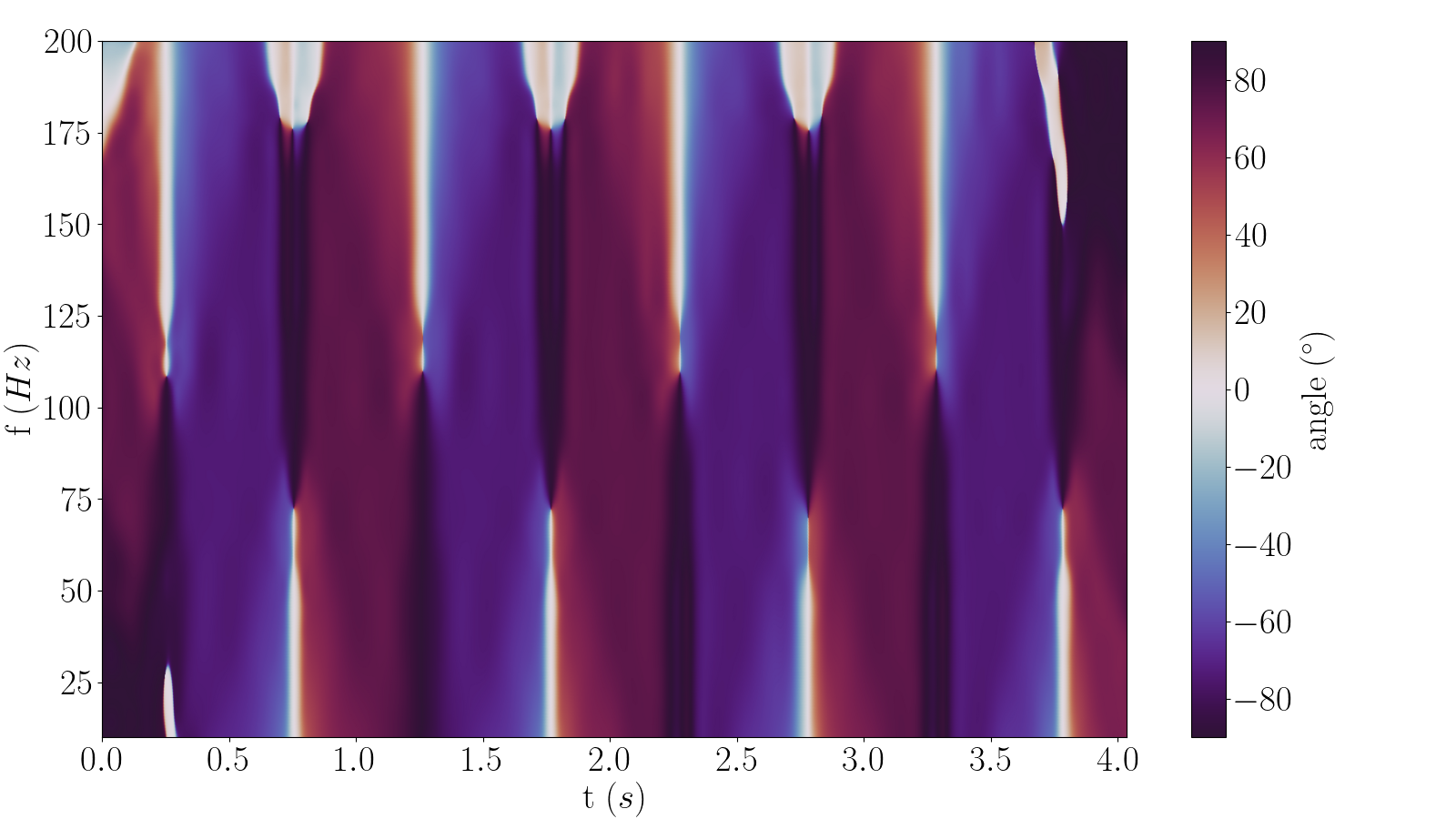}
    \end{minipage}
  }
  \hspace{-3mm}\subfloat[]{
    \begin{minipage}[t]{0.245\textwidth}\centering
      \includegraphics[width=\linewidth,trim={0cm 4mm 0cm 0cm},clip]{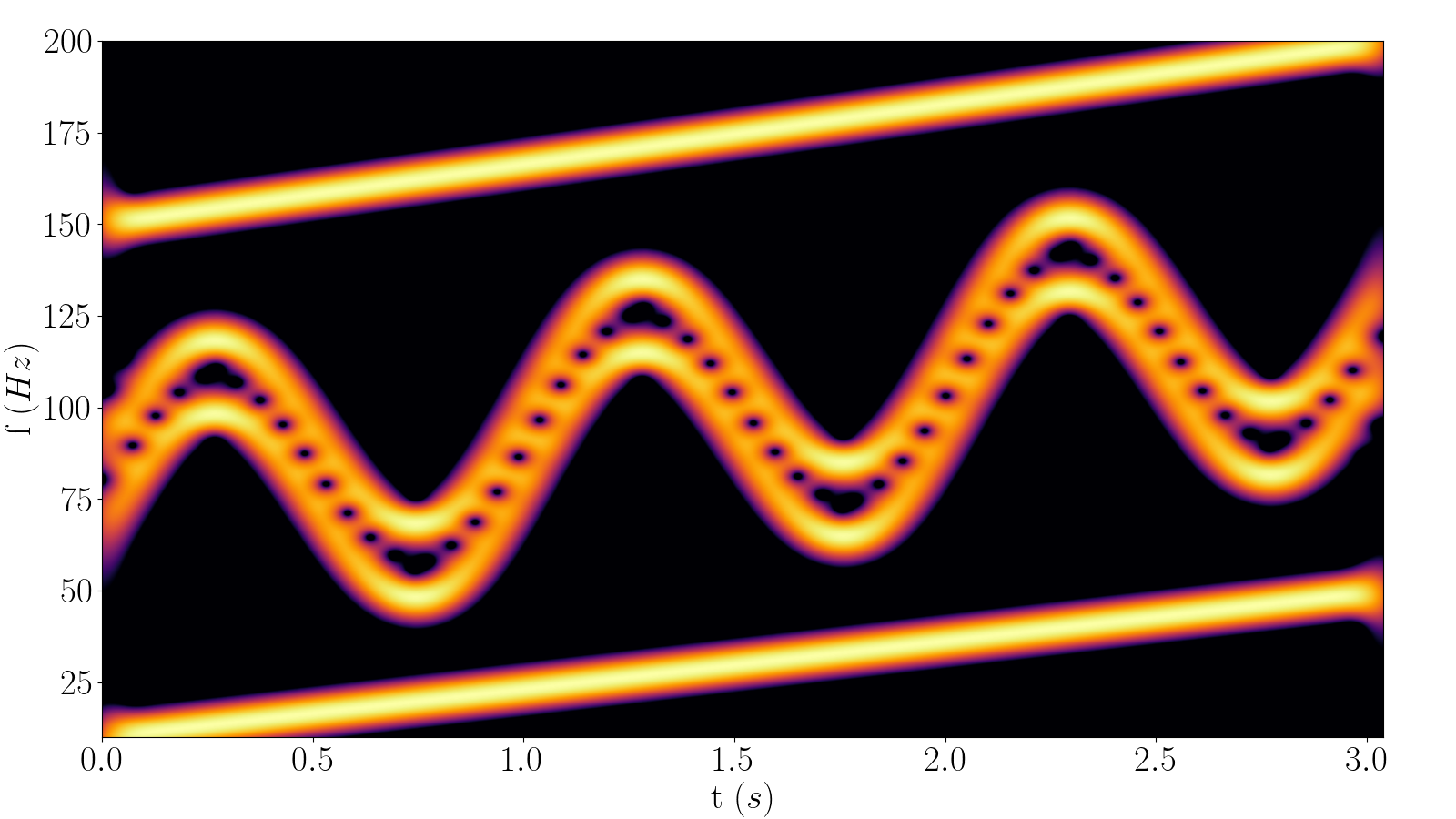}\\[-2mm]
      \includegraphics[width=\linewidth,trim={0cm 4mm 0cm 0cm},clip]{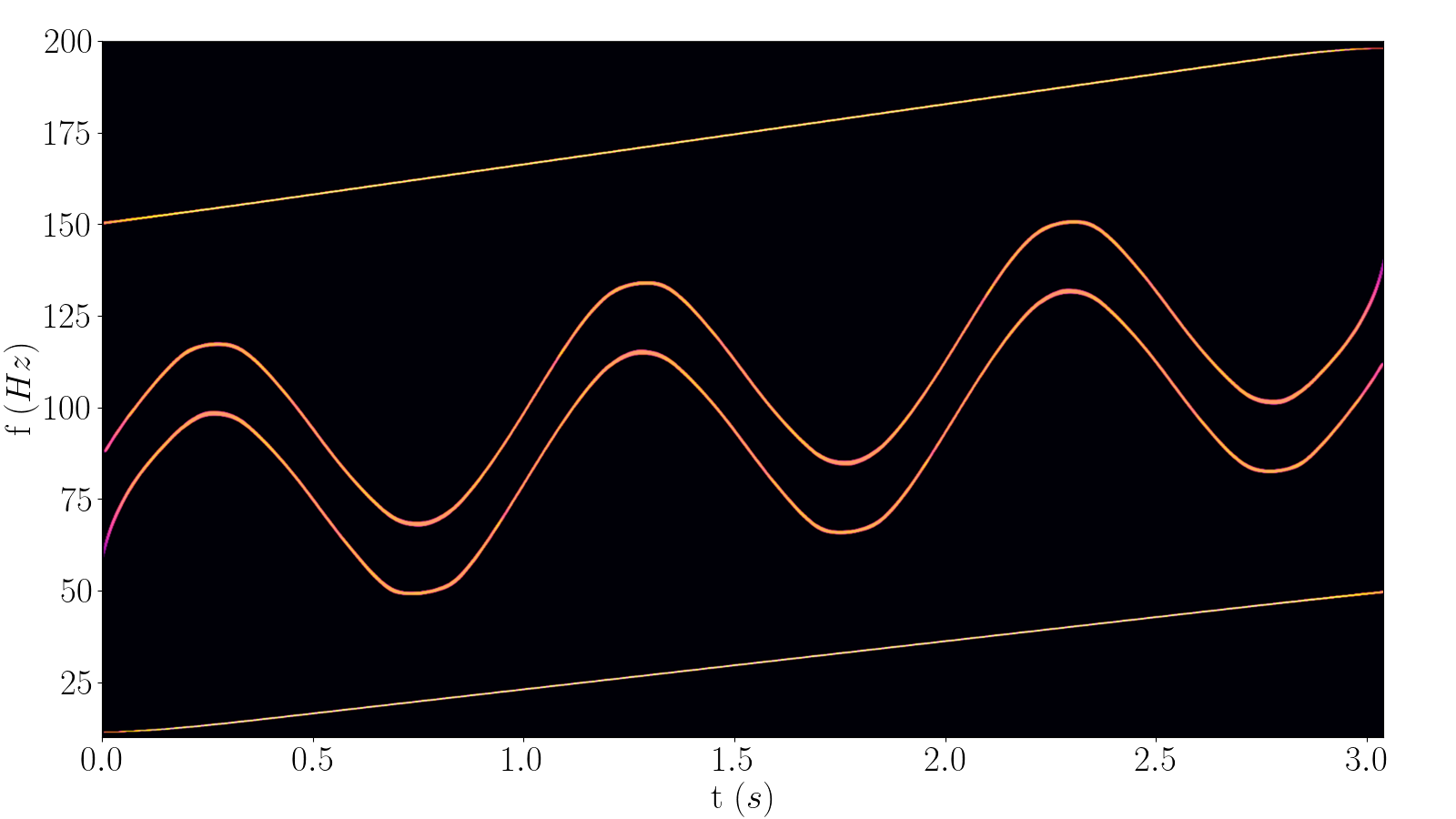}\\[-2mm]
      \includegraphics[width=\linewidth,trim={0cm 4mm 0cm 0cm},clip]{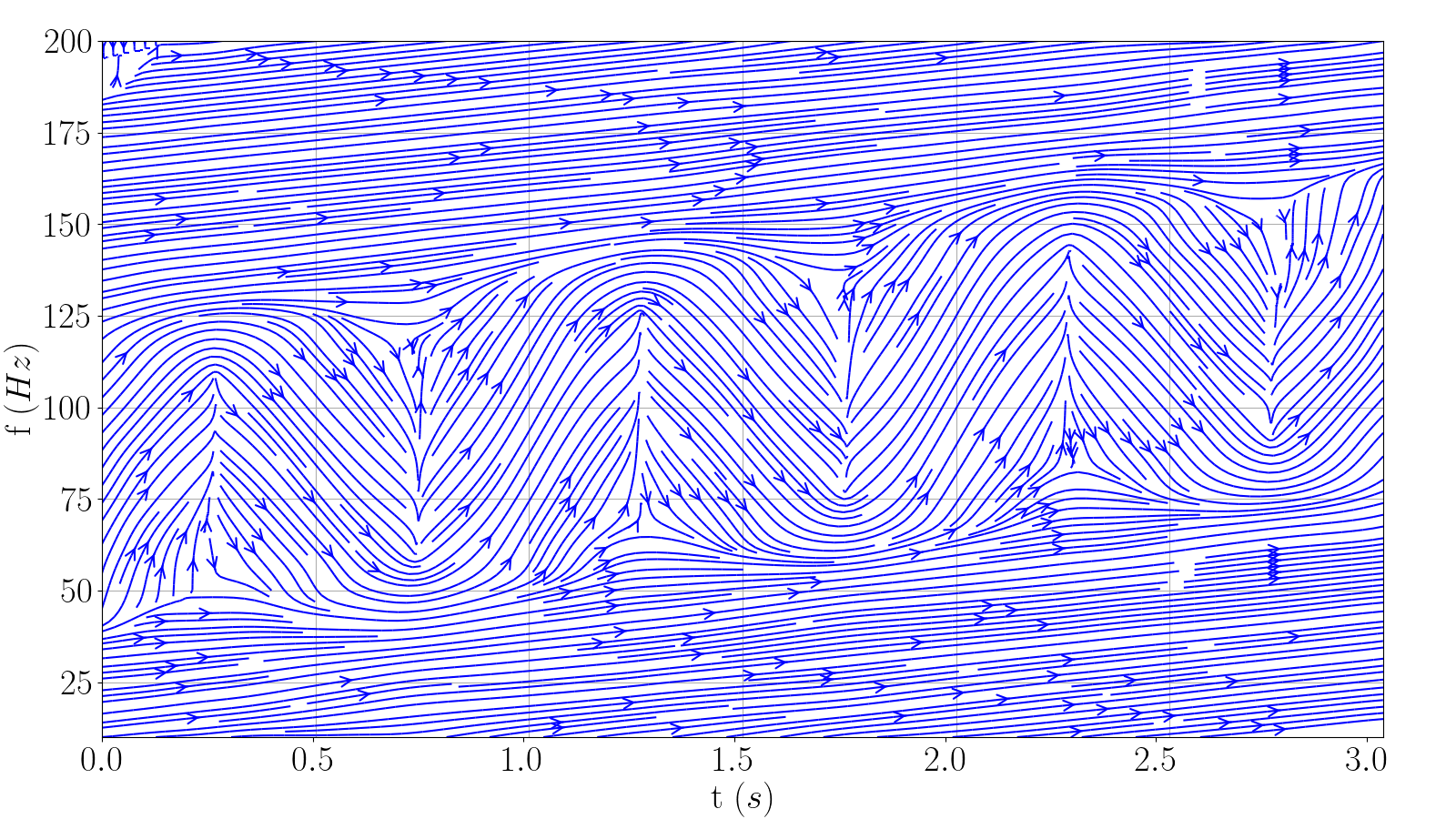}\\[-2mm]
      \includegraphics[width=\linewidth,trim={0cm 0mm 0cm 0cm},clip]{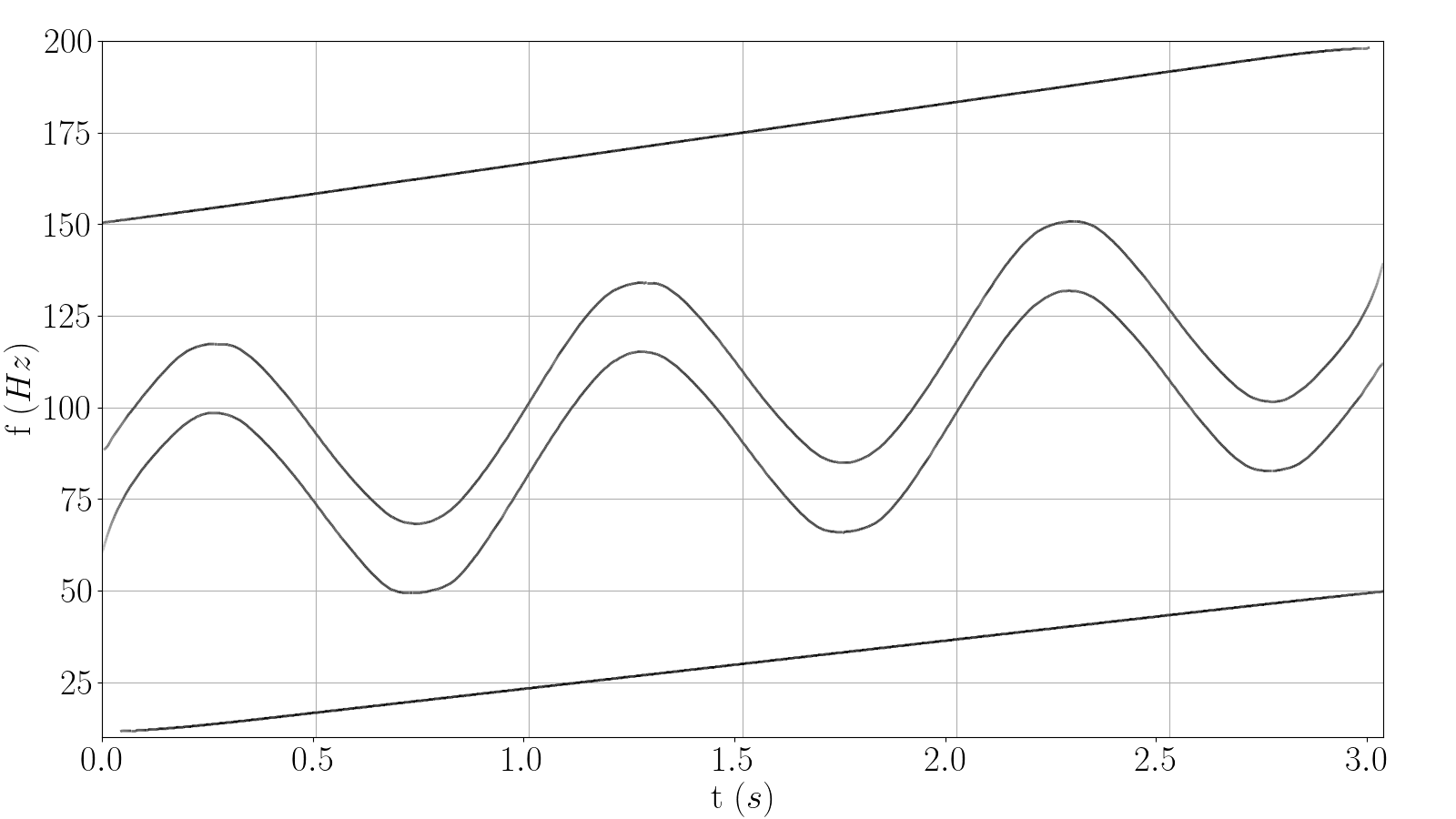}\\[-0mm]
      \includegraphics[width=\linewidth,trim={0cm 2mm 0cm 0cm},clip]{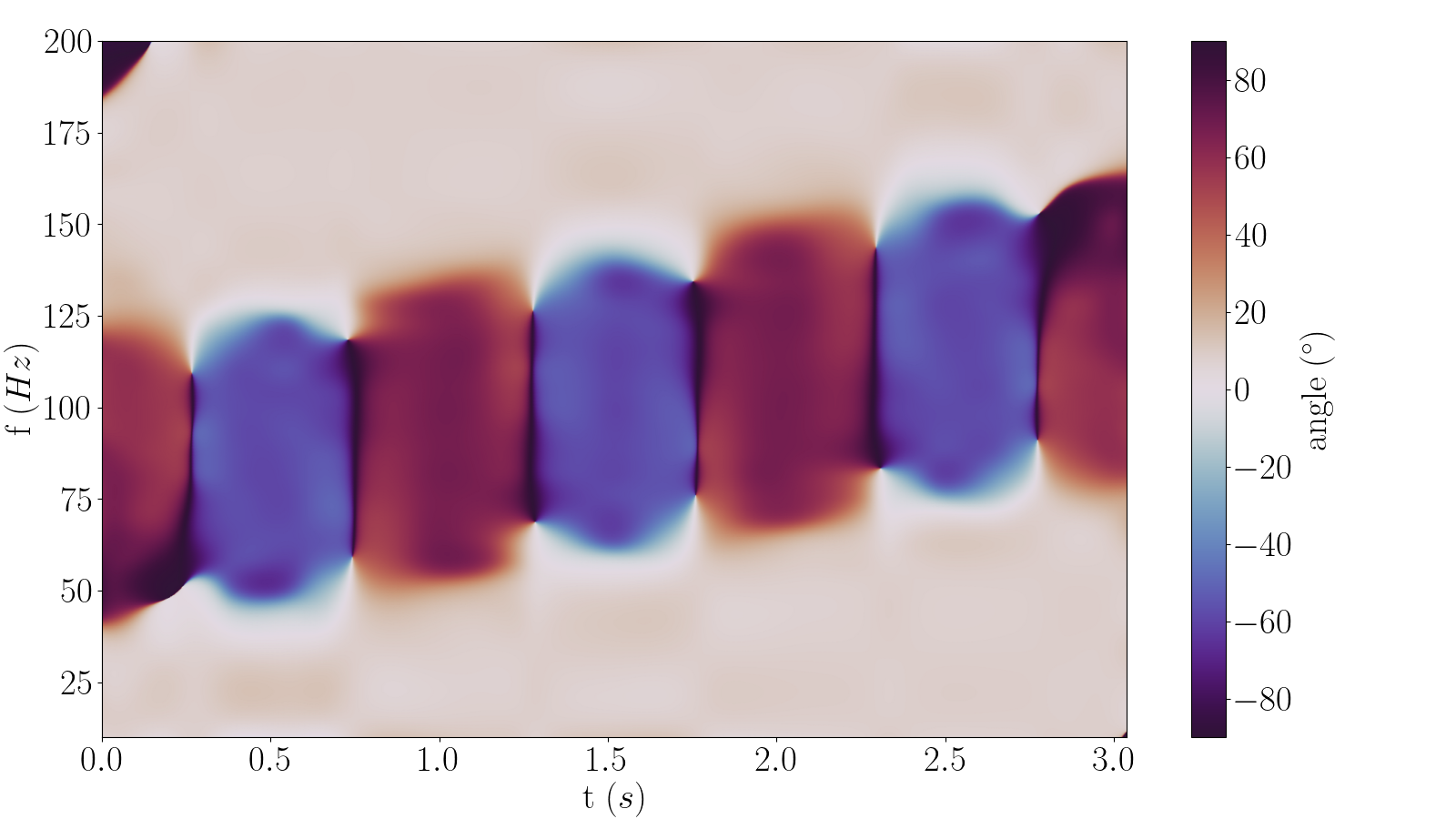}
    \end{minipage}
  }
  \hspace{-3mm}\subfloat[]{
    \begin{minipage}[t]{0.245\textwidth}\centering
      \includegraphics[width=\linewidth,trim={0cm 4mm 0cm 0cm},clip]{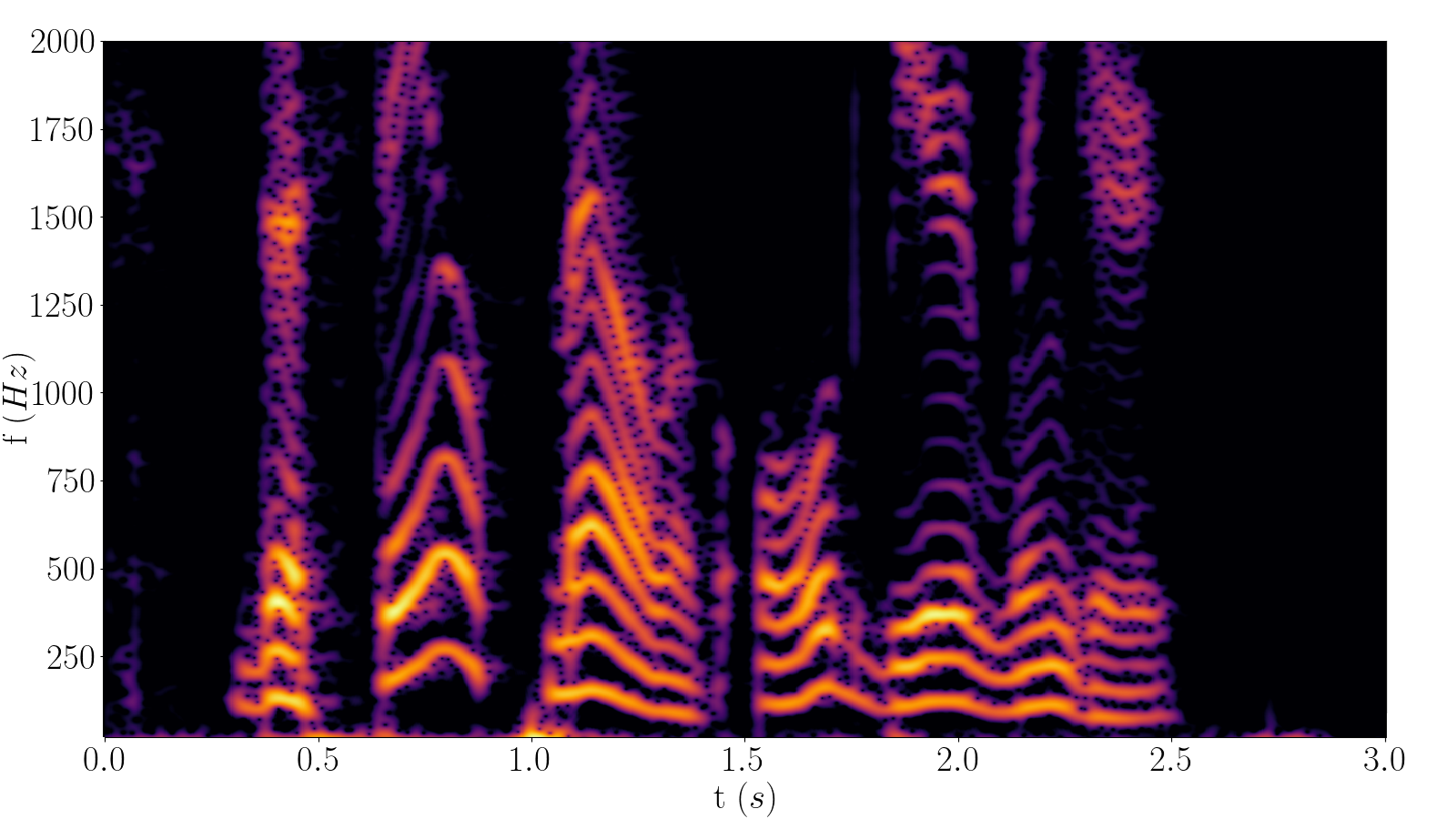}\\[-2mm]
      \includegraphics[width=\linewidth,trim={0cm 4mm 0cm 0cm},clip]{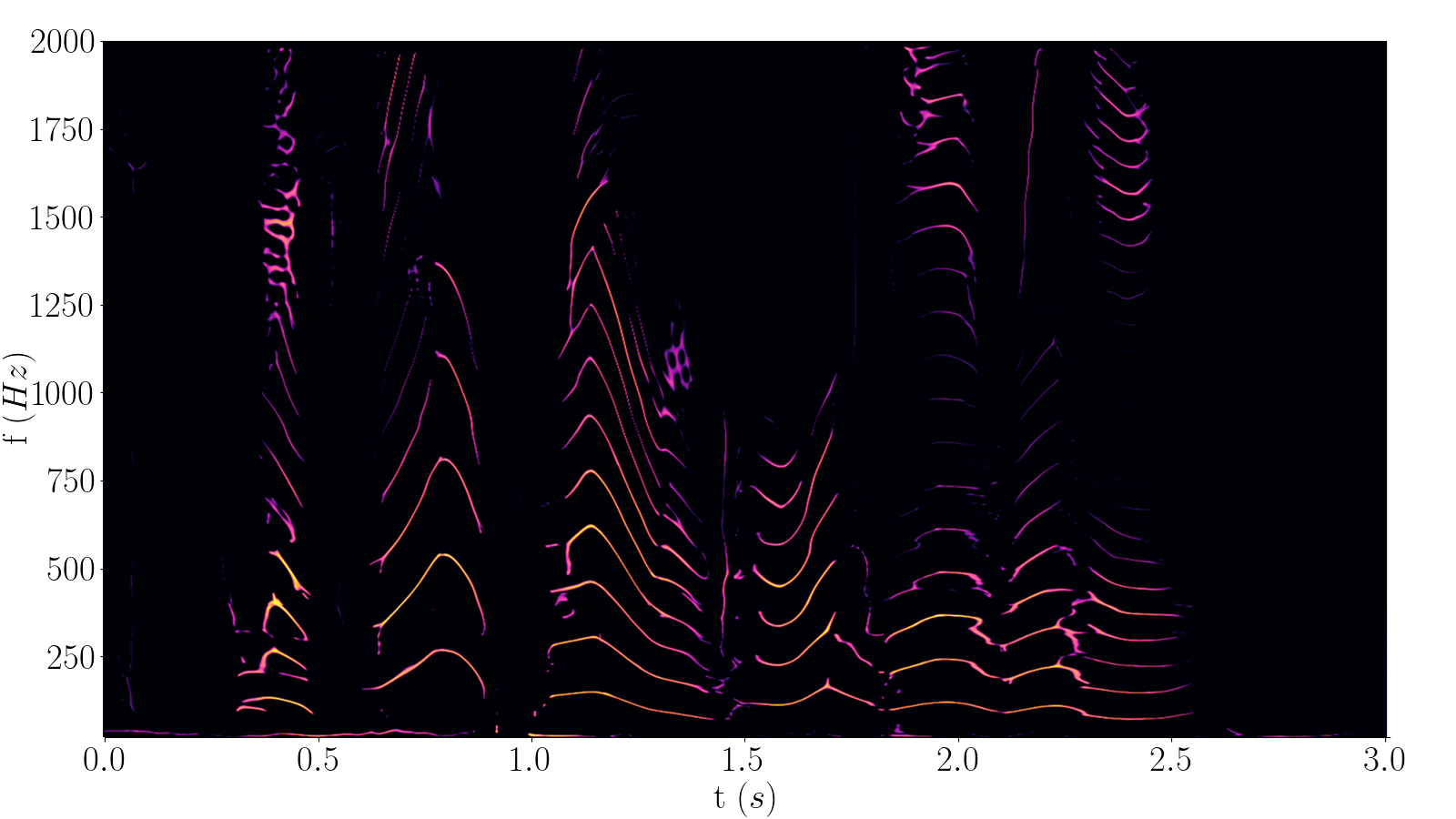}\\[-2mm]
      \includegraphics[width=\linewidth,trim={0cm 4mm 0cm 0cm},clip]{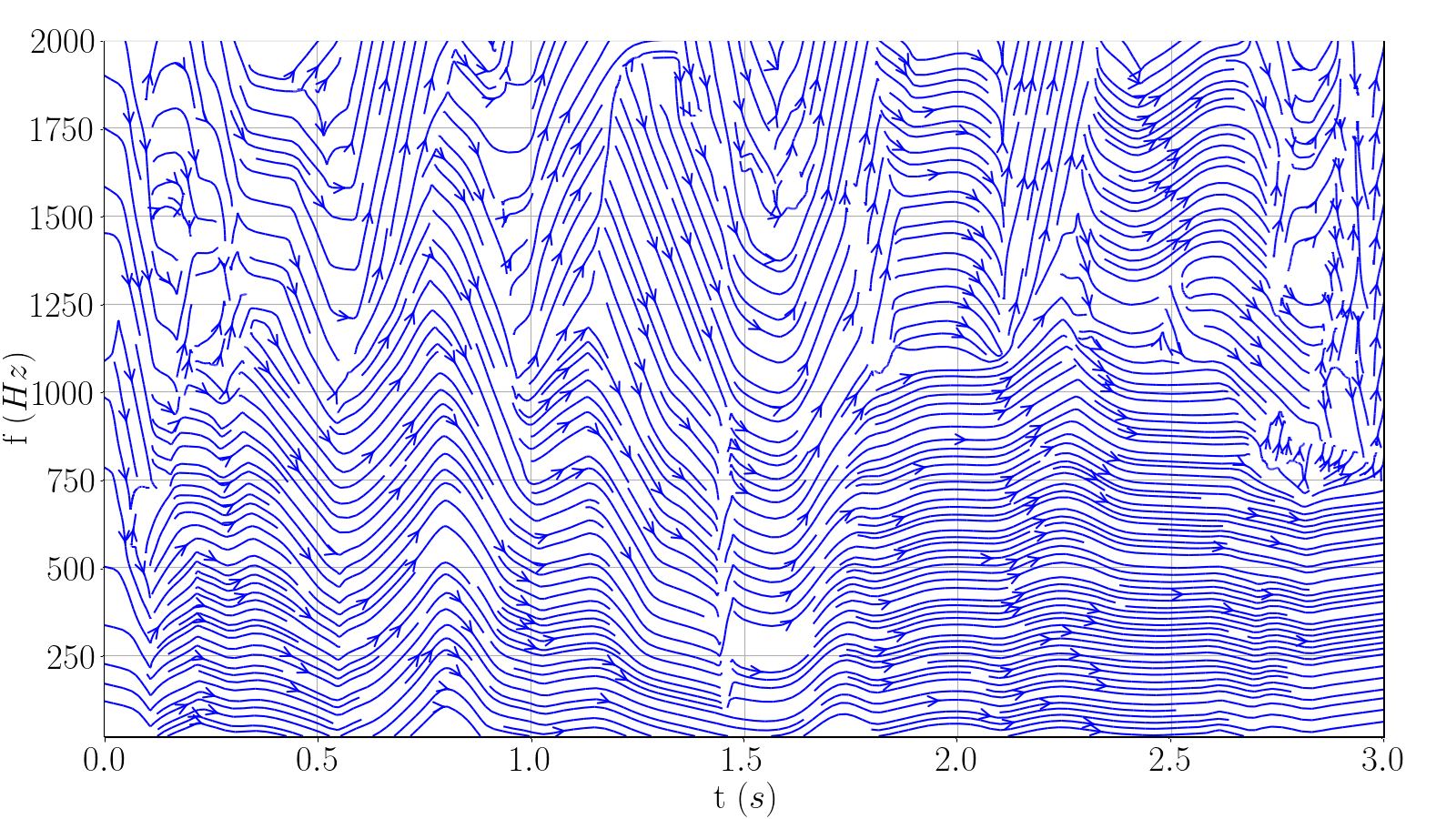}\\[-2mm]
      \includegraphics[width=\linewidth,trim={0cm 0mm 0cm 0cm},clip]{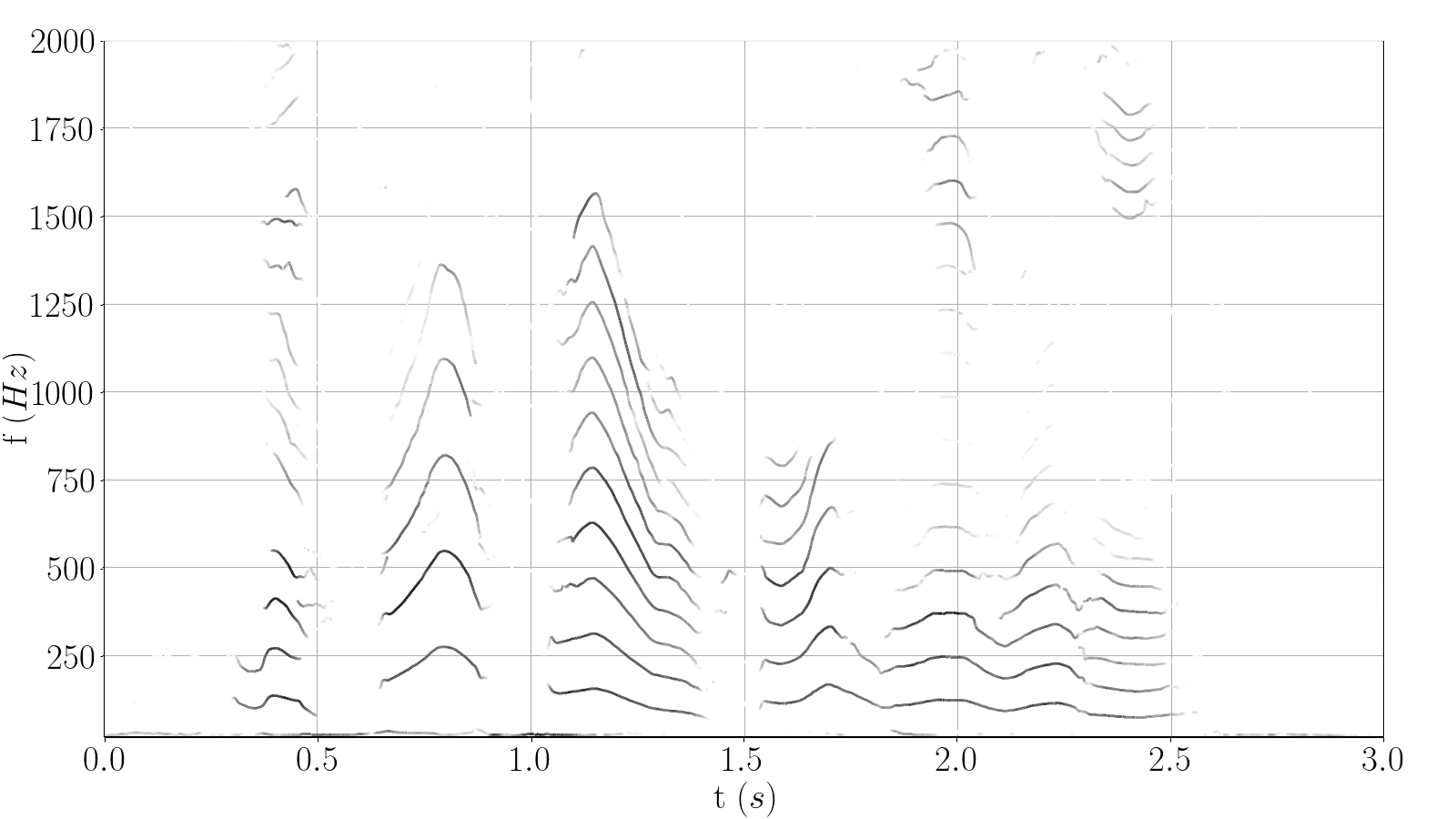}\\[-0mm]
      \includegraphics[width=\linewidth,trim={0cm 2mm 0cm 0cm},clip]{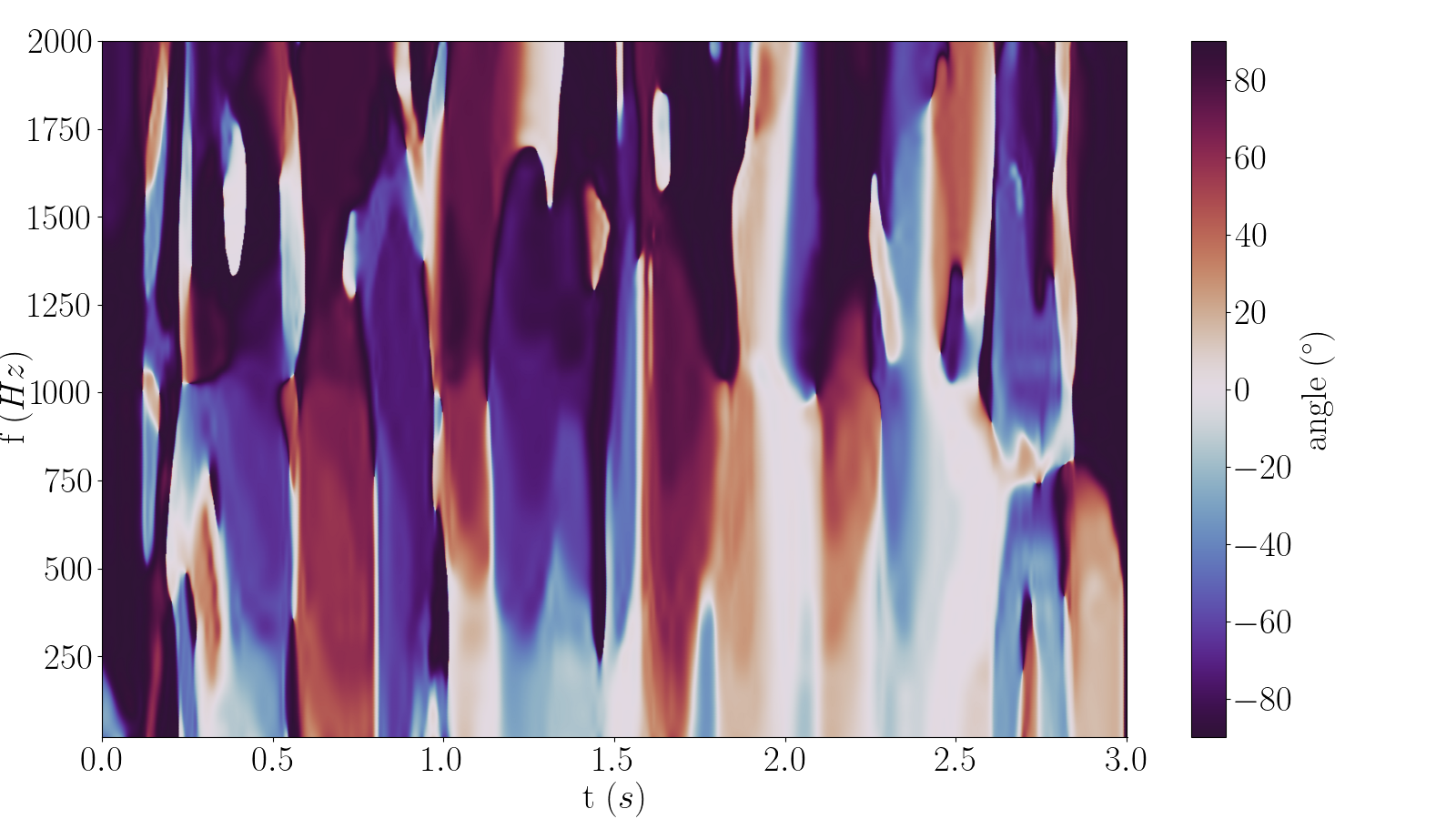}
    \end{minipage}
  }  
  \hspace{-3mm}\subfloat[]{
    \begin{minipage}[t]{0.245\textwidth}\centering
      \includegraphics[width=\linewidth,trim={0cm 4mm 0cm 0cm},clip]{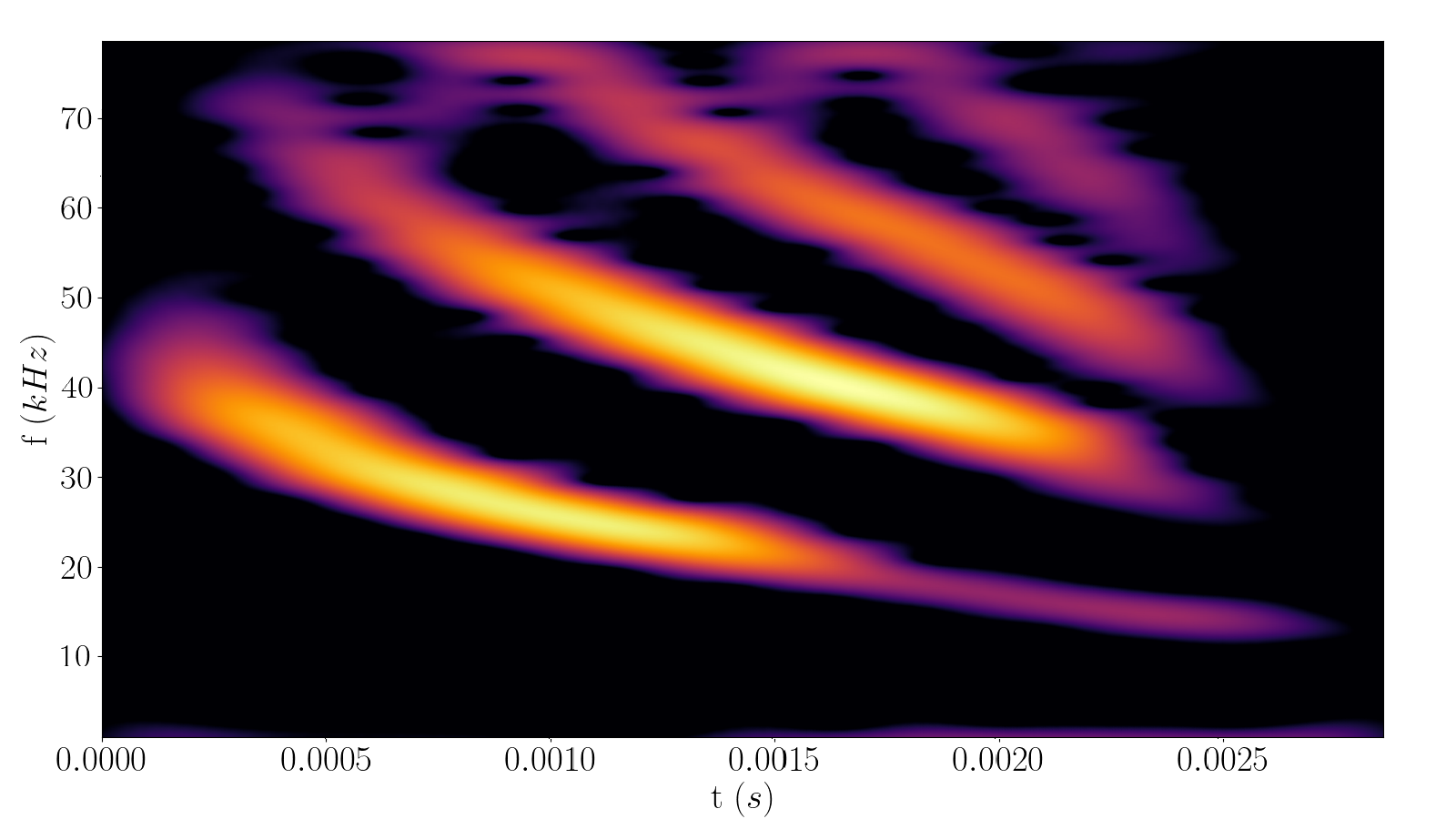}\\[-2mm]
      \includegraphics[width=\linewidth,trim={0cm 4mm 0cm 0cm},clip]{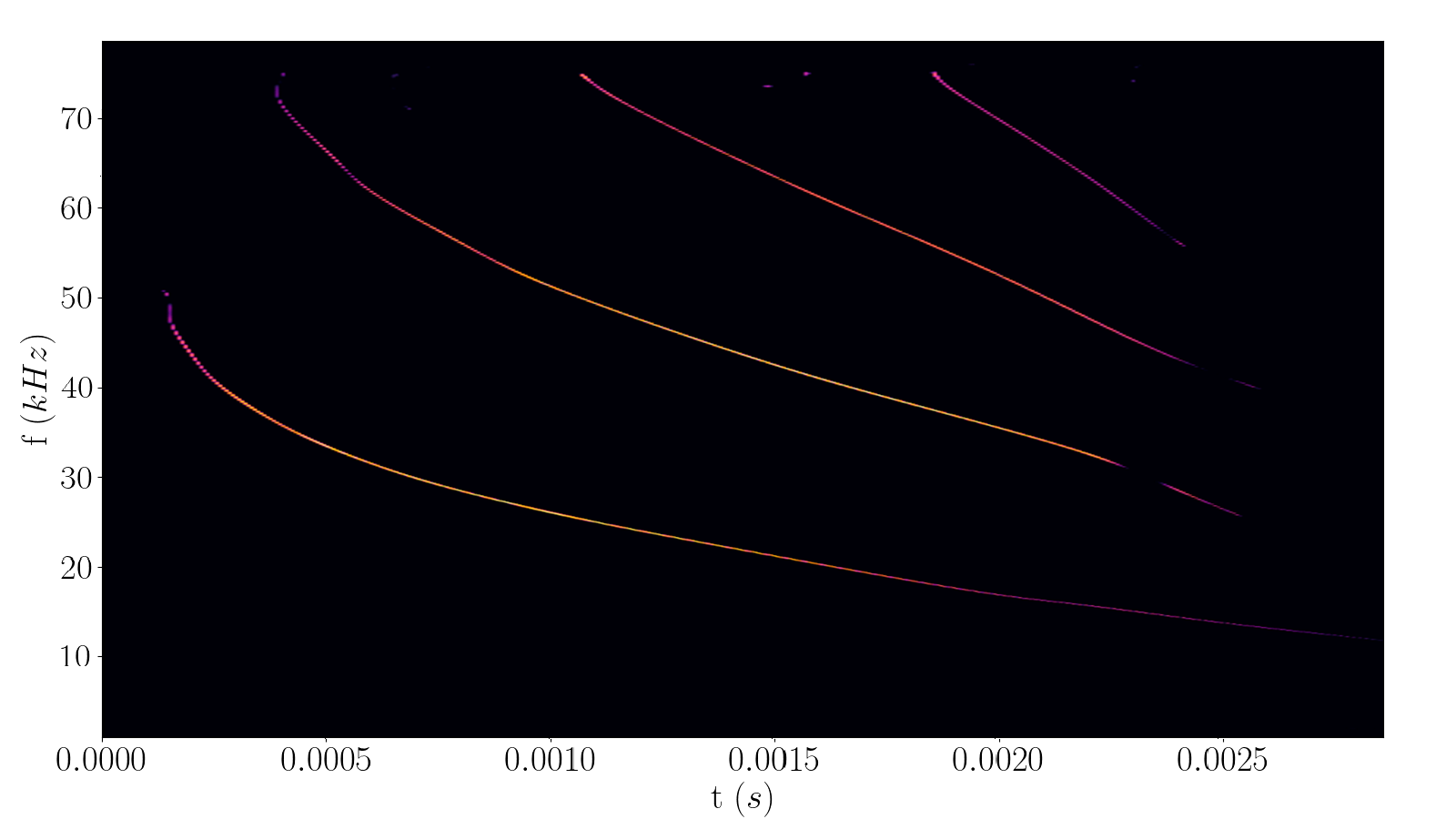}\\[-2mm]
      \includegraphics[width=\linewidth,trim={0cm 4mm 0cm 0cm},clip]{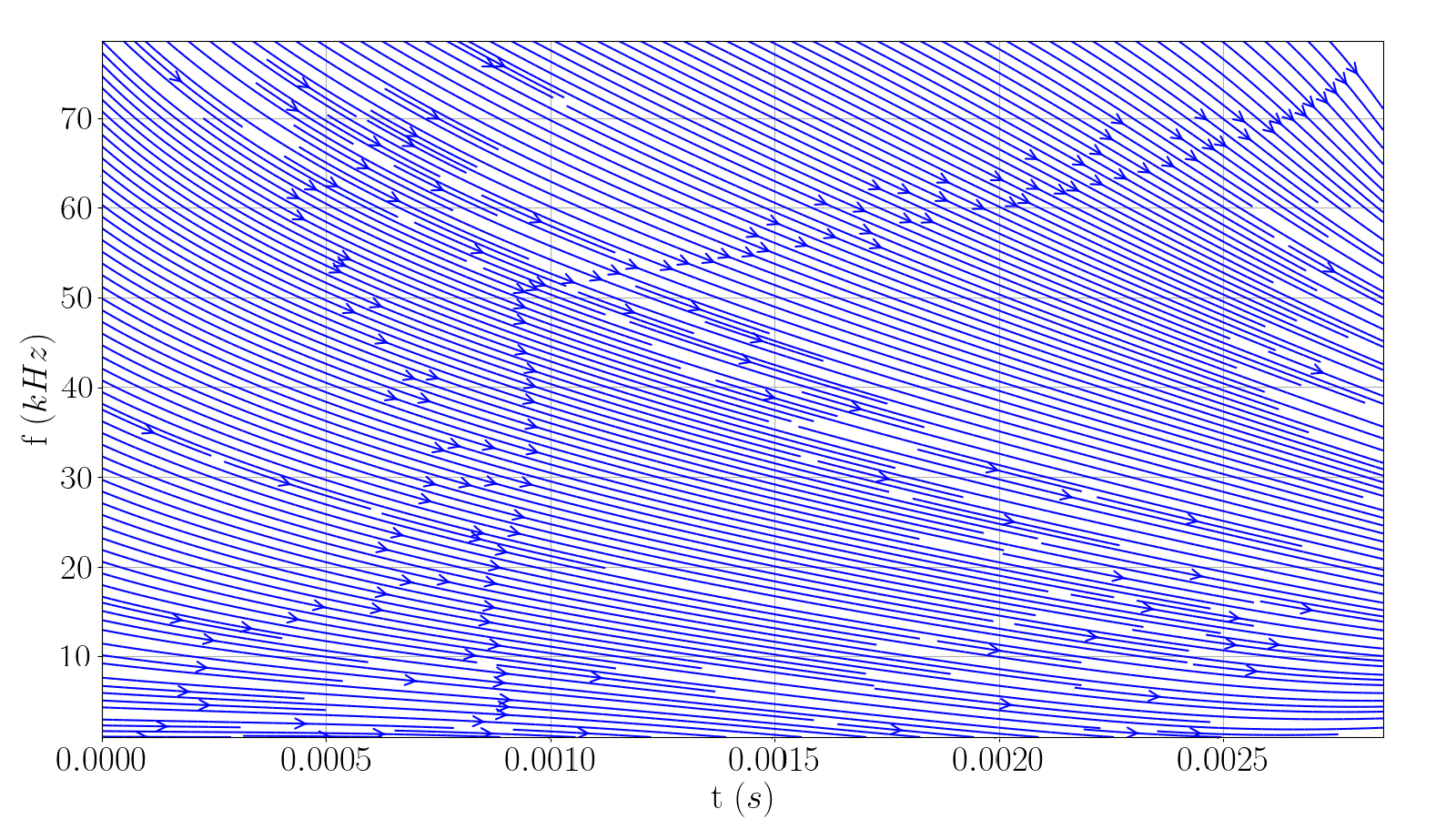}\\[-2mm]
      \includegraphics[width=\linewidth,trim={0cm 0mm 0cm 0cm},clip]{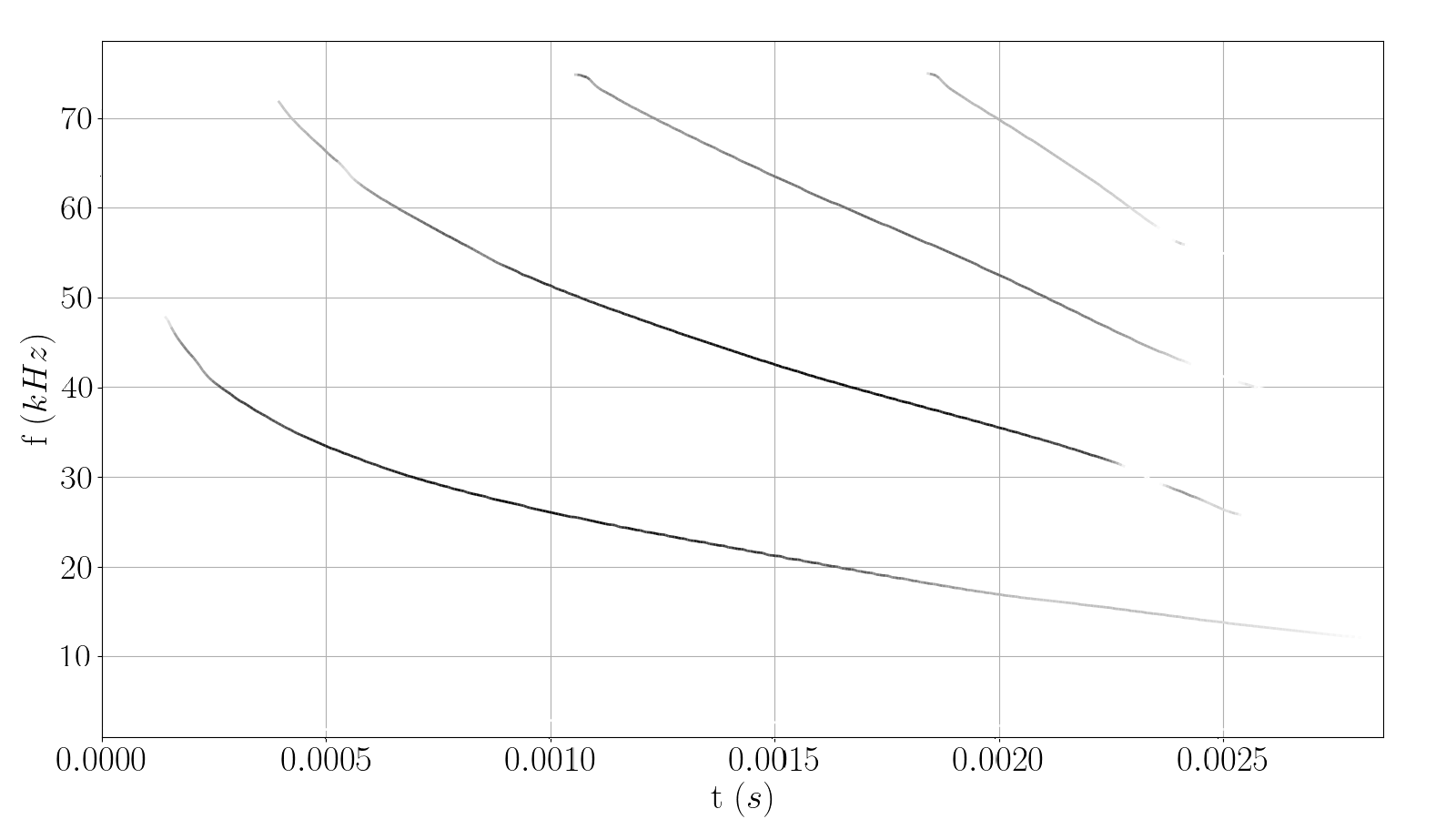}\\[-0mm]
      \includegraphics[width=\linewidth,trim={0cm 2mm 0cm 0cm},clip]{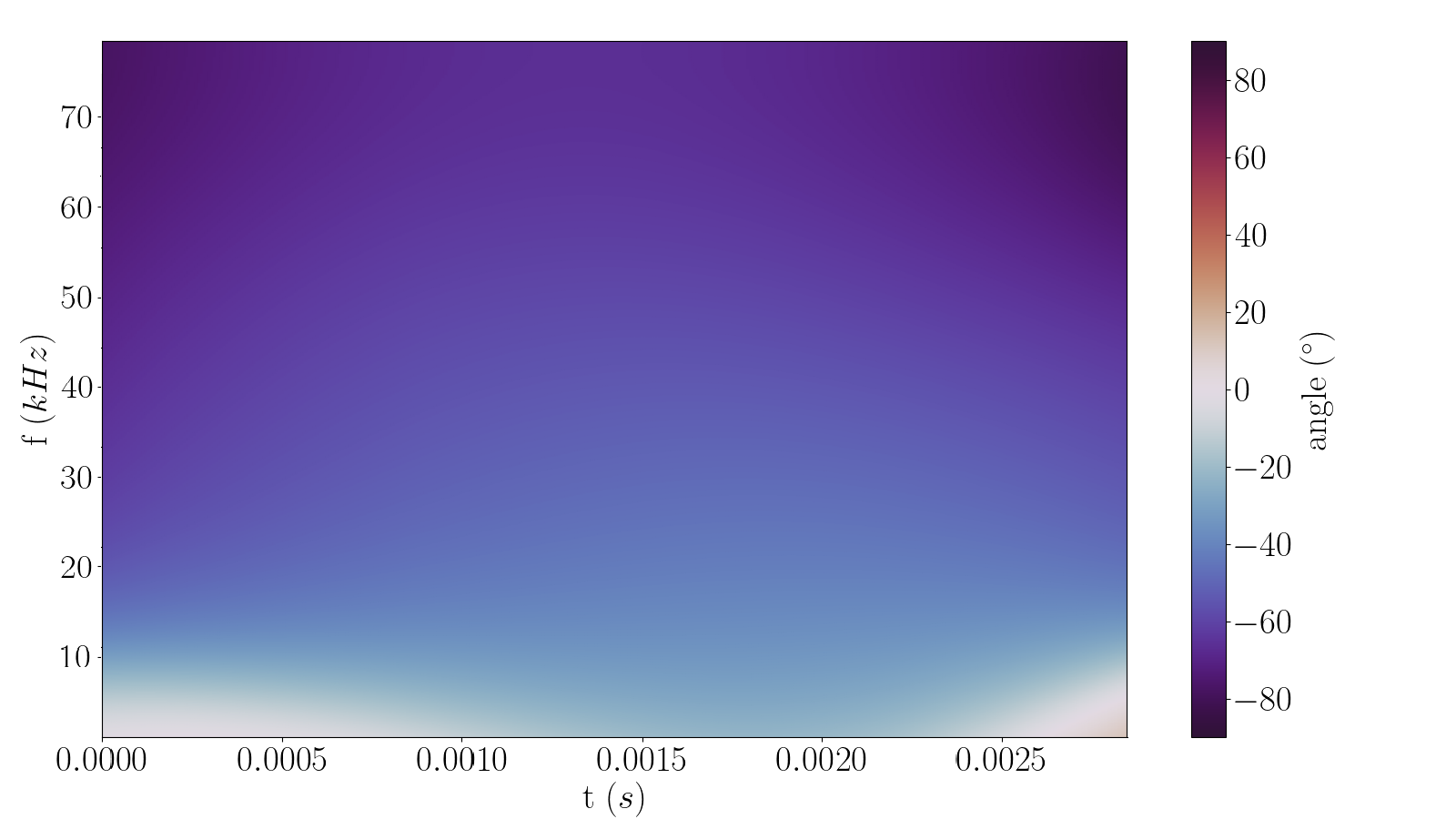}
    \end{minipage}
  }
  \caption{RIFT results across four examples (columns). \textbf{Rows} (top to bottom): 
  (1) Reference CWT evaluated with a kernel using the isotropic standard deviation $\sigma_{\text{iso}}$; 
  (2) RIFT result; 
  (3) time–frequency streamlines, determined via RK4 integration on the IPD field; 
  (4) Kalman-tracked trajectories.
  (5) smoothed Instantaneous Phase Direction (IPD) field;  
  \textbf{Columns}: 
  (a) $x_1(t)$, Eq.~\eqref{x_1}; 
  (b) $x_4(t)$, Eq.~\eqref{x_4}; 
  (c) speech extract from the ``Harvard sentences'' database~\cite{speech} with the phrase ``The stale smell of old beer lingers''; 
  (d) bat echolocation signal consisting of a 2.5\,\textmu s echolocation pulse emitted by the Large Brown Bat, \textit{Eptesicus fuscus}. \emph{The authors wish to thank Curtis Condon, Ken White, and Al Feng of the Beckman Institute of the University of Illinois for the bat data and for permission to use it in this paper.}\vspace{-0mm}}
  \label{fig:all_examples_grid}
\end{figure*}

The following section presents a series of comparative visualisations to demonstrate the capabilities of the proposed RIFT. Figure~\ref{fig:all_examples_grid} presents the various components of the RIFT for a variety of signals. For each example, a reference CWT is provided with an isotropic Gaussian Cohen's class kernel (relative to the T--F grid) with standard deviation $\sigma_{\text{iso}}$. This is followed by the outputted RIFT, the smoothed Instantaneous Phase Direction (IPD) field, the corresponding T--F streamlines (determined by applying the RK4 method to the IPD field), and the subsequently Kalman-tracked signal component trajectories. The algorithm is evaluated on two cross-term heavy synthetic waveforms; a parallel pair of sinusoidal frequency modulated signals, Eq.~\eqref{x_1} shown in Fig. \ref{fig:all_examples_grid}(a), and the following multicomponent waveform, shown in Fig. \ref{fig:all_examples_grid}(b):

\begingroup
\setlength{\abovedisplayskip}{-5pt}
\setlength{\belowdisplayskip}{5pt}
\vspace{-3.85mm}\small \begin{align}
\label{x_4}
    x_4\left(t\right) &= \sin\left( 2\pi \int_0^t \left[110 + 30 \sin\left(2\pi \tau\right) + \tfrac{50\tau}{3} - 25 \right] d\tau \right) \notag\\
    &\hspace{5mm}+ \sin\left( 2\pi \int_0^t \left[90 + 30 \sin\left(2\pi \tau\right) + \tfrac{50\tau}{3} - 25 \right] d\tau \right) \notag\\
    &\hspace{10mm} + \sin\left( 2\pi \int_0^t \left[150 + \tfrac{50\tau}{3} \right] d\tau \right) \notag\\
    &\hspace{15mm}+ \sin\left( 2\pi \int_0^t \left[10 + \tfrac{40\tau}{3} \right] d\tau \right).
\end{align} \normalsize\vspace{-1.85mm}
\endgroup

Likewise, real-world results include a speech extract from the ``Harvard sentences'' database \cite{speech} are shown in Fig. \ref{fig:all_examples_grid}(c), in addition to a bat echolocation signal in Fig. \ref{fig:all_examples_grid}(d), a commonly used demonstration example in T--F analysis. To enhance visibility due to signal sparsity, figure brightness has been increased (visualisation only).

For comparative evaluation, two dual-component synthetic signals are tested; the cross-term heavy signal Eq.~\eqref{x_1}, and the following intersecting component signal:

\vspace{-1.85mm} \small \begin{align}
\label{x_6}
    x_6\left(t\right) &= \sin\left( 2\pi \int_0^t \left[60 + 30 \sin\left(\tfrac{\pi}{2}\tau\right)\right] d\tau \right) \notag\\
    &\hspace{5mm}+ \sin\left( 2\pi \int_0^t \left[30 + 15\tau \right] d\tau \right).
\end{align} \normalsize\vspace{-2.25mm}

For each example, a reference spline ITFR is provided, followed by the RIFT result, Spline-RIFT result, Synchrosqueezing Transform \cite{Thakur2013}, Reassignment algorithm \cite{reassignment}, Adaptive Optimal-Kernel (AOK) algorithm \cite{jones2}, S-Method \cite{stankovic1994smethod}, Choi-Williams distribution \cite{ChoiWilliamsDistribution}, WVD \cite{wvd}, and the Synchroextracting Transform \cite{yu2017set}. All reassignment methods are computed with respect to the CWT with an isotropic Gaussian kernel relative to the normalised T–F grid. 

\begin{figure*}[!t]
  \centering
  \subfloat[]{\includegraphics[width=\linewidth, trim={0cm 2mm 0cm 1mm}, clip]{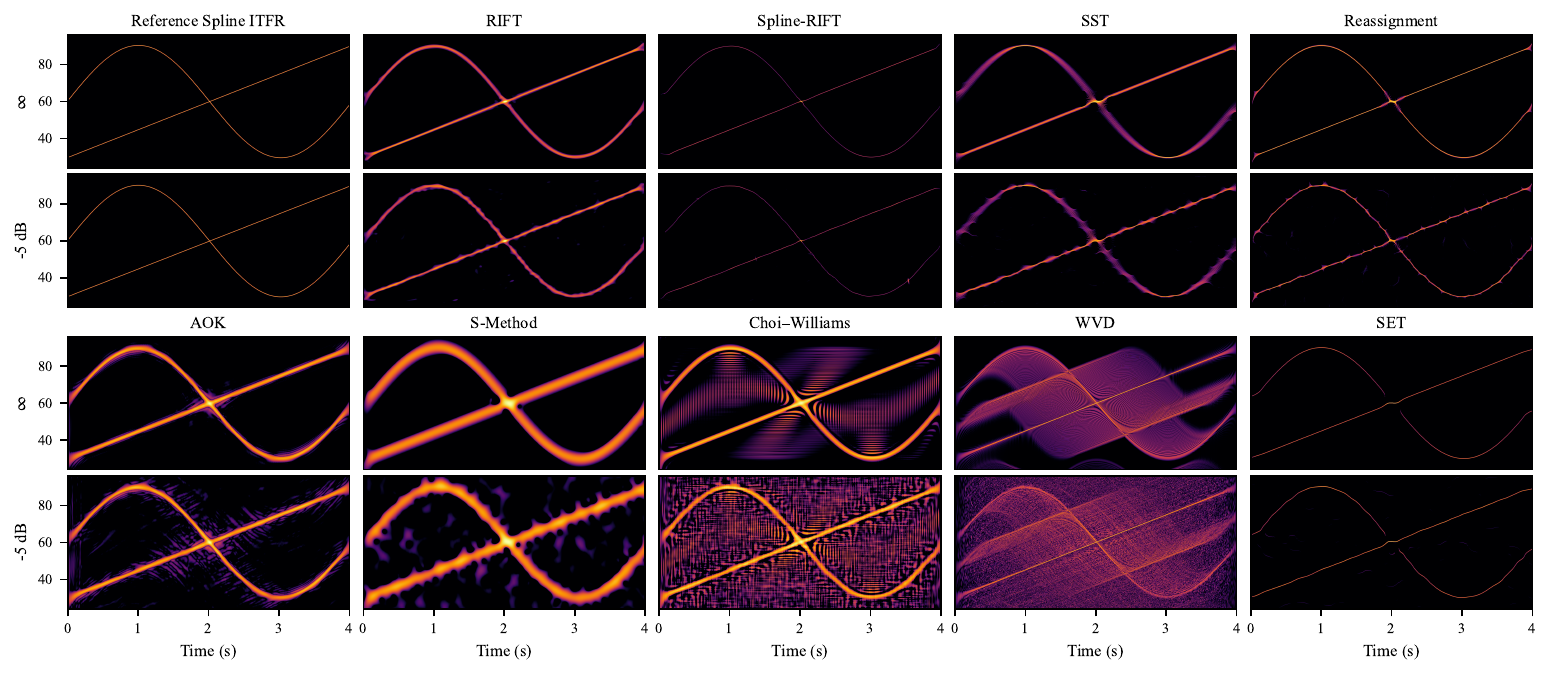}}
  \\
\vspace{-3mm}\subfloat[]{\includegraphics[width=\linewidth, trim={0cm 2mm 0cm 1mm}, clip]{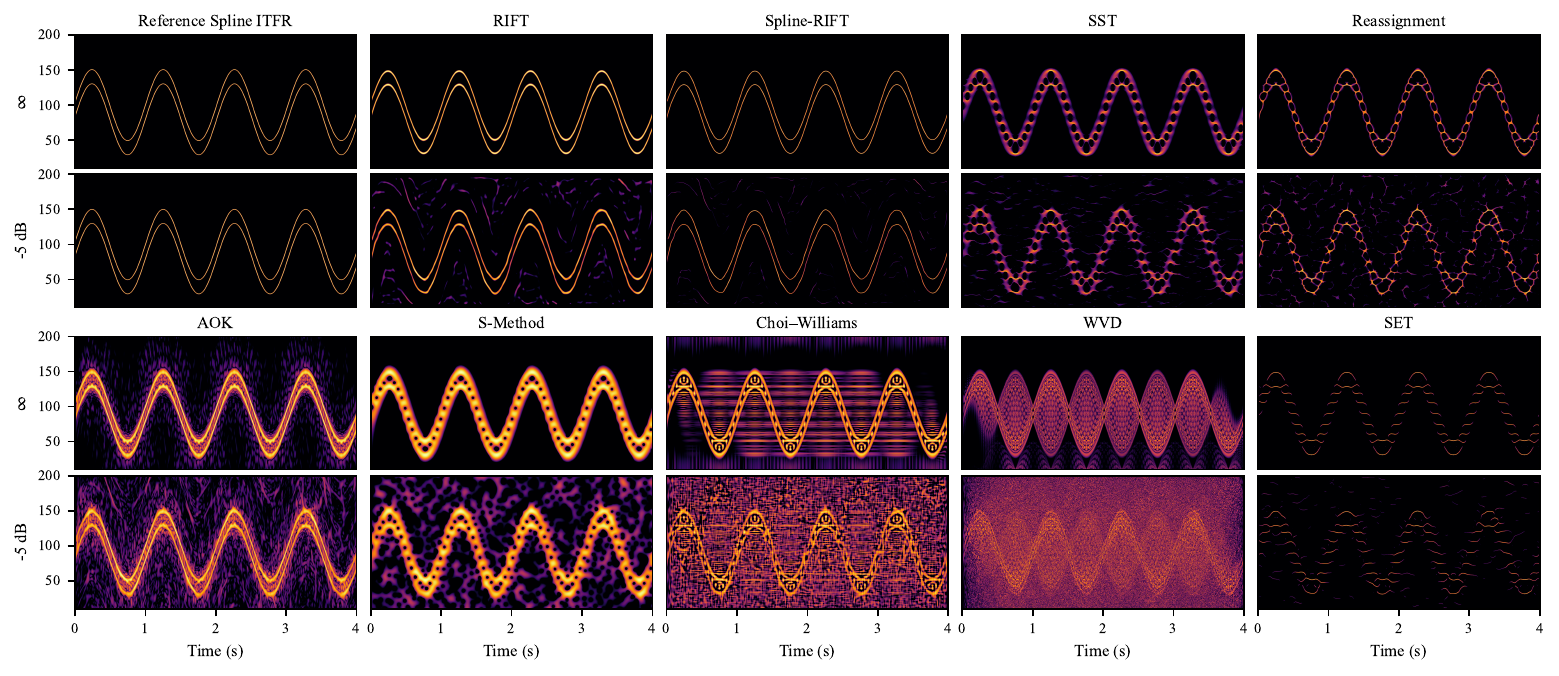}}
  \vspace{-1mm}\caption{T--F representations for (a) $x_6(t)$ and (b) $x_1(t)$. Two method blocks (columns = methods; rows = SNR). Shown SNRs: $\infty$ and $-5$\,dB. Axes (Time in seconds; Frequency in Hz).\vspace{-8mm}}
  \label{fig:montage_main}
\end{figure*}

Let \(R=|\text{T--F Representation}|\), and \(C=\text{ITFR}_{ref}\). The reference ITFR, $\text{ITFR}_{ref}$, is rendered as per the ITFR model in Section \ref{sec:itfr}, rasterised with Xiaolin Wu's antialiased line algorithm \cite{wu1991aa} and lightly convolved with an isotropic Gaussian ($\sigma_{I}=1.5$px) to form a tolerance tube to ensure metric stability against discretisation and sub-pixel jitter; note that ablating $\sigma_{I}$ confirms metric ranking insensitivity to this parameter (see Table \ref{tab:combined_sigma} in the SM). Let \(P=R/\sum_{t,f} R\) and \(Q=C/\sum_{t,f} C\). The following metrics are reported: the \textit{Bhattacharyya overlap} quantifies global probabilistic overlap between \(P\) and \(Q\): \(\mathrm{BC}=\sum_{t,f}\sqrt{P\,Q}\)~\cite{Bhattacharyya1943} (higher is better). The \textit{Jensen--Shannon divergence} measures the symmetric, finite KL discrepancy between \(P\) and \(Q\): \(\mathrm{JS}=\tfrac12\mathrm{KL}(P\|M)+\tfrac12\mathrm{KL}(Q\|M)\), \(M=\tfrac12(P+Q)\)~\cite{Lin1991} (lower is better). The \textit{Ridge Energy Ratio} captures on-ridge concentration as the fraction of raw energy inside the tube weights \(W=C/\max C\): \(\mathrm{RER}=\sum_{t,f} P\,W\) ~\cite{SlepianPollak1961I, Stankovic2001} (higher is better). The metrics are reported for varying additive white Gaussian noise (AWGN) signal-to-noise ratios (SNRs) to test noise tolerance: $-10, -5, 0, 5$, and $\infty \, \text{dB}$ (noise-free signal). Results for SNRs $-5$ and $\infty \, \text{dB}$ are shown for all evaluated methods in Fig. \ref{fig:montage_main}, and metrics for all SNRs are reported in Table \ref{tab:metrics_combined}. To ensure readability at publication scale, we apply a small isotropic Gaussian anti-aliasing prefilter ($\sigma = 1$px) for visualisation only, uniformly across methods and SNRs; all quantitative metrics are computed from the unfiltered TFRs.

As observed in Fig. \ref{fig:all_examples_grid}(a) and \ref{fig:all_examples_grid}(b), where the cross-terms present in the CWTs significantly degrade overall accuracy, the entropic filtering scheme integrated into the RIFT correctly identifies the signal trajectory alignments across the T--F plane, thereby eliminating cross-terms and isolating auto-terms. Moreover, the integration of the Lucy--Richardson Deconvolution algorithm recovers the original WVD-level resolution and yields results that are often visually indistinguishable from the theoretical ITFRs (for instance, Fig. \ref{fig:montage_main}(b)). Likewise, the IPD field effectively visualises the instantaneous directionality of the signal components, as further demonstrated by the T--F streamlines which closely follow the signal trajectories. Furthermore, the Kalman tracking scheme extracts the signal components, successfully avoiding cross-term interference. Similarly, the speech and bat echolocation extract results presented in figures \ref{fig:all_examples_grid}(c) and \ref{fig:all_examples_grid}(d), respectively, also exhibit significant cross-term reduction compared to the CWT, while maintaining WVD-level resolution. In both cases, the RIFT effectively tracks the evolving contours of these signals, extracting the signal components with high precision despite high T--F trajectory proximity.

As observed in table \ref{tab:metrics_combined}, across both examples and all SNRs, the RIFT achieves the best distributional fidelity, with a higher Bhattacharyya overlap (↑) and lower Jensen–Shannon divergence (↓). In contrast, the Spline-RIFT attains the highest Ridge Energy Ratio (↑). This pattern is expected: when the instantaneous frequency contains higher-order terms (\(\phi''(t)\!\neq\!0\); e.g., sinusoidal IFs), the Wigner–Ville auto-term has a non-zero instantaneous bandwidth, so the ridge widens in proportion to the local curvature. In conjunction with the Lucy–Richardson TV regularisation, this affords the RIFT a small tolerance to localisation error around the ITFR, while the Spline-RIFT, by construction, concentrates energy strictly along the spline-tracked ridges, maximising concentration in the ITFR region. In both examples, the RIFT accurately resolves component intersections and effectively mitigates cross terms, as most evident in Fig.~\ref{fig:montage_main}(b) when compared with competing methods. 

\vspace{-0mm}\section{Conclusion}
In conclusion, the Reconstructive Ideal Fractional Transform (RIFT) offers a robust, probabilistic framework for estimating ideal T--F representations of complex nonstationary signals, attaining high resolution auto-term resolution comparable to that of the WVD whilst effectively suppressing cross-terms. By integrating a proposed constellation of Continuous Fractional Wavelet Transforms, spatially-varying Lucy--Richardson deconvolution, and an entropic-based weighting optimisation scheme, the method provides a comprehensive approach to refining T--F representations. The resulting optimised Instantaneous Phase Direction field provides localised curvature visualisation and enables accurate signal component tracking and the construction of the Spline-RIFT. Collectively, these capabilities, supported by evaluations on synthetic and real-world signals, demonstrate RIFT’s potential as a powerful tool for high-resolution, cross-term-free T--F analysis across a wide range of applications, paving the way for advanced signal manipulation, visualisation, and synthesis.

\vspace{-3mm}\begin{table*}[!t]
\centering
\scriptsize
\setlength{\tabcolsep}{9pt}
\begin{tabular}{l|cccccccccc}
\hline
\multicolumn{11}{l}{\textbf{Bhattacharyya overlap (↑)}}\\
\cline{1-11}
SNR (dB) & AOK & CWT & Choi–Williams & RIFT & Reassignment & S-Method & SET & SST & Spline-RIFT & WVD \\ \hline
-10 dB & 0.4369 & 0.3436 & 0.2577 & \textbf{0.5428} & 0.2854 & 0.3273 & 0.1487 & 0.2876 & 0.2644 & 0.1702 \\
-5 dB & 0.5423 & 0.3893 & 0.3600 & \textbf{0.6951} & 0.4406 & 0.3751 & 0.2421 & 0.3736 & 0.4364 & 0.1999 \\
0 dB & 0.5807 & 0.4115 & 0.4275 & \textbf{0.7054} & 0.5075 & 0.3973 & 0.2485 & 0.4075 & 0.4462 & 0.2299 \\
5 dB & 0.5966 & 0.4193 & 0.4605 & \textbf{0.7041} & 0.5549 & 0.4051 & 0.2773 & 0.4260 & 0.4589 & 0.2473 \\
$\infty$ & 0.6022 & 0.4227 & 0.4825 & \textbf{0.7061} & 0.5637 & 0.4083 & 0.2892 & 0.4299 & 0.4556 & 0.2663 \\
\hline\hline
\multicolumn{11}{l}{\textbf{Jensen--Shannon divergence (↓)}}\\
\cline{1-11}
SNR (dB) & AOK & CWT & Choi–Williams & RIFT & Reassignment & S-Method & SET & SST & Spline-RIFT & WVD \\ \hline
-10 dB & 0.4365 & 0.5057 & 0.5585 & \textbf{0.3557} & 0.5147 & 0.5155 & 0.6099 & 0.5108 & 0.5358 & 0.6201 \\
-5 dB & 0.3575 & 0.4733 & 0.4912 & \textbf{0.2443} & 0.4116 & 0.4824 & 0.5513 & 0.4520 & 0.4224 & 0.6015 \\
0 dB & 0.3287 & 0.4573 & 0.4436 & \textbf{0.2372} & 0.3675 & 0.4666 & 0.5478 & 0.4291 & 0.4179 & 0.5818 \\
5 dB & 0.3168 & 0.4517 & 0.4195 & \textbf{0.2380} & 0.3355 & 0.4610 & 0.5272 & 0.4163 & 0.4085 & 0.5701 \\
$\infty$ & 0.3126 & 0.4492 & 0.4033 & \textbf{0.2362} & 0.3367 & 0.4588 & 0.5186 & 0.4134 & 0.4108 & 0.5569 \\
\hline\hline
\multicolumn{11}{l}{\textbf{Ridge Energy Ratio (↑)}}\\
\cline{1-11}
SNR (dB) & AOK & CWT & Choi–Williams & RIFT & Reassignment & S-Method & SET & SST & Spline-RIFT & WVD \\ \hline
-10 dB & 0.0654 & 0.0392 & 0.0256 & 0.1201 & 0.0918 & 0.0373 & 0.1148 & 0.0750 & \textbf{0.1984} & 0.0114 \\
-5 dB & 0.1009 & 0.0497 & 0.0437 & 0.1971 & 0.1693 & 0.0479 & 0.1963 & 0.1082 & \textbf{0.3464} & 0.0158 \\
0 dB & 0.1167 & 0.0556 & 0.0596 & 0.2160 & 0.2089 & 0.0536 & 0.2041 & 0.1194 & \textbf{0.3602} & 0.0209 \\
5 dB & 0.1234 & 0.0576 & 0.0688 & 0.2191 & 0.2583 & 0.0556 & 0.2416 & 0.1240 & \textbf{0.3761} & 0.0243 \\
$\infty$ & 0.1259 & 0.0586 & 0.0757 & 0.2214 & 0.3019 & 0.0565 & 0.2557 & 0.1264 & \textbf{0.3734} & 0.0285 \\
\hline\hline
\end{tabular}
\vspace{1mm}\caption{T--F Evaluation metrics across SNRs: -10, -5, 0, +5, and $\infty$ (dB). Bold indicates the best value per row; arrows denote the preferred direction (↑ = higher, ↓ = lower). Cells report the mean across the two examples (equal weight). \vspace{-8mm}}
\label{tab:metrics_combined}
\end{table*}

\vspace{-0mm}

\vspace{-4mm}\section{Biography}

\vspace{-13mm}\begin{IEEEbiography}[{\includegraphics[width=1in,height=1.25in,clip,keepaspectratio]{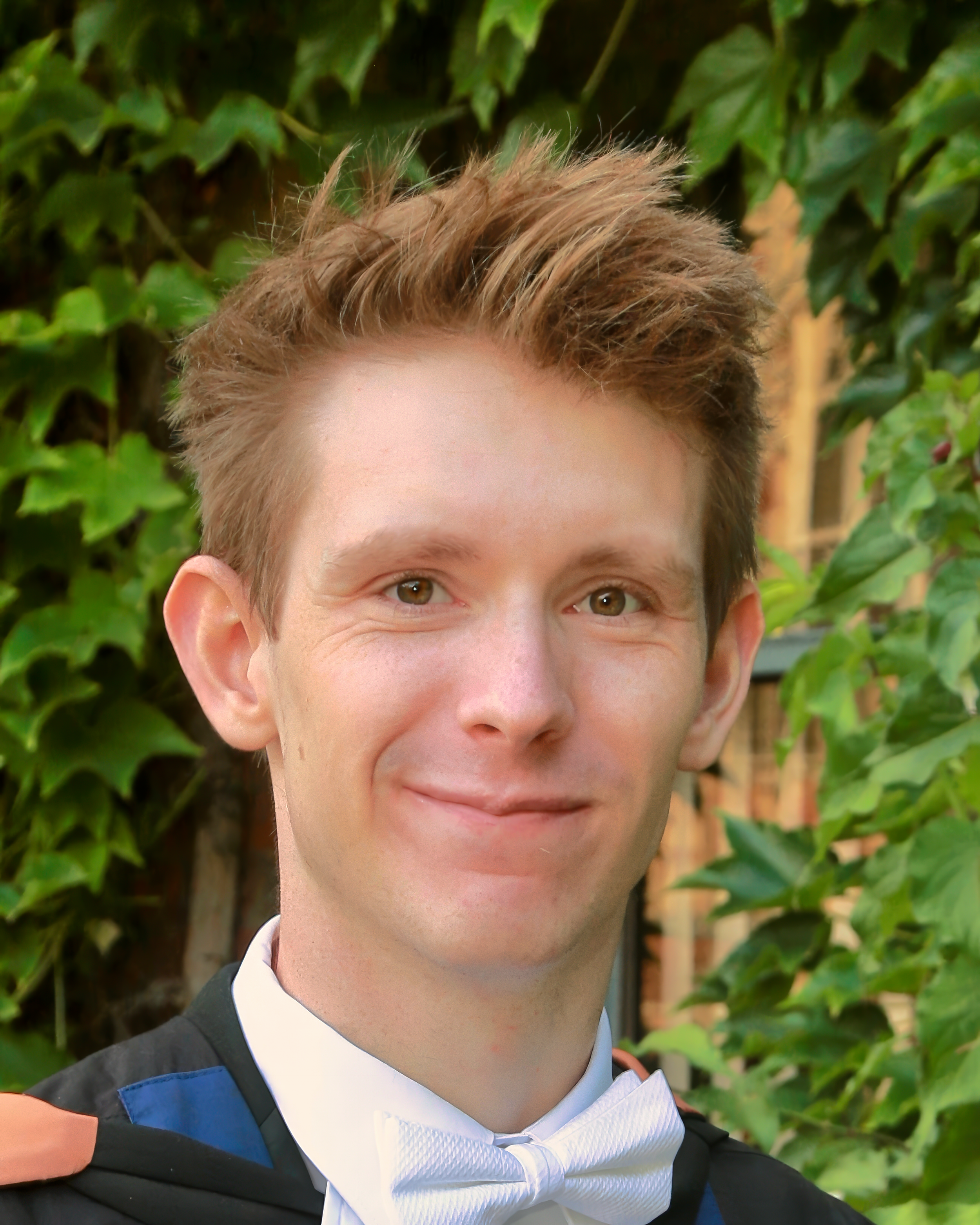}}]{James M. Cozens} (Member, IEEE) is a Ph.D. candidate in the Statistical Signal Inference (SSigInf) Group at the University of Cambridge. He holds an M.Eng. in Information and Computer Engineering from the University of Cambridge. His research focuses on statistical signal processing and machine learning methods in the context of audio and music processing, including high-resolution time--frequency analysis, music transcription, beat tracking, and signal decomposition. Additional interests involve multi-object tracking and deep learning-based hierarchical generative models for music visualisation and composition, with an emphasis on achieving expressive and structural consistency. His work has been featured in leading IEEE publications and presented at top-tier international conferences.
\end{IEEEbiography}

\vspace{-10mm}\begin{IEEEbiography}[{\includegraphics[width=1in,height=1.25in,clip,keepaspectratio]{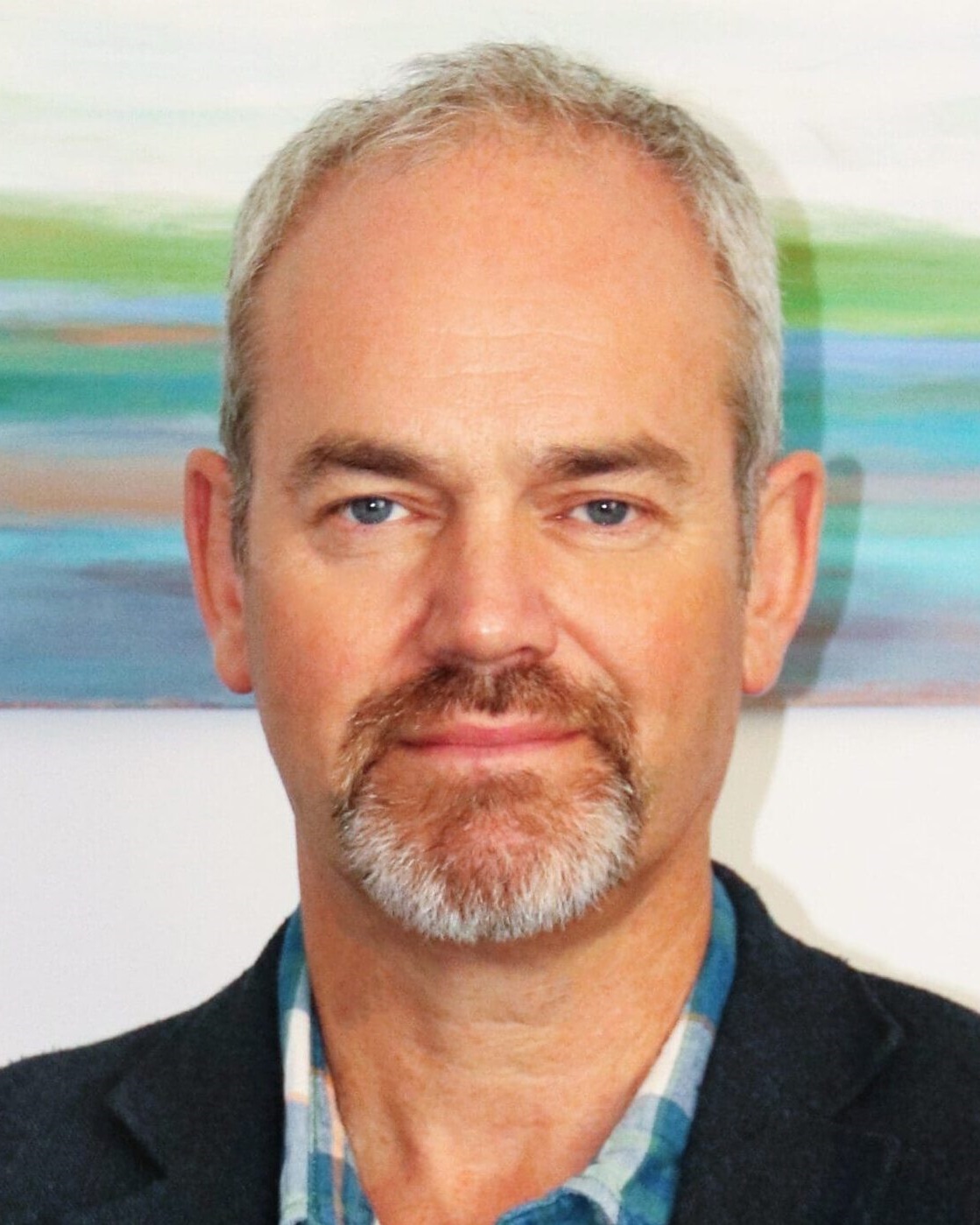}}]{Simon J. Godsill} (FIEEE) received the undergraduate and Ph.D. degrees from the University of Cambridge, Cambridge, U.K., while being a member of Selwyn College, Cambridge, in 1988 and 1994, respectively. He is currently Professor of Statistical Signal Processing and Head of Information Engineering at the Department of Engineering, Cambridge University, Cambridge, U.K. He is also a Professorial Fellow at Corpus Christi College Cambridge, Cambridge. He coordinates an active research group in Statistical Signal Inference (SSigInf) and its applications within the Probabilistic Systems, Information, and Inference Group ($\psi^2$), University of Cambridge, specialising in Bayesian computational methodology, multiple object tracking, audio and music processing, and time series modelling. A particular methodological theme over many years has been the development of novel techniques for optimal Bayesian filtering and smoothing, using sequential Monte Carlo or particle filtering methods. He has authored extensively in journals, books, and international conference proceedings, and regularly delivers invited and plenary addresses at conferences.
\end{IEEEbiography}

\section*{Supplementary Material}
\renewcommand{\thesubsection}{\Alph{subsection}}

\subsection{Derivation of the convolutional relationship between a Continuous Wavelet Transform with a \textit{complex-valued} window function and the WVD}

\label{sm:A}

Let $\Phi{\left(\omega, t\right)}$ be a Continuous Wavelet Transform (CWT) \cite{wavelet} with a \textit{complex-valued} window function, such that:

\vspace{-2.25mm} \small \begin{align}
    \Phi{\left(\omega, t\right)} &= \left |\left[z * W_{\omega}\right]{\left(t\right)} \right |^2
    = \left |\left[z\left(\tau\right) * \Omega^*{\left(\tau\right)} e^{j \omega \tau} \right](t)\right |^2,
\end{align} \normalsize \vspace{-2.25mm} 

\noindent where $W_{\omega}{\left(t\right)}$ and $\Omega{\left(t\right)}$ are the CWT wavelet and window functions respectively for angular frequency $\omega$. Thus:

\vspace{-2.25mm} \small \begin{align}
    \Phi{\left(\omega, t\right)} &= \left[z\left(\tau\right) * \Omega^*{\left(\tau\right)} e^{j \omega \tau} \right](t) \cdot \left(\left[z\left(\tau\right) * \Omega^*{\left(\tau\right)} e^{j \omega \tau} \right](t)\right)^* \notag \\
    &= \int_{\mathbb{R}^2} z\left(t_1\right)z^*\left(t_2\right)\Omega^*{\left(t - t_1\right)} \Omega{\left(t - t_2\right)}
    \notag \\& \hspace{10mm} e^{j \omega \left(t - t_1\right)}e^{-j \omega \left(t - t_2\right)} \, d t_1\, d t_2. 
\end{align} \normalsize \vspace{-2.25mm} 

\noindent Substituting the average time, $T=(t_1 + t_2)/2$, and the time lag, $\tau = t_1 - t_2$, the Jacobian for the substitution becomes:

\vspace{-2.25mm} \small \begin{align}
    &\left |\mathbf{J}\right | = \begin{vmatrix}
        \frac{\partial \tau}{\partial t_1} & \frac{\partial \tau}{\partial t_2} \vspace{1mm}\\
        \frac{\partial T}{\partial t_1} & \frac{\partial T}{\partial t_2}
    \end{vmatrix}
    = \begin{vmatrix}
        1 & -1 \\
        \frac{1}{2} & \frac{1}{2}
    \end{vmatrix} = 1.
\end{align} \normalsize \vspace{-2.25mm} 

\noindent Thus:

\vspace{-2.25mm} \small \begin{align}
    \Phi{\left(\omega, t\right)} &= \int_{\mathbb{R}^2} z\left(T + \frac{\tau}{2}\right)z^*\left(T - \frac{\tau}{2}\right)\Omega^*{\left(t - \left(T + \frac{\tau}{2}\right)\right)} 
    \notag\\& \hspace{10mm} \cdot \Omega{\left(t - \left(T - \frac{\tau}{2}\right)\right)}e^{-j \omega \left(T + \frac{\tau}{2}\right)}e^{j \omega \left(T - \frac{\tau}{2}\right)} \, d \tau\, d T 
    \notag\\&=\int_{\mathbb{R}^2} R_z\left(T, \tau\right)R_{\Omega}\left(t-T, \tau\right)e^{-j \omega \tau} \, d \tau\, d T,
\end{align} \normalsize \vspace{-2.25mm} 

\noindent where $R_z\left(t, \tau\right)$ is the instantaneous correlation of the waveform $z(t)$ at time $t$ for time lag $\tau$:

\vspace{-2.25mm} \small \begin{align}
    R_z\left(t, \tau\right) = z\left(t + \frac{\tau}{2}\right)z^*\left(t - \frac{\tau}{2}\right).
\end{align} \normalsize \vspace{-2.25mm} 

\noindent Note that:

\vspace{-2.25mm} \small \begin{align}
    WVD_z\left(\omega, \, t\right) &= \frac{1}{\sqrt{2\pi}}\mathcal{F}\left\{R_z\left(t, \tau\right)\right\}\\
    \to R_z\left(t, \tau\right) &= \sqrt{2\pi}\mathcal{F}^{-1}\left\{WVD_z\left(\omega, \, t\right)\right\}
    \notag \\&= \int_{\mathbb{R}} WVD_z\left(\omega, \, t\right) e^{j \omega \tau}\, d \omega.
\end{align}\normalsize \vspace{-2.25mm} 

\noindent Thus:

\vspace{-2.25mm} \small  \begin{align}
    \Phi{\left(\omega, t\right)} &= \int_{\mathbb{R}^4} e^{j \omega_1 \tau} WVD_z\left(\omega_1, T\right)e^{j \omega_2 \tau} \notag\\
    & \hspace{10mm} \cdot WVD_{\Omega}\left(\omega_2, t - T\right) e^{-j \omega \tau} \, d \tau \, d T \, d \omega_1 \, d \omega_2 \notag\\
    &= \int_{\mathbb{R}^3} WVD_z\left(\omega_1, T\right) WVD_{\Omega}\left(\omega_2, t - T\right) \notag\\
    & \hspace{10mm} \cdot \left(\int_{\mathbb{R}} e^{j \tau\left(\omega_1 + \omega_2 - \omega\right)} \, d \tau \right)\, d T \, d \omega_1 \, d \omega_2.
\end{align} \normalsize \vspace{-2.25mm} 

\noindent Note that:

\vspace{-2.25mm} \small  \begin{align}
    \int_{\mathbb{R}} e^{j \tau\left(\omega_1 + \omega_2 - \omega\right)} \, d \tau &= 2 \pi \delta \left(\omega_1 + \omega_2 - \omega\right).
\end{align} \normalsize \vspace{-2.25mm} 

\noindent Thus:

\vspace{-2.25mm} \small  \begin{align}
\label{WVD_CWT_REL2}
    \Phi{\left(\omega, t\right)}
    &= 2 \pi \int_{\mathbb{R}^3} WVD_z\left(\omega_1, T\right) WVD_{\Omega}\left(\omega_2, t - T\right) \notag\\
    & \hspace{10mm} \cdot \delta \left(\omega_1 + \omega_2 - \omega\right) \, d T \, d \omega_1 \, d \omega_2 \notag\\
    &=  2 \pi \int_{\mathbb{R}^2} WVD_z\left(\omega_1, T\right) WVD_{\Omega}\left(\omega - \omega_1, t - T\right) \, d T \, d \omega_1 \notag \\
    &= 2 \pi \left[WVD_z * WVD_{\Omega} \right] \left(\omega, t\right)
\end{align} \normalsize\vspace{-2.65mm}

\vspace{-2mm}\subsection{Derivation of the Cohen's class time--frequency convolution kernel}
\label{sm:kernel_derivation}

Let $\bar{\Pi}_{\sigma, \theta}\left(\omega, \, t\right) = WVD{\left\{\Omega_{\sigma, \theta}{\left(t\right)}\right\}}$. Thus: 

\vspace{-2.25mm} {\allowdisplaybreaks \small \begin{align}
    \bar{\Pi}_{\sigma, \theta}\left(\omega, \, t\right) &=\frac{1}{2\pi}\int_{\mathbb{R}} \Omega_{\sigma, \theta}{\left(t + \frac{\tau}{2}\right)}\Omega_{\sigma, \theta}{\left(t-\frac{\tau}{2}\right)}^*e^{-j\tau\omega}\,d\tau
    \notag\\&=\frac{1}{2\pi}\int_{\mathbb{R}} \frac{1}{\sqrt[4]{\pi\sigma^2}}e^{-\frac{1}{2\sigma^2}\left(t+\frac{\tau}{2}\right)^2}e^{j\frac{\tan{\theta}}{2}\left(t+\frac{\tau}{2}\right)^2}
    \notag\\& \hspace{5mm} \cdot\left(\frac{1}{\sqrt[4]{\pi\sigma^2}}e^{-\frac{1}{2\sigma^2}\left(t-\frac{\tau}{2}\right)^2}e^{j\frac{\tan{\theta}}{2}\left(t-\frac{\tau}{2}\right)^2}\right)^*e^{-j\tau\omega}\,d\tau
    \notag\\&=\frac{1}{2\sqrt{\pi^3\sigma^2}}e^{-\left[\frac{t^2}{\sigma^2} +\sigma^2\left(\omega -\tan{\left(\theta\right)} \, t\right)^2\right]}
    \notag\\& \hspace{5mm} \cdot \int_{\mathbb{R}} e^{-\frac{1}{4\sigma^2}\left[\left(\tau + 2\sigma^2 j\left[\omega -\tan{\left(\theta\right)} \, t\right]\right)^2\right]}\,d\tau
    \notag\\&=\frac{1}{2\sqrt{\pi^3\sigma^2}}e^{-\left[\frac{t^2}{\sigma^2} +\sigma^2\left(\omega -\tan{\left(\theta\right)} \, t\right)^2\right]}2\sqrt{\pi\sigma^2}
    \notag\\&=\frac{1}{\pi}e^{-\left[\frac{t^2}{\sigma^2} +\sigma^2\left(\omega -\tan{\left(\theta\right)} \, t\right)^2\right]}.
\end{align}} \normalsize\vspace{-2.65mm}

\setcounter{page}{1}
\vspace{-2mm}\subsection{Derivation of the Cohen's class kernel principal axis angular and standard deviation offset}

Let $\sigma_0$ and $\kappa$ be the implemented wavelet parameters in Eq.~\eqref{omega_simple}, i.e.,  $\Omega_{\sigma_0, \kappa}(t)$ with corresponding Cohen's class kernel $\bar{\Pi}_{\sigma_0, \kappa}\left(\omega, \, t\right)$.

\subsubsection{Standard deviation}
\label{sm:kernel_parameters_offset_sd}

The true standard deviation along the Cohen's class kernel principal axis of the resulting skewed Gaussian can be evaluated by extracting the eigenvalues, $\lambda_{1,2}$, of $\mathbf{\Sigma}$:

\vspace{-2.25mm} {\small
\begin{align}
    \det{\left(\mathbf{\Sigma} - \lambda_{1, 2}\mathbf{I}\right)} &= 0 \\
    \to \lambda_{1, 2} &= \frac{1}{a+c \pm\sqrt{\left(a-c\right)^2+b^2}},
\end{align}
} \normalsize \vspace{-2.65mm}

\noindent with $a, b, c$ taken from Eq.~\eqref{a, b, c} from the main paper. Thus, the \emph{variance} (eigenvalue) along the principal axis, $\lambda_1$, is

\vspace{-2.25mm} \small
\begin{align}
    \lambda_1
            &=\frac{1}{\frac{1}{\sigma_0^2} + \sigma_0^2 \sec^2{\kappa} - \sqrt{\left(\frac{1}{\sigma_0^2} - 2\sigma_0^2 + \sigma_0^2 \sec^2{\kappa}\right)^2+4\sigma_0^4\tan^2{\kappa}}}.
            \label{lambda}
\end{align}
\normalsize \vspace{-2.65mm}

To now redefine the standard deviation of the kernel in terms of the standard deviation along the principal axis, Eq.~\eqref{lambda} can be solved for $\sigma_0$, keeping $\lambda_1$ constant:

\vspace{-2.25mm} \small
\begin{align}
    \sigma_0^2 &= \frac{2 \lambda_1 + \frac{1}{2\lambda_1} \pm \sqrt{\left(2\lambda_1+\frac{1}{2\lambda_1}\right)^2-4\sec^2{\kappa}}}{2\sec^2{\kappa}}.
\end{align}
\normalsize \vspace{-2.65mm}

Let $\sigma$ be the standard deviation along the \emph{principal axis} of the kernel (divided by $\sqrt{2}$ given that the WVD corresponds to the squared modulus of the time--frequency representation, such that $\sigma=\sqrt{\lambda_1/2}$). Thus, the required implemented wavelet parameters $(\sigma_0,\kappa)$ to achieve the desired \emph{principal-axis} standard deviation, $\sigma$, is given by:

\vspace{-2.25mm} \small
\begin{align}
\label{sigma_{theta}2}
        \sigma_{0}(\sigma,\kappa)
  \!&=\! \left[\!
      \frac{ (\sigma^{2}\!+\!\sigma^{-2})
        +\operatorname{sgn}(\sigma\!-\!1) \sqrt{(\sigma^{2}\!+\!\sigma^{-2})^{2}\! -\! 4\,\sec^{2}\kappa}
      }{ 2\,\sec^{2}\kappa }
     \!\right]^{\!\tfrac{1}{2}}
\end{align}
\normalsize \vspace{-2.65mm}

\noindent valid for $\sigma \neq 1$ and when:

\vspace{-2.25mm} \small
\begin{align}
    \left(\frac{1}{\sigma^{2}}+\sigma^{2}\right)^{2}-4\sec^{2}\kappa \ge 0\,,
    \, \text{i.e.} \,
    \left| \kappa\right| \leq \arccos\left(\frac{2\sigma^{2}}{\sigma^{4}+1}\right).
\end{align}
\normalsize \vspace{-2.65mm}

\noindent Note that $\operatorname{sgn}(x)$ denotes the sign function, with $\operatorname{sgn}(x) = -1\ (x<0),\ 0\ (x=0),\ 1\ (x>0)$. The sign function here ensures that the correct root is selected for the two cases $\sigma <1$, and $\sigma > 1$. 

\subsubsection{Angle}
\label{sm:kernel_parameters_offset_angle}

The true angle, $\varphi$, of the principal axis of the resulting skewed Gaussian can be determined by computing the angular rotation ($\varphi$) of the exponent required to remove the non-diagonal elements (terms with $\omega t$ indexed $(1, 0)$ or $(0, 1)$):

\vspace{-2.25mm} \small
\begin{align}
    \left(\mathbf{R_{\varphi}}^T\mathbf{\Sigma}^{-1}\mathbf{R_{\varphi}}\right)_{(1, 0)} &= 0
\end{align}
\normalsize \vspace{-2.65mm}

\noindent where:

\vspace{-2.25mm} \small
\begin{align}
    \mathbf{R_{\varphi}} &= \begin{bmatrix}
        \cos{\varphi} & \sin{\varphi} \\
        -\sin{\varphi} & \cos{\varphi}
    \end{bmatrix}.
\end{align}
\normalsize \vspace{-2.65mm}

\noindent Therefore, rearranging for $\varphi$:

\vspace{-2.25mm} \small
\begin{align}
    \varphi\left(\sigma, \kappa\right) &= \frac{1}{2}\arctan{\left(\frac{b}{a - c}\right)} \notag\\
    &= -\frac{1}{2}\arctan{\left(\frac{2\tan{\kappa}}{\frac{1}{\sigma_{0}{\left(\sigma, \kappa\right)}^4} + \tan^2{\kappa} - 1}\right)}.
\end{align}
\normalsize \vspace{-2.65mm}

\noindent However, note that the function switches to different axes at the denominator asymptote points, $\kappa_{\sigma}$:

\vspace{-2.25mm} \small
\begin{align}
    \frac{1}{\sigma_{0}{\left(\sigma, \kappa_{\sigma} \right)}^4} + \tan^2{\kappa_{\sigma}} - 1 &= 0
\end{align}
\normalsize \vspace{-2.65mm}

\noindent Substituting $\sigma_0$ from Eq.~\eqref{sigma_{theta}2} and rearranging for $\kappa_{\sigma}$ yields:

\vspace{-2.25mm} \small
\begin{align}
    \kappa_{\sigma} &= \pm \arccos\left(\sqrt{\frac{\left(\sigma^2 + \frac{1}{\sigma^2}\right)^2}{2\left(\sigma^4 + \frac{1}{\sigma^4}\right)}}\right).
\end{align}
\normalsize \vspace{-2.65mm}

\noindent Thus, to effectively unwrap the function, the following expression can be employed:

\vspace{-2.25mm} \small
\begin{align}
\label{solution2}
    \varphi\left(\sigma, \kappa\right)
    &= -\frac{1}{2}\arctan{\!\left(\frac{2\tan{\kappa}}{\sigma_{0}{\left(\sigma, \kappa\right)}^{-4}\! +\! \tan^2{\kappa}\! - \!1}\right)}\! + \!K_\sigma\left(\kappa\right),
\end{align}
\normalsize \vspace{-2.65mm}

\noindent where:

\vspace{-2.25mm} \small
\begin{align}
\label{axis-switch}
    K_{\sigma}\!\left(\kappa\right)\!&=\frac{\pi}{4}s(\sigma)\!\left(\left\lfloor\! \frac{1}{\pi}\left(\kappa\! +\! \kappa_{\sigma}\right)\!\right\rfloor\! +\!\left\lfloor \!\frac{1}{\pi}\left(\kappa\! -\! \kappa_{\sigma}\right)\!\right\rfloor\! +\! 1\right),
\end{align}
\normalsize \vspace{-2.65mm}

\noindent where $s(\sigma)=(1\!+\!\operatorname{sgn}(\sigma\! -\! 1))$. Note that in the limit

\begin{figure}
    \centering
    \subfloat[]{{\includegraphics[width=0.45\columnwidth, trim={0cm 1mm 0cm 0cm}, clip]{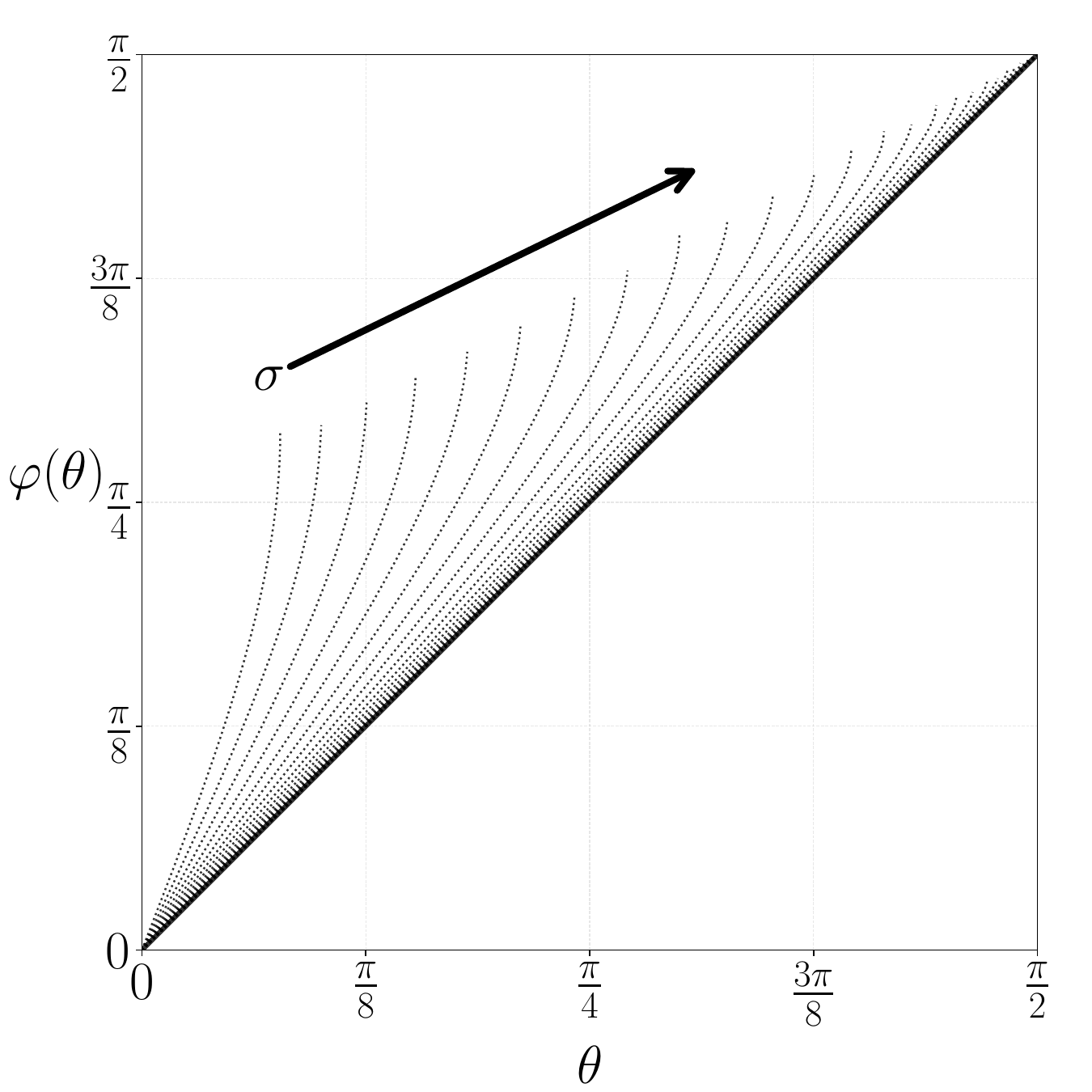}}}%=
    \subfloat[]{{\hspace{3mm}\includegraphics[width=0.45\columnwidth, trim={0cm 1mm 0cm 0cm}, clip]{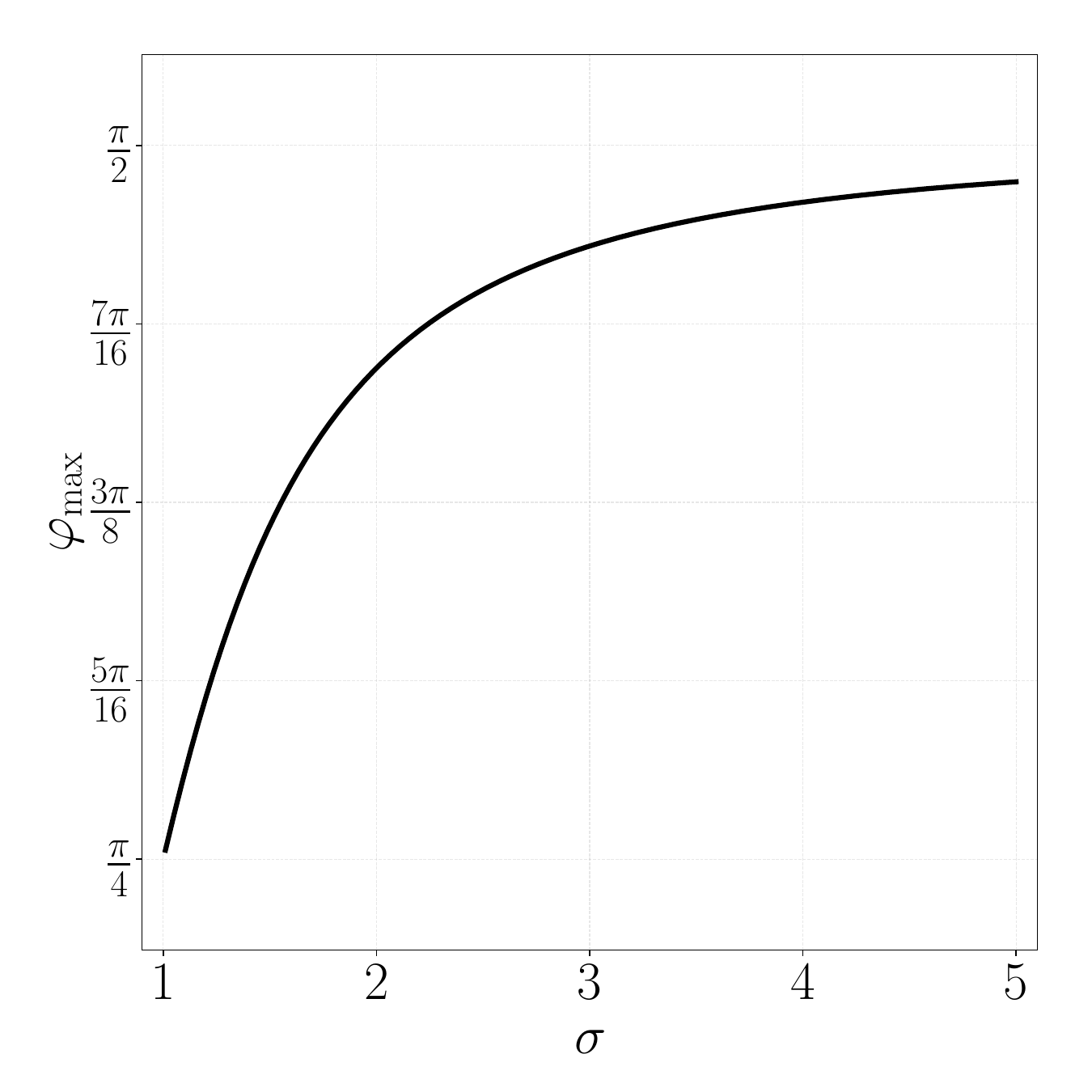}}}%=
    \vspace{-1mm}\caption{(a) Visualisation of the extent of the angular deviation plotted for a range of $\sigma$, with $\theta$ being the implemented wavelet input angle, and $\varphi\left(\theta\right)$, the actual kernel angle. As $\sigma \to \infty$, the plot approaches the ideal $\varphi\left(\theta\right)=\theta$ line. (b) Maximum permitted kernel principal axis angle, $\varphi\left(\sigma,\arccos\left(\frac{2\sigma^{2}}{\sigma^{4}+1}\right)\right)$, for a given principal axis standard deviation, $\sigma$. \vspace{-4mm}}%
    \label{fig:max_phi}
\end{figure}

\vspace{-2.25mm} \small
\begin{align}
    \lim_{\sigma \to \infty}\left\{\varphi\left(\sigma, \kappa\right)\right\} &= \frac{1}{2}\arctan{\left(-\frac{2\tan{\kappa}}{\tan^2{\kappa} - 1}\right)}\\&=\frac{1}{2}\arctan{\left(\tan{2\kappa}\right)} = \kappa,
\end{align}
\normalsize \vspace{-2.65mm}

as expected, whereas as $\sigma \to 1^+$, $\varphi\left(\sigma, \kappa\right)$ increasingly deviates from $\kappa$. Figure~\ref{fig:max_phi} presents a visualisation of the extent of this angular deviation. Thus, to account for the discrepancy between the expected and resulting angle, let $\theta = \varphi\left(\kappa, \sigma\right)$ in Eq.~\eqref{solution2} (where $\theta$ is the intended angle of the principal axis), then $\kappa\left(\sigma, \theta\right) = \varphi^{-1}\left(\sigma, \theta\right)$. The derived wavelet and the corresponding Cohen's class kernel are provided in Eqs. \eqref{window_function_CFWT} and \eqref{pi_relationship}, respectively in the main text.

\vspace{-0mm}

\vspace{-2mm}\subsection{Derivation of the Reconstructive Ideal Fractional Transform}
\label{sm:deriving_RIFT}

 Taking the probabilistic model and discretisation provided in the main text:

\vspace{-2.25mm} \small  \begin{align}
    p\left(\mathbf{\Phi} \mid \text{\bf{ITFR}}\right) &\propto \prod_{n=1}^{N} \prod_{m=1}^{M} \prod_{i=1}^{I} \prod_{j=1}^{J} e^{-\bar{P}_{i, j}^{(n, m)} \left(R_{i, j}^{(n, m)} - \Phi_{i, j}^{(n, m)}\right)^2} \notag\\
    &=\prod_{n, m, i, j} e^{-\bar{P}_{i, j}^{(n, m)} \left(\left[\text{ITFR} * \Pi^{(n, m)}\right]_{i, j} - \Phi_{i, j}^{(n, m)}\right)^2},
\end{align}\normalsize \vspace{-2.25mm} 

\noindent the negative log-likelihood becomes:

\vspace{-2.25mm} \small  \begin{align}
    \mathcal{L}\left(g_{i, j}\right) &= \sum_{n, m, i, j} \bar{P}_{i, j}^{(n, m)} \left(\left[g * \Pi^{(n, m)}\right]_{i, j} - \Phi_{i, j}^{(n, m)}\right)^2,
\end{align}\normalsize \vspace{-2.25mm} 

\noindent with $g_{i,j} = \text{ITFR}_{i, j}$ for convenience. Taking gradients yields:

\vspace{-2.25mm} \small  \begin{align}
    &\frac{\partial \mathcal{L}\left(g_{u, v}\right)}{\partial g_{u, v}}  \propto \sum_{n, m, i, j} \bar{P}_{i, j}^{(n, m)}\left(\left[g * \Pi^{(n, m)}\right]_{i, j} - \Phi_{i, j}^{(n, m)}\right) \notag\\& \hspace{15mm}\times \frac{\partial}{\partial g_{u, v}} \left\{\left[g * \Pi^{(n, m)}\right]_{i, j} - \Phi_{i, j}^{(n, m)}\right\} \notag\\
    &= \sum_{n, m, i, j} \bar{P}_{i, j}^{(n, m)}\left(\left[g * \Pi^{(n, m)}\right]_{i, j} - \Phi_{i, j}^{(n, m)}\right) \Pi_{i-u, j-v}^{(n, m)}.
\end{align}\normalsize \vspace{-2.25mm} 

\noindent To simplify the expression, it can be assumed that the entropic weighting function, $\bar{P}_{i, j}^{(n, m)}$, remains constant for each convolution with respect to pixel $(u, v)$, such that $\bar{P}_{i, j}^{(n, m)} = \bar{P}_{u, v}^{(n, m)}$. This is reasonable given that the weighting function is known to vary smoothly in $(i, j)$ relative to the kernel standard deviations. Given also that $\Pi_{i, j}^{(n, m)} = \Pi_{-i, -j}^{(n, m)}$ (the Gaussian kernel is symmetric):

\vspace{-2.25mm} \small  \begin{align}
    &=  \sum_{n, m, i, j}\bar{P}_{u, v}^{(n, m)}\left(\left[g * \Pi^{(n, m)}\right]_{i, j} - \Phi_{i, j}^{(n, m)}\right) \Pi_{u - i, v - j}^{(n, m)} \notag\\
    &=  \sum_{n, m}\bar{P}_{u, v}^{(n, m)}\left(\left[g * \Pi^{(n, m)} * \Pi^{(n, m)}\right]_{u, v} \right. \notag \\
    &\hspace{25mm} \left. - \left[\Phi^{(n, m)} * \Pi^{(n, m)}\right]_{u, v}\right).
\end{align}\normalsize \vspace{-2.25mm} 

Thus, setting the gradient to $0$:

\vspace{-2.25mm} \small  \begin{align}
    &\sum_{n, m}\bar{P}_{u, v}^{(n, m)}\left[\hat{g} * \Pi^{(n, m)} * \Pi^{(n, m)}\right]_{u, v} =  \notag \\ &\hspace{15mm}\sum_{n, m}\bar{P}_{u, v}^{(n, m)}\left[\Phi^{(n, m)} * \Pi^{(n, m)}\right]_{u, v},
\end{align} \normalsize \vspace{-2.25mm} 

\noindent where $\hat{g}$ is the unconstrained maximum likelihood solution for the model. Thus, the following normal equations are obtained: 

\vspace{-2.25mm} \small  \begin{align}
    \left[\hat{g} *_{SV} \Psi\right]_{u, v} &= \Phi_{T, u, v}, \label{spatial_varying_sol}
\end{align} \normalsize \vspace{-2.25mm} 

\noindent where $*_{SV}$ denotes a spatially varying convolution, with:

\vspace{-2.25mm} \small  \begin{align}
\left[\hat{g} *_{SV} \Psi\right]_{u, v} &= \sum_{i,j}\hat{g}_{u - i, v - j} \, \Psi_{i, j; u, v} \\
    \Psi_{i, j; u, v} &= \sum_{n, m}\bar{P}_{u, v}^{(n, m)}\left[\Pi^{(n, m)} * \Pi^{(n, m)}\right]_{i, j} \\ 
    \Phi_{T, u, v} &= \sum_{n, m}\bar{P}_{u, v}^{(n, m)}\left[\Phi^{(n, m)} * \Pi^{(n, m)}\right]_{u, v}\,.
\end{align}\normalsize\vspace{-2.65mm}

\begin{figure*}
    \centering
    \subfloat[]{{\hspace{-0mm}\includegraphics[width=0.22\textwidth, trim={0cm 3mm 0cm 0cm}, clip]{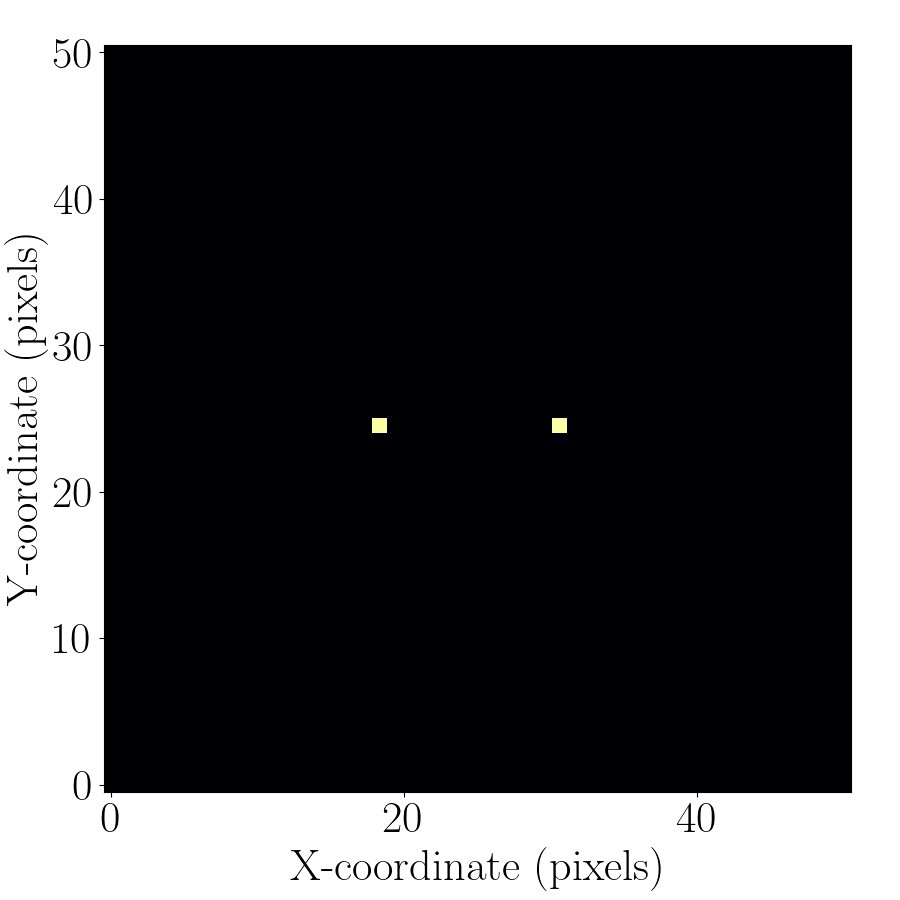}}}%=
    \subfloat[]{{\hspace{3mm}\includegraphics[width=0.22\textwidth, trim={0cm 3mm 0cm 0cm}, clip]{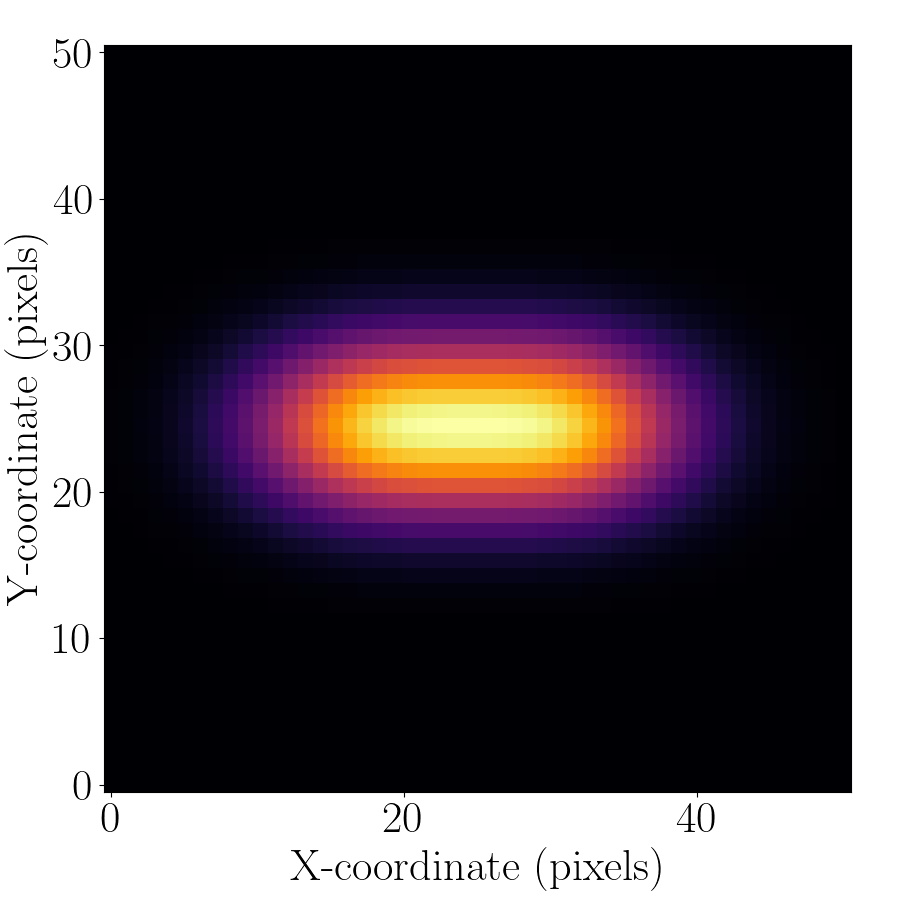}}}%=
    \subfloat[]{{\hspace{3mm}\includegraphics[width=0.22\textwidth, trim={0cm 3mm 0cm 0cm}, clip]{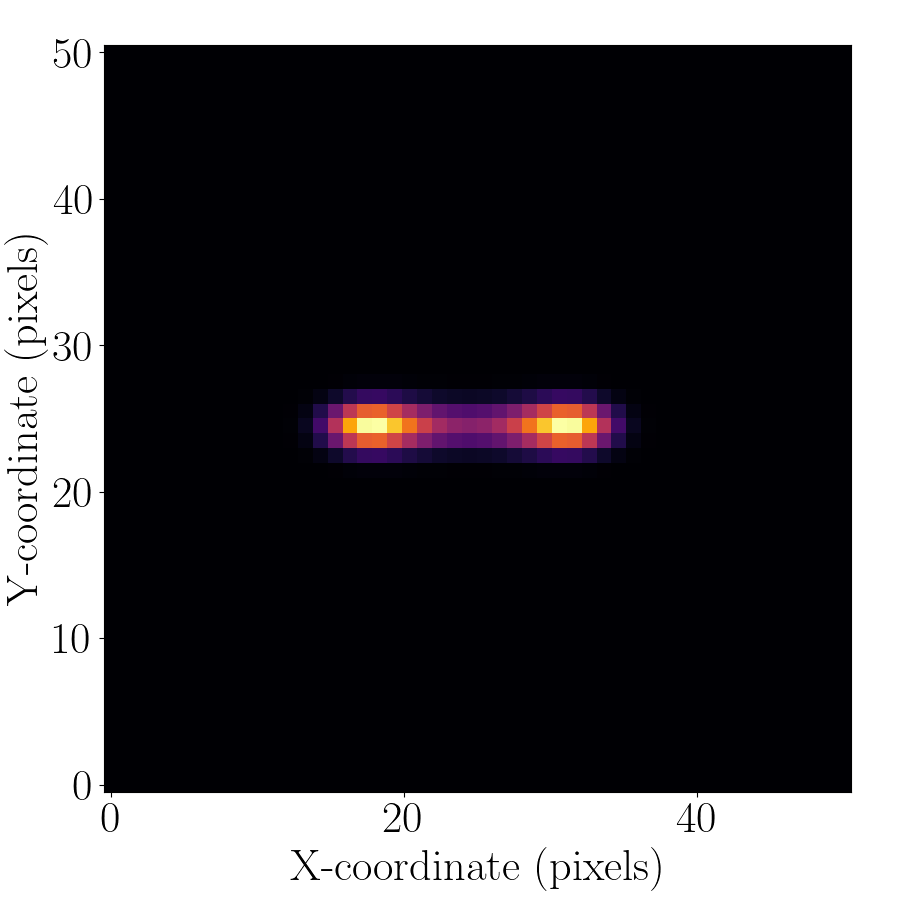}}}%=
    \subfloat[]{{\hspace{3mm}\includegraphics[width=0.22\textwidth, trim={0cm 3mm 0cm 0cm}, clip]{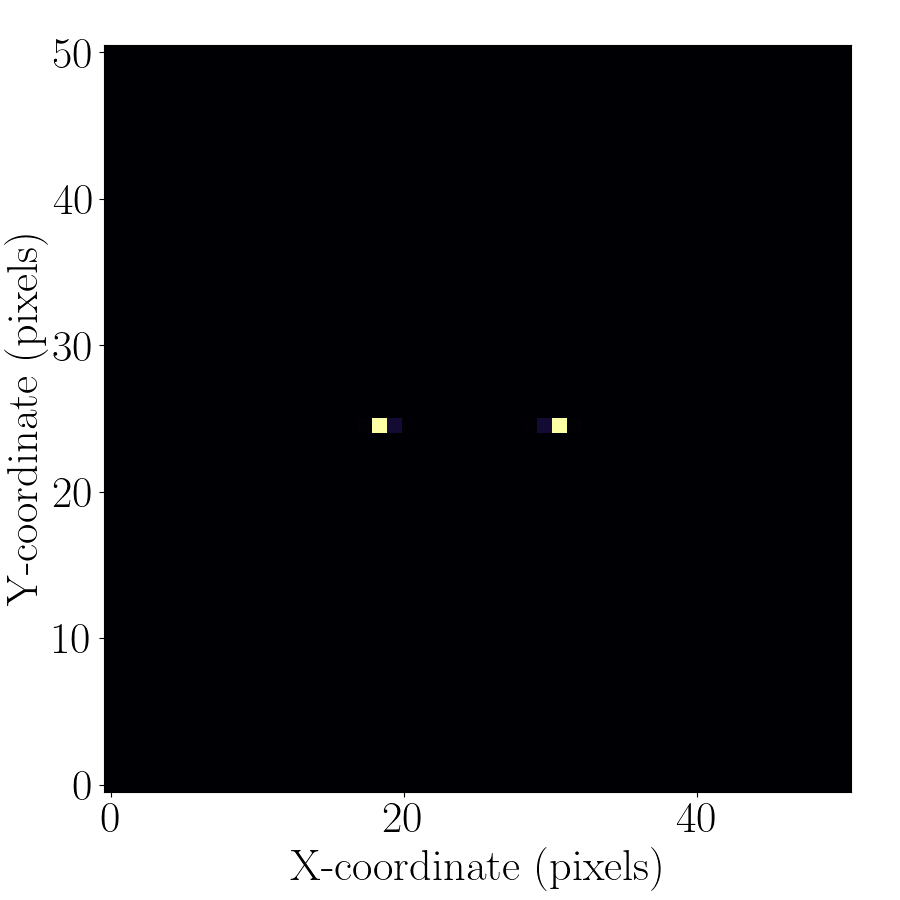}}}%=
    \vspace{-1mm}\caption{A demonstration of the iterative Lucy--Richardson algorithm for two closely spaced delta functions (T--F events), $\text{ITFR} = \delta\left(x-18, y-24\right) + \delta\left(x-30, y-24\right)$, where $\delta\left(x-a, y-b\right)$ is the discrete delta function ($1$ for $x=a, y=b$, and $0$ otherwise), and with a Gaussian PSF with $\sigma_x = 6$, and $\sigma_y = 4$. (a) is the ITFR, (b) is the result of the convolution between the Gaussian PSF and the ITFR (representative of the CFWT), (c) corresponds to 100 iterations, and (d) corresponds to the reconstructed ITFR after 10000 iterations. \vspace{-3mm}}%
    \label{fig:LR_figure}
\end{figure*}

The final derived generalised form is provided in Section~\ref{section:probability} of the main text. Likewise, Fig.~\ref{fig:LR_figure} demonstrates the iterative behaviour of the Lucy--Richardson deconvolution algorithm for a simple Gaussian PSF example. 

\vspace{-0mm}

\vspace{-2mm}\subsection{Block-wise Lucy--Richardson implementation details}
\label{block-wise implementation}

Let $(l, h)$, with $l \in \{1, ...,L\}$ and $h \in \{1, ...,H\}$, index each segment (block), and let $(u,v) \in B^{(l,h)}$ denote that $(u,v)$ lies inside block $(l,h)$. For any array $\mathbf{X}=[X_{u,v}]$, we write $\mathbf{X}^{(l,h)}$ for its restriction to block $(l,h)$. Each segment $(l, h)$ is computed according to:

\vspace{-2.65mm}
\small
\begin{align}
\label{rift_solution_block}
\hat{\mathbf{ITFR}}_{k+1}^{(l,h)}
&=
\left(
\frac{\mathbf{\hat{\Phi}_T}^{(l,h)}}{
      \hat{\mathbf{ITFR}}_{k}^{(l,h)} * \hat{\mathbf{\Psi}}^{(l,h)}}
*
\tilde{\hat{\mathbf{\Psi}}}^{(l,h)}
\right)
\odot
\hat{\mathbf{ITFR}}_{k}^{(l,h)} \notag
\\
&\hspace{20mm}\odot\;
\left[
1 - \lambda
\operatorname{div}\!\left(
\tfrac{\nabla \hat{\mathbf{ITFR}}_{k}^{(l,h)}}
     {\|\nabla \hat{\mathbf{ITFR}}_{k}^{(l,h)}\|_2}
\right)
\right]^{-1},
\end{align}
\normalsize
\vspace{-2.65mm}

\noindent where $\hat{\mathbf{\Psi}}^{(l,h)} = [\hat{\Psi}_{i,j}^{(l,h)}]_{1\le i\le I_{\Psi}, 1\le j\le J_{\Psi}}$ is the locally
space-invariant PSF for that block, and $\tilde{\hat{\mathbf{\Psi}}}^{(l,h)}$ is the T--F flipped PSF defined by $\tilde{\hat{\Psi}}_{i,j}^{(l,h)} := \hat{\Psi}_{-i,-j}^{(l,h)}$, with:

\vspace{-0.65mm}
\small
\begin{align}
\label{eq:psihat_block_compact}
\hat{\Psi}_{i, j}^{(l, h)} &=
\sum_{n, m}
\bigl\langle \bar{P}^{(n,m)} \bigr\rangle^{(l,h)}
\left[\Pi^{(n, m)} * \Pi^{(n, m)}\right]_{i, j},
\end{align}
\normalsize
\vspace{-1.65mm}

\noindent and:

\vspace{-2.65mm}
\small
\begin{align}
\label{eq:block_mean_P}
\bigl\langle \bar{P}^{(n,m)} \bigr\rangle^{(l,h)}
=
\frac{1}{|B^{(l,h)}|}
\sum_{(u,v)\in B^{(l,h)}} \bar{P}_{u, v}^{(n, m)},
\end{align}
\normalsize
\vspace{-2.65mm}

\noindent where $|B^{(l,h)}|$ is the number of pixels in block $(l,h)$. $\hat{\boldsymbol{\Phi}}_T^{(l,h)}$ is the data term in \eqref{kernels}, but restricted to block $(l,h)$ such that $\hat{\boldsymbol{\Phi}}_T^{(l,h)} := \boldsymbol{\Phi}_T$, for $(u,v)\in B^{(l,h)}$. Note that if $\bar{P}_{u, v}^{(n,m)}$ is sufficiently sparse at each pixel
(i.e., concentrated on a small subset of $(n,m)$), then $\hat{\Phi}_{T, u, v}$ can be approximated as:

\vspace{-1.65mm}
\small
\begin{align*}
\hat{\Phi}_{T, u, v}^{(l,h)}
&=
\sum_{n, m}
\bigl\langle \bar{P}^{(n,m)} \bigr\rangle^{(l,h)}
\left[\Phi^{(n, m)} * \Pi^{(n, m)}\right]_{u, v}, \, (u,v)\in B^{(l,h)}.
\end{align*}
\normalsize
\vspace{-1.65mm}

In addition, if $\bar{P}_{u, v}^{(n,m)}$ is approximately spatially
invariant over the whole image, then the LR process can be
approximated by a single spatially invariant LR deconvolution (i.e., with $L=H=1$).

\vspace{-0mm}

\subsection{Proof of RIFT's time and frequency shift equivariance}
\label{sm:time/freq_shift}

RIFT is built from the CFWT bank $\Phi_{\sigma,\theta}(\omega,t)
= |[z * W_{\sigma,\theta,\omega}](t)|^2$ followed by local entropic weighting
and a positivity-constrained LR--TV inference step. We provide a brief proof
sketch of shift equivariance of the CFWT stage; the remaining stages preserve
this property by construction up to discretisation.

\noindent\textit{Time shift.}
Let $z_\tau(t)=z(t-\tau)$. Then
$[z_\tau * W_{\sigma,\theta,\omega}](t)=[z * W_{\sigma,\theta,\omega}](t-\tau)$,
hence
$\Phi^{(\tau)}_{\sigma,\theta}(\omega,t)=\Phi_{\sigma,\theta}(\omega,t-\tau)$.

\noindent\textit{Frequency shift.}
Let $z_\nu(t)=z(t)e^{j\nu t}$. Using the modulation property inside the CFWT
definition yields
$[z_\nu * W_{\sigma,\theta,\omega}](t)=e^{j\nu t}[z * W_{\sigma,\theta,\omega-\nu}](t)$,
and therefore
$\Phi^{(\nu)}_{\sigma,\theta}(\omega,t)=\Phi_{\sigma,\theta}(\omega-\nu,t)$. 

Thus, the RIFT output $\widehat{I}(\omega,t)$ is time- and frequency-shift
equivariant up to discretisation.

\begin{table*}[!t]
\centering
\scriptsize
\setlength{\tabcolsep}{9pt}
\begin{tabular}{l|cccccccccc}
\hline
\multicolumn{11}{l}{\textbf{Ranking stability across $\mathbf{\sigma_I}$
 (ITFR modelling ablation)}}\\
\cline{1-11}
 $\sigma_I$ & AOK & CWT & Choi–Williams & RIFT & Reassignment & S-Method & SET & SST & Spline-RIFT & WVD \\ \hline
 0.0 & 0.4747 & 0.1905 & 0.2097 & 0.7675 & 0.7766 & 0.1737 & 0.4745 & 0.3094 & \textbf{0.9629} & 0.0000 \\ 
 0.5 & 0.5192 & 0.2193 & 0.2378 & 0.8142 & 0.7257 & 0.2004 & 0.3940 & 0.3275 & \textbf{0.8593} & 0.0000 \\ 
 1.0 & 0.5513 & 0.2513 & 0.2658 & \textbf{0.8392} & 0.6349 & 0.2308 & 0.2992 & 0.3460 & 0.7197 & 0.0000 \\ 
 1.5 & 0.5800 & 0.2863 & 0.2936 & \textbf{0.8539} & 0.5823 & 0.2636 & 0.2455 & 0.3636 & 0.6257 & 0.0000 \\ 
 2.0 & 0.6128 & 0.3283 & 0.3248 & \textbf{0.8677} & 0.5517 & 0.3032 & 0.2131 & 0.3849 & 0.5589 & 0.0058 \\ 
 2.5 & 0.6510 & 0.3820 & 0.3642 & \textbf{0.8799} & 0.5413 & 0.3546 & 0.2072 & 0.4158 & 0.5200 & 0.0289 \\ 
 3.0 & 0.6899 & 0.4385 & 0.4036 & \textbf{0.8905} & 0.5409 & 0.4088 & 0.2139 & 0.4485 & 0.4976 & 0.0582 \\ 
\hline\hline
\end{tabular}
\vspace{1mm}\caption{Per-$\sigma_I$ combined scores. For each $\sigma_I$, we first sum each metric across SNRs, then min--max normalise across methods within that $\sigma_I$ (inverting JS since lower is better), and finally average the three normalised metrics to obtain the combined score. Bold indicates the best method per $\sigma_I$.}
\label{tab:combined_sigma}
\end{table*}

\begin{figure*}[!t]
  \centering

  \vspace{-0mm}
  \vspace{-5mm}\subfloat[]{{\hspace{-0mm}\includegraphics[width=0.25\textwidth, trim={0cm 3mm 2cm 0cm}, clip]{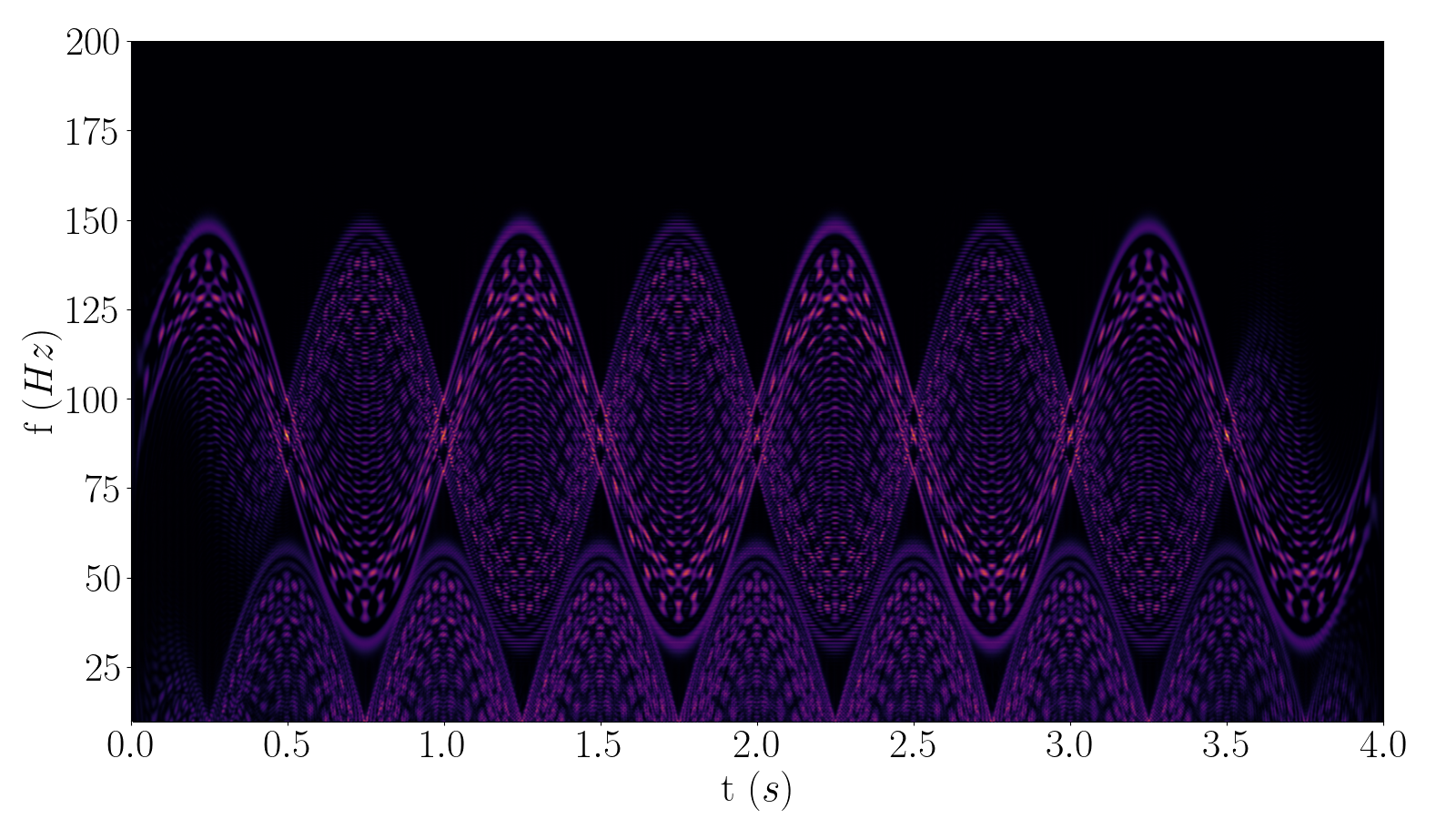}}}%=
  \subfloat[]{{\hspace{-0mm}\includegraphics[width=0.25\textwidth, trim={0cm 3mm 2cm 0cm}, clip]{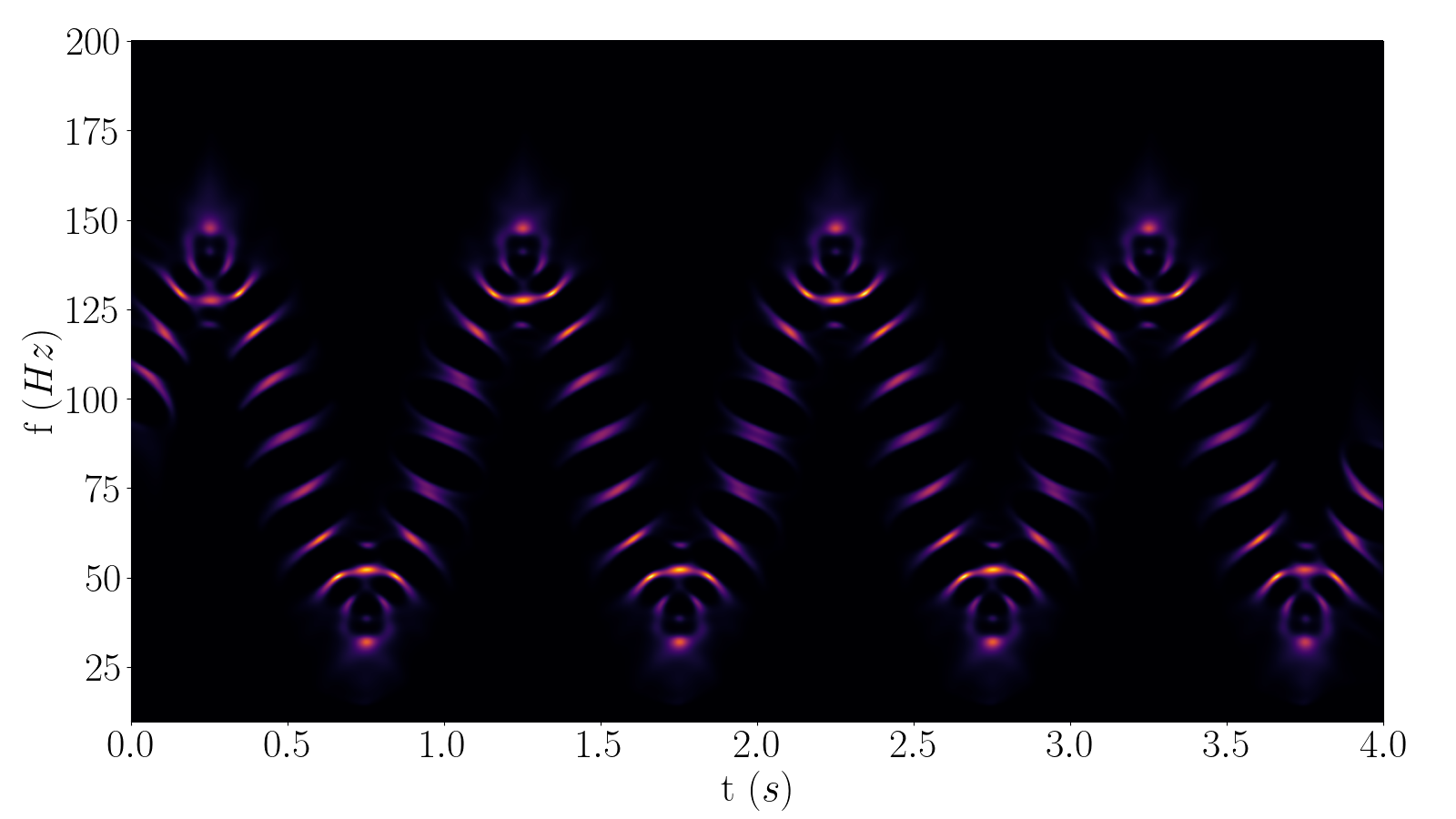}}}%=
  \subfloat[]{{\hspace{-0mm}\includegraphics[width=0.25\textwidth, trim={0cm 3mm 2cm 0cm}, clip]{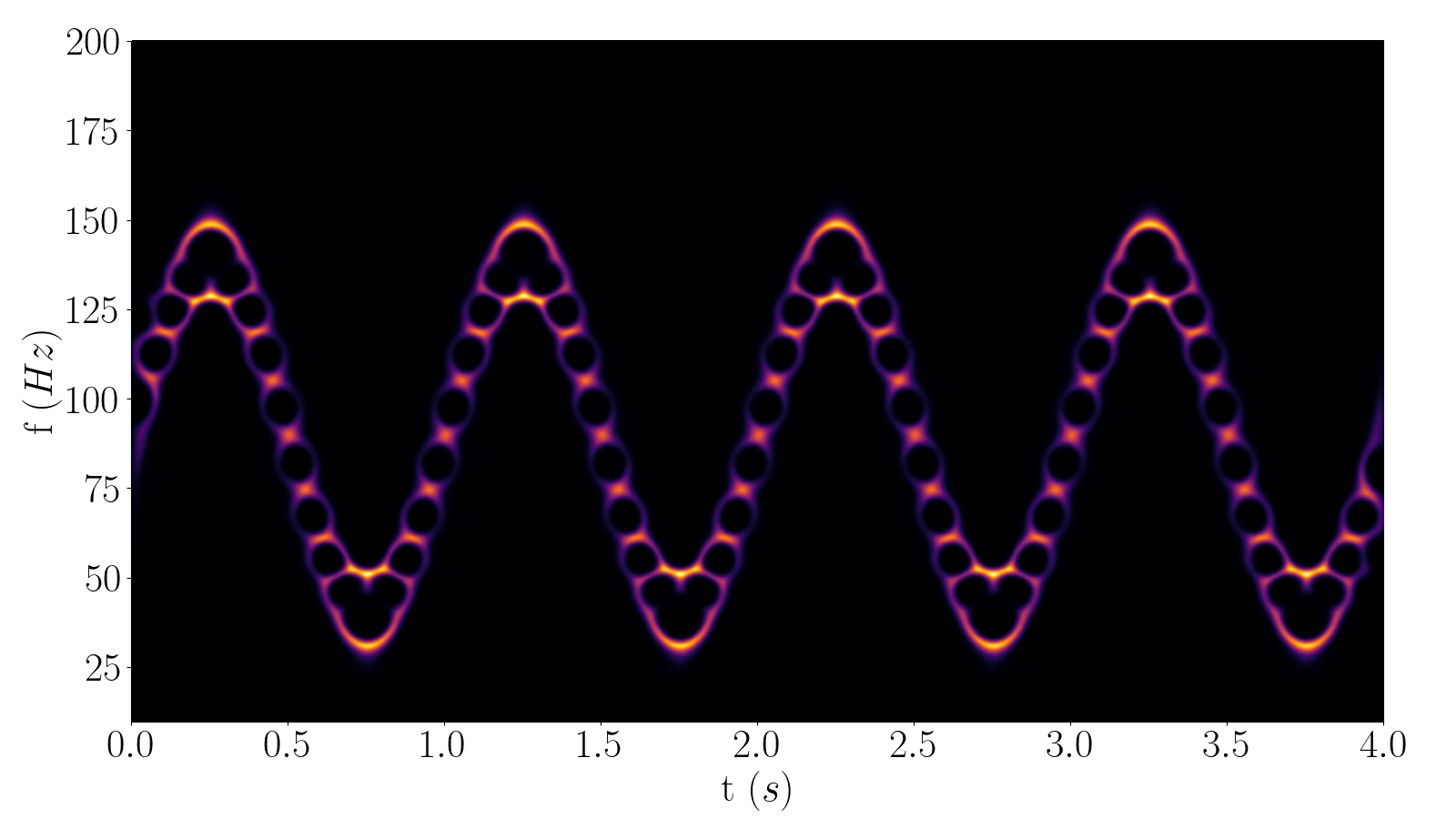}}}%=
  \subfloat[]{{\hspace{-0mm}\includegraphics[width=0.25\textwidth, trim={0cm 3mm 2cm 0cm}, clip]{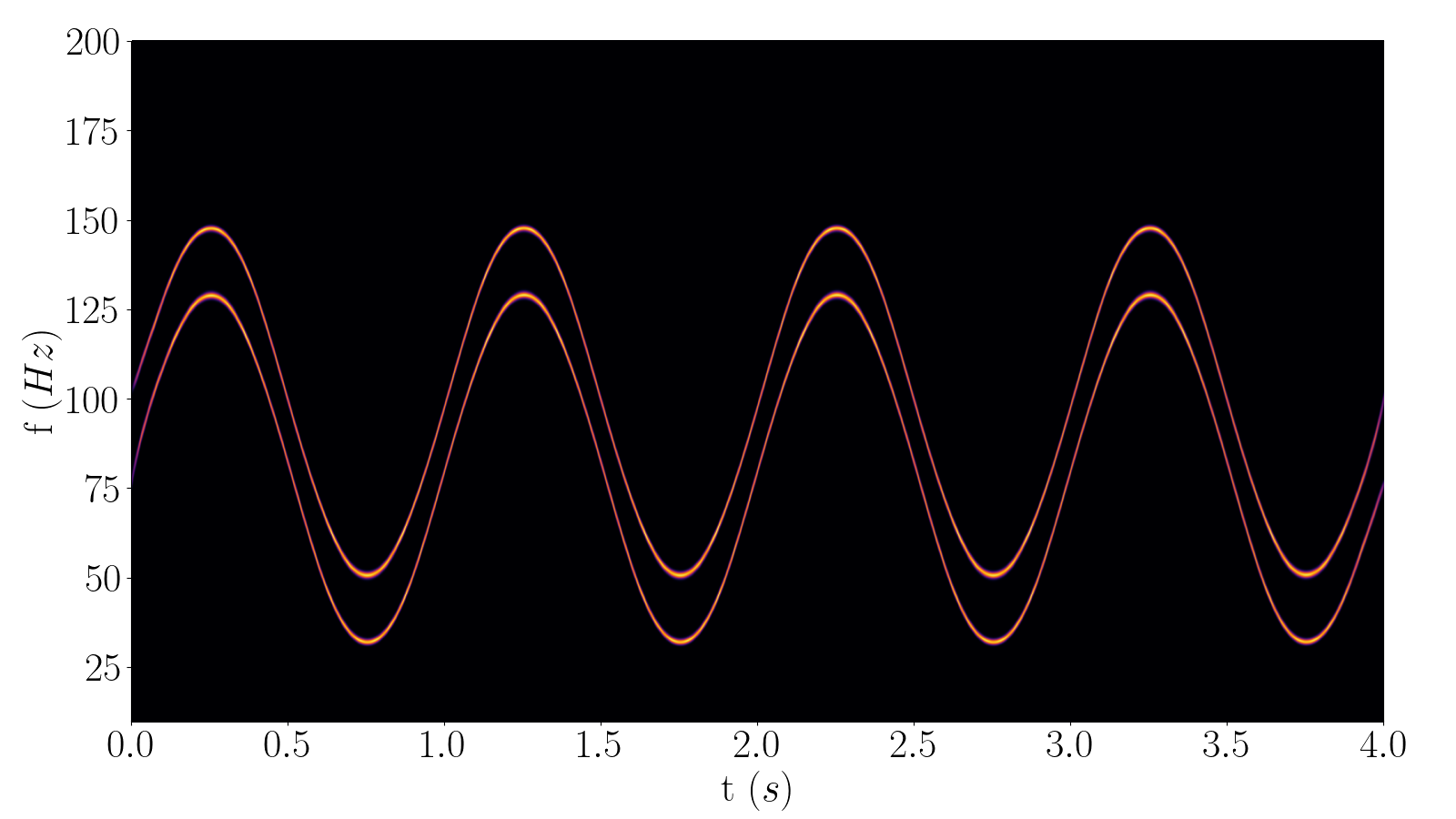}}}%=
  \vspace{-1mm}

  \par\vspace{2mm}
  \centering
  \scriptsize
  \setlength{\tabcolsep}{4pt}
  \begin{tabular}{l|ccccccccccc}
  \hline
  \multicolumn{12}{l}{\textbf{Alpha sweep on $x_1(t)$ at $\infty$ dB}}\\
  \cline{1-12}
  Metric & $\alpha=-25$ & $\alpha=-20$ & $\alpha=-15$ & $\alpha=-10$ & $\alpha=-5$ & $\alpha=0$ & $\alpha=5$ & $\alpha=10$ & $\alpha=15$ & $\alpha=20$ & $\alpha=25$ \\ \hline
  Bhattacharyya overlap $(\uparrow)$ & 0.225 & 0.237 & 0.256 & 0.288 & 0.354 & 0.561 & 0.720 & 0.704 & 0.690 & 0.669 & 0.652 \\
  Jensen--Shannon divergence $(\downarrow)$ & 0.560 & 0.552 & 0.540 & 0.520 & 0.479 & 0.346 & 0.232 & 0.239 & 0.246 & 0.261 & 0.272 \\
  Ridge Energy Ratio $(\uparrow)$ & 0.047 & 0.049 & 0.053 & 0.059 & 0.074 & 0.135 & 0.230 & 0.269 & 0.293 & 0.296 & 0.294 \\
  \hline
  Combined $(\uparrow)$ & 0.000 & 0.019 & 0.050 & 0.100 & 0.206 & 0.562 & 0.912 & 0.945 & \textbf{0.960} & 0.936 & 0.910 \\
  \hline\hline
  \end{tabular}

  \caption{Top: a visualisation of the effect of varying $\alpha$ on the entropic filtering process for the signal $x_1(t)$. (a) presents the WVD ($WVD > 0$ shown for convenience), and (b) presents the RIFT T--F representation for $\alpha=-15$, isolating cross-term components. (c) presents the RIFT T--F representation for $\alpha=0$, illustrating the result if cross-term suppression is not employed, and (d) employs $\alpha=15$, resulting in the desired cross-term free estimated ITFR. Bottom: metric T--F evaluation on $x_1(t)$: Bhattacharyya overlap$(\uparrow)$, Jensen--Shannon divergence$(\downarrow)$, Ridge Energy Ratio $(\uparrow)$, and Combined (↑) across $\alpha$ (best shown in bold).}
  \label{fig:ALPHA_SWEEP}
\end{figure*}

\end{document}